\documentclass[aps,prx,citeautoscript,footinbib,eqsecnum,twocolumn]{revtex4-2}
\usepackage{feynmp}

\usepackage{subcaption}
\usepackage{amsmath}
\usepackage{amssymb}
\usepackage{bbm}
\usepackage{braket}
\usepackage{xcolor}
\usepackage{pifont}
\usepackage{slashed}
\usepackage[mathscr]{euscript}
\usepackage[shortlabels]{enumitem}
\usepackage{tikz-feynman}
\usetikzlibrary{arrows}
\usepackage[papersize={8.5in,11in}]{geometry}

\allowdisplaybreaks

\usepackage{graphicx}
\usepackage{subcaption}
\usepackage[colorlinks=true]{hyperref}
\usepackage{comment}
\usepackage{multirow}

\renewcommand{\approx}{\simeq}
\renewcommand{\Re}{\text{Re}}
\renewcommand{\Im}{\text{Im}}

\renewcommand{\vec}[1]{\boldsymbol{#1}}
\newcommand{\vn}[1]{|\boldsymbol{#1}|}

\definecolor{wrongultramarine}{rgb}{1,0.5,0}

\newcommand{\rd}{{\rm d}}
\newcommand{\sgn}{{\rm sgn\,}}
\newcommand{\Tr}{{\rm \, Tr\,}}
\newcommand{\calN}{{\mathcal N}}
\newcommand{\calG}{{\mathcal G}}
\newcommand{\calF}{{\mathcal F}}

\newcommand{\calH}{{\mathcal H}}
\newcommand{\calO}{{\mathcal O}}
\newcommand{\verteq}{\rotatebox{90}{$\,=$}}
\newcommand{\vertdots}{\rotatebox{90}{$\,\dots$}}
\newcommand{\update}[1]{{ {#1}}}

\hypersetup{
    bookmarks=true,         
    unicode=false,          
    pdftoolbar=true,        
    pdfmenubar=true,        
    pdffitwindow=false,     
    pdfstartview={FitH},    
    pdfsubject={},   
    pdfcreator={},   
    pdfproducer={}, 
    pdfkeywords={} {} {}, 
    pdfnewwindow=true,      
    colorlinks=true,       
    linkcolor=blue, 
    citecolor=blue,        
    filecolor=magenta,      
    urlcolor=blue           
}

\tikzset{
  mid arrow/.style={postaction={decorate,decoration={
        markings,
        mark=at position .575 with {\arrow[#1]{stealth}}
      }}},
  near arrow/.style={postaction={decorate,decoration={
        markings,
        mark=at position .275 with {\arrow[#1]{stealth}}
      }}},
   far arrow/.style={postaction={decorate,decoration={
        markings,
        mark=at position .800 with {\arrow[#1]{stealth}}
      }}},
   boson/.style={decorate, draw=black,
    decoration={snake,amplitude=1pt, segment length=5pt},
      },
   mid triangle/.style={postaction={decorate,decoration={
        markings,
        mark=at position .575 with {\arrow[#1]{triangle 45}}
      }}}
}

\begin{document}

\title{Fluctuation Spectrum of Critical Fermi Surfaces}
\begin{abstract}
  We investigate the low-energy effective theory of a Fermi surface coupled to an Ising-nematic quantum critical point in (2+1) spacetime dimensions with translation symmetry. We formulate the system using the large $N$ Yukawa-SYK model, whose saddle point is described by the Migdal-Eliashberg equations. The low-energy physics can be revealed by studying the Gaussian fluctuation spectrum around the saddle point, which is generated by the Bethe-Salpeter kernel $K_\text{BS}$. Based on the Ward identities, we propose an inner product on the space of two point functions, which reveals a large number of soft modes of $K_\text{BS}$. These soft modes parameterize deformation of the Fermi surface, and their fluctuation eigenvalues describe their decay rates. We analytically compute these eigenvalues for a circular Fermi surface, and we discover the odd-parity modes to be parametrically longer-lived than the even-parity modes, due to the kinematic constraint of fermions scattering on a convex FS. The sign of the eigenvalues signals an instability of the Ising-nematic quantum critical point at zero temperature for a convex Fermi surface. At finite temperature, the system can be stabilized by thermal fluctuations of the critical boson. We derive an effective action that describes the soft-mode dynamics, and it leads to a linearized Boltzmann equation, where the real part of the soft-mode eigenvalues can be interpreted as the collision rates. The structure of the effective action is similar to the theory of linear bosonization of a Fermi surface. As an application of the formalism, we investigate the hydrodynamic transport of non-Fermi liquid. Analyzing the Boltzmann equation, we obtain a conventional hydrodynamic transport regime and a tomographic transport regime. In both regimes, the conductance of the system in finite geometry can be a sharp indicator for the soft-mode dynamics and non-Fermi liquid physics.
\end{abstract}

\author{Haoyu Guo}
\affiliation{Laboratory of Atomic and Solid State Physics, Cornell University,
142 Sciences Drive, Ithaca NY 14853-2501, USA}

\date{\today}

\maketitle
\tableofcontents

\section{Introduction}

    Landau's Fermi liquid (FL) theory \cite{LDLandau1956,AAAbrikosov1963} is the textbook theory for describing the low-energy limit of interacting fermionic systems at finite density. The central assumption of FL theory is the existence and longevity of quasiparticles, which is established for short-range interactions \cite{PColeman2015}. However, when the FL is subject to long-range fluctuations, such as critical fluctuations near a quantum critical point (QCP) \cite{JAHertz1976,AJMillis1993} or emergent gauge fields  in spin liquid or quantum Hall systems \cite{PALee1989,BIHalperin1993}, the scattering rate $\Gamma_k$ of the would-be quasiparticles can become parametrically larger than its energy $\varepsilon_k$, and quasiparticles are destroyed. The strongly correlated metal that arises is called a non-Fermi liquid (NFL).

   The critical Fermi surface is a toy model for studying NFLs in (2+1) spacetime dimensions, consisting of a Fermi surface (FS) coupled to a critical (gapless) bosonic field \cite{SSLee2018,PALee1989,AJMillis1993,JPolchinski1994, BIHalperin1993,YBKim1994a,CNayak1994,SSLee2009,MAMetlitski2010,
    DFMross2010,SSur2014,MAMetlitski2015,SAHartnoll2014,
    AEberlein2017,THolder2015,THolder2015a,ALFitzpatrick2014,
    JADamia2019,JADamia2020,JADamia2021,SPRidgway2015,AAPatel2018b,
    DChowdhury2018c,EGMoon2010,AAbanov2020,YMWu2020,AVChubukov2020,
    XWang2019,AKlein2020,OGrossman2021,DChowdhury2020,IEsterlis2019,
    DHauck2020,YWang2020a,EEAldape2022,AAPatel2017a,AAPatel2019,
    VOganesyan2001,AAPatel2018,AVChubukov2017,DLMaslov2017,
    SLi2023,Iesterlis2021,HGuo2022a,HGuo2024,YBKim1995a,LVDelacretaz2022a,
    SEHan2023,UMehta2023,DVElse2021a,DVElse2021,ZDShi2022,ZDShi2023,TPark2023,IMandal2022,KRIslam2023}. It is believed to capture essential physics related to half-filled Landau level, quantum spin liquids with spinon FS, and metallic quantum criticality.
     Starting from the seminal work by Sung-Sik Lee \cite{SSLee2009}, various large $N$ approaches have been proposed to analytically control the model (see \cite{SSLee2018} for a review). The leading order results of the different approaches conform to the Migdal-Eliashberg (ME) theory \cite{ABMigdal1958,GMEliashberg1960,FMarsiglio2020,MProtter2021}, which was originally proposed by Migdal and Eliashberg to describe the electron-phonon interaction. ME theory proposes a self-consistent set of integral equations to compute the fermion and the boson self-energies by ignoring the vertex corrections. The original justification of the approximation is that phonons move much slower than electrons. Similar ideas have been applied to the critical FS by assuming that the Landau-damped critical boson is also slow compared to the fermions \cite{PALee1989,BIHalperin1993,YBKim1994a,MAMetlitski2010,DFMross2010,EEAldape2022,Iesterlis2021,HGuo2022a,
    ZDShi2022,ZDShi2023,DLMaslov2010a,SSZhang2024}, and the results are consistent with Monte-Carlo simulations \cite{AKlein2020,XYXu2020}.  Recently, ME theory can be formulated as a large $N$ saddle point of Yukawa-SYK models \cite{DChowdhury2022a,Iesterlis2021,HGuo2022a,AAPatel2023,HGuo2024,ZDShi2022,ZDShi2023,EEAldape2022} by utilizing random couplings in the flavor space inspired by the Sachdev-Ye-Kitaev models \cite{SSachdev1993,AYKitaev2015,DChowdhury2022a}. In that regard, the Yukawa-SYK theory can be viewed as a systematic generalization of ME theory.

     In this paper, we study the fluctuation spectrum of the translational invariant critical FS within the Yukawa-SYK framework. From this, we try to address several important questions related to the critical FS due to coupling to a $\vec{Q}=0$ gapless boson arising from a QCP.

     \textbf{What is the effective low-energy degrees of freedom (DOF) of the critical FS and what is their dynamics?} In general, the effective degrees of freedom of a theory in the infrared (IR) can be very different from the defining degrees of freedom in the ultraviolet (UV). For example, in a system of free fermions coupled to disorders, the low-energy dynamics is described by a sigma model of the bosonic diffusons \cite{FEvers2008}.
      Returning to the critical FS, it has recently been proposed that the low-energy DOF is the local (in momentum space) density $n_\theta$ on the FS or equivalently the shape fluctuations of the FS. Recently, there are top-down approaches that try to start from this assumption and infer information about the critical FS.
      Else {\it et al.} argued that the dynamics of these deformations can be constraint by 'tHooft anomalies of $\rm{LU}(1)$ symmetry \cite{DVElse2021a,ZDShi2022,ZDShi2023,DVElse2023}. One limitation of the anomaly argument is the assumption that $n_\theta$ is exactly conserved at every point on the FS, which only holds in the IR, where the inter-patch scattering that relaxes $n_\theta$ are treated as irrelevant operators. However, these scattering events are the leading-order contributions to dissipation which differentiate between FL and NFL, and it is unclear what the crossover energy scale is above which scattering becomes important. Another recent approach to capture the effective dynamics of $n_\theta$ is to utilize co-adjoint orbit bosonization \cite{LVDelacretaz2022a,UMehta2023,SEHan2023}, which starts from the collisionless Boltzmann equation for $n_\theta$ and quantizes it, but the systematic inclusion of scattering effects into the formalism is still an ongoing effort. 

    \textbf{Is the QCP and the NFL stable?} In previous literature, it is shown that the SU(2) ferromagnetic QCP (which couples to the spins of the fermions) is unstable in (2+1)D \cite{JRech2006,AVChubukov2009}. When approaching the QCP from a FL \cite{AVChubukov2009}, the instability manifests as a Pomeranchuk instability that preempts the divergence of the spin susceptibility. At the critical point \cite{AVChubukov2009}, the instability manifests as more singular corrections to the static spin susceptibility with a wrong sign. When the coupling between the itinerant fermion and the critical fluctuation is in the charge channel, it is shown \cite{JRech2006} that the particular instability considered in \cite{JRech2006,AVChubukov2009} is absent, but it is unclear whether there can be other instability present.  

    \textbf{How to detect a translational-invariant NFL?}  Experimentally, many NFL candidates are discovered through transport measurements such as conductivity. Although there is a belief that translational invariant NFL captures the essential physics of some correlated phenomena \cite{DVElse2021a,DVElse2021,ZDShi2022,ZDShi2023}, it is difficult to identify through transport experiments that measure the uniform conductivity $\sigma(\omega)=\omega(\omega,\vec{p}=0)$. This is because $\sigma(\omega)$ is highly constrained by momentum conservation.  The non-zero overlap between the current operator and the momentum operator implies that the DC conductivity $\sigma(0)$ diverges in a translationally invariant system with a single FS. The finite-frequency optical conductivity $\sigma(\omega)$ is constrained by statements such as Kohn's theorem \cite{WKohn1961}. In particular, the proposed $|\omega|^{-2/3}$ correction \cite{YBKim1994a} to the Drude conductivity is canceled for a Galilean invariant band structure \cite{DLMaslov2011,HKPal2012,HGuo2022a,ZDShi2022,SLi2023,YGindikin2024} and the magnitude of the next-order term is much smaller and depends on the details of the dispersion.
     Therefore, efficient identification of the translationally invariant NFL requires the removal of the infinite DC Drude peak. A proposal that does not involve breaking translational symmetry is to utilize critical drag \cite{DVElse2021}, but this only reduces but does not eliminate the Drude peak. The mostly considered option that breaks translational symmetry is by umklapp or disorder, but they have their drawbacks: Umklapp is only active above a certain temperature threshold \cite{PALee2021,XWang2019}. Disorder is shown to be a relevant perturbation to the translational invariant ME saddle point, and the low-energy theory is instead a marginal Fermi liquid \cite{Iesterlis2021,HGuo2022a,AAPatel2023}, and we lose the translational invariant ME saddle point that we wish to detect. Even if we try to retain the ME saddle point by making the disorder perturbative, the DC resistivity might still be suppressed by the FS geometry \cite{DLMaslov2011}.


     In this paper, we address the key questions raised above via a bottom-up approach by studying the Yukawa-SYK model \cite{DChowdhury2022a,Iesterlis2021,HGuo2022a,AAPatel2023,HGuo2024,ZDShi2022,ZDShi2023,EEAldape2022} of the Ising-nematic quantum critical point. Our strategy is analogous to the SYK model \cite{AYKitaev2015,JMaldacena2016c}, where the $1/N$ fluctuation reveals the Schwarzian action of gravity as the low-energy theory of the model.  Conceptually, our study consists of the following steps: First, we start from the large $N$ limit, and obtain the saddle point (Migdal-Eliashberg equations). Second, we compute the $1/N$ fluctuations around the saddle, which requires diagonalizing the fluctuation kernel $K_\text{BS}$. In this step, we need to first find an inner product for the fluctuation and then compute the eigenvalues with respect to that inner product. After that, we discover that there are a large number of soft modes that describe the low-energy physics of the system, and we derive an effective action that captures the soft-mode dynamics.

     Being more specific to our problem, the kernel $K_\text{BS}$ that we need to diagonalize is the one that generates the Bethe-Salpeter equations.  The soft modes that emerge are those that parameterize the deformation of the FS, and physically they are ``soft" or slow-decaying because of the dominance of small-angle scattering near the QCP.
      The dissipation rates of these soft modes determine the stability of ME theory for the QCP.
      The Gaussian effective action for these soft modes leads to a linearized kinetic equation, which interpolates between  Landau's kinetic equation \cite{LDLandau1956,AAAbrikosov1963} in the FL far from the QCP and the Prange-Kadanoff kinetic equation \cite{REPrange1964,YBKim1995a,SEHan2023,IMandal2022,KRIslam2023} near the QCP. The kinetic equation describes the propagation of the soft modes in real space.

     Based on the kinetic equation, we propose that hydrodynamic transport \cite{RNGurzhi1968,RJaggi1991,MJMdeJong1995,MMueller2008,MMuller2009,AVAndreev2011,ALucas2016a,BNNarozhny2015,APrincipi2016,LLevitov2016,GFalkovich2017,DABandurin2016,
    JCrossno2016,LVDelacretaz2019,HGuo2017,HGuo2017a,HGuo2018,RKrishnaKumar2017} can be an efficient probe for translationally invariant NFL. Hydrodynamics naturally emerges in a translationally invariant NFL due to the large scattering rate, which conserves momentum. Furthermore, the boundary of the system provides the desired translational symmetry breaking without interfering with the bulk physics. Because momentum dissipation only happens at the boundary, the conductance scales nontrivially with the size of system cross section. In the conventional hydrodynamic regime, only momentum is conserved in the system and the conductance of the system is expected to scale as $W^2$ where $W$ is the cross-section of the system (in 2D it is a length) \cite{HGuo2017a}.     Due to the propagation of the long-lived soft modes, there can be a tomographic transport regime \cite{PLedwith2019}, where the total conductance of the system scales nontrivially with temperature $T$ and the system cross section $W$, providing a direct observable consequence of the soft modes and an unambiguous signature of NFL.

    \subsection{Organization of the paper and summary of results}\label{sec:summary}

    In Sec.~\ref{sec:model}, we review the Yukawa-SYK formulation of the Ising-nematic QCP. We consider four regimes near the QCP as shown in the $T=0$ phase diagram (Fig.~\ref{fig:pd}), where we solve the ME equations to obtain the saddle point self-energies. Regime A is a strongly coupled NFL with fermion self-energy $\Sigma(i\omega)\propto |\omega|^{2/3}\gg|\omega|$ at $T=0$, where $z$ is the dynamical exponent of the boson.  Regime $B$ is a perturbative NFL (PNFL) obtained by extending the NFL to higher energy scales such that $\Sigma(i\omega)\ll|\omega|$. Regime C is a FL near the QCP where the boson correlation length $\xi_c=m_b^{-1}$ ($m_b$ is the mass of the critical boson) is still longer than the Fermi wavelength $\lambda_F$. Regime D is a FL far from the QCP where $\xi_c< \lambda_F$.

    In Sec.~\ref{sec:fluctuation}, we formulate the fluctuation spectrum of the critical FS by studying the Gaussian $1/N$ fluctuations of the theory around the ME saddle point. The fluctuation kernel $K_\text{BS}$ is the one that generates the Bethe-Salpeter ladder, and it can be specified by Feynman diagrams, which in the perturbative regime can be related to the conventional density-of-states (DOS), Maki-Thompson (MT) and Aslamazov-Larkin (AL) diagrams.

    The kernel $K_\text{BS}$ can be diagonalized when the center-of-mass (CoM) 3-momentum $p=(i\Omega,0)$, where the Ward identities of particle and momentum conservation help to identify a pre-conditioner $M$ that transforms $K_\text{BS}$ to a kinetic operator $L\sim K_\text{BS}M$, and the density and the momentum vertices are exact zero modes of $L$. The eigenvalues of $L$ can be interpreted as the characteristic collision rates that appear on the right-hand side of a kinetic equation. Due to strong small-angle scattering, $L$ contains a large number of approximate zero modes that can be interpreted as parameterizing the shape deformations of the FS.

    In Sec.~\ref{sec:soft}, we perform a careful analysis on these approximate zero modes and resolve their eigenvalues, so they become soft modes. We present a qualitative picture which models scattering of fermions on the FS as random walk events, which can capture some aspects of the soft-mode dynamics.
    The soft eigenvalues are then calculated using perturbation theory based on a double expansion of the small scattering angle $\delta\theta\sim q/k_F$ ($q$ is the typical bosonic momentum) and the small fermion dispersion $\xi_k/(v_F k_F)$. We find that for a circular or convex FS, there is a scale separation between the even-angular-harmonic eigenvalues and the odd-angular-harmonic eigenvalues. The odd-harmonic eigenvalues are much smaller due to the kinematic constraint \cite{DLMaslov2011,HKPal2012,PJLedwith2019,PLedwith2019,XWang2019,HGuo2022a,SLi2023,YGindikin2024,JHofmann2023,JHofmann2022,ENilsson2024}  of fermion scattering on a convex FS.

    In Sec.~\ref{sec:stability}, we use the soft modes to diagnose the stability of the QCP. By relating the soft-mode eigenvalues to the Pomeranchuk order parameter of the FS, we find that the eigenvalues, as a function of the external Matsubara frequency $i\Omega$, must have positive real part after continuing to the real axis $i\Omega\to \omega+i0$ for the QCP to be stable. However, this is not satisfied for the NFL associated with Ising-Nematic QCP. The problem lies in the odd-angular-harmonic soft modes of a convex FS, whose decay rate scales as $|\Omega|^{8/3}$, which continues to a negative real part. We therefore conclude that the Ising-Nematic QCP is unstable at $T=0$. However, we find that thermal fluctuations of the QCP help to stabilize the NFL at finite temperatures, and we propose a phase diagram of the QCP in Fig.~\ref{fig:pdT}. Another possible scenario of a stable NFL is that the dynamical exponent $z_b$ of the boson is renormalized, and the stability bound is $2<z_b<8/3$.

    In Sec.~\ref{sec:eft}, we generalize the soft-mode dynamics to finite CoM momentum $p=(i\Omega,\vec{p})$ by deriving a Gaussian effective action for the soft modes. The action we obtain is similar to the result of linear bosonization \cite{FDMHaldane2005,AHoughton2000,LVDelacretaz2022a,UMehta2023,SEHan2023}, but with an additional term due to the soft-mode eigenvalues which incorporates scattering effects. The equation of motion (EoM) of the action is a linearized kinetic equation or the Boltzmann equation.  When we are deep in the FL phase, we can recover Landau's kinetic equation. When we move close to the QCP, we obtain a kinetic equation similar to the one first derived by Prange and Kadanoff \cite{REPrange1964,YBKim1995a,SEHan2023,IMandal2022,KRIslam2023}.

    In Sec.\ref{sec:hydro}, we demonstrate that these soft modes can have measurable consequences on hydrodynamic transport.
    We apply the kinetic equation from Sec.~\ref{sec:eft}  to study the hydrodynamic transport of the NFL and the FL near the Ising-nematic QCP.  We obtain a conventional hydrodynamic regime and a tomographic transport regime which features a scale-dependent viscosity. The tomographic regime appears because of the scale separation between the even-harmonic and the odd-harmonic eigenvalues.
     We infer the scaling of the conductance in a finite geometry by computing the non-local DC conductivity $\sigma(\vec{p})$ as a function of wave vector $\vec{p}$, and then the conductance $G$ can be inferred by $G\sim \sigma(|\vec{p}|\sim 1/W)$ where $W$ is the cross section of the system, and the result is summarized in Table.~\ref{tab:conductance}.

    In the appendices, we present various calculation details not elaborated in the main text. As a side story, in Appendix.~\ref{sec:conductivity} we apply the fluctuation spectrum to compute the optical conductivity of the system. The optical conductivity of a convex FS turns out to be insensitive to the soft-mode eigenvalues and is therefore not a good probe for the soft-mode physics.

  \section{The Model and its saddle point} \label{sec:model}

        We consider a concrete model of a FS described by field $\psi$ coupled to an Ising-nematic quantum critical point described by real bosonic field $\phi$ in (2+1)D spacetime. The action of the model is
        \begin{equation}\label{eq:Lagrangian}
        \begin{split}
          \mathcal{S}&=\int \rd\tau \sum_{\vec{k}} \psi_{\vec{k}}^\dagger(\tau)\left[\partial_\tau+\xi_{\vec{k}}\right]\psi_{\vec{k}}(\tau)\\
          &+ \frac{1}{2}\int\rd \tau \sum_{\vec{q}} \phi_{-\vec{q}}(\tau)\left[\omega_{\vec{q}}^2+r\right]\phi_{\vec{q}}(\tau)\\
          &+g\int\rd \tau \sum_{\vec{k},\vec{q}} \phi_{\vec{q}}(\tau) S_{\vec{k},\vec{q}} \psi^\dagger_{\vec{k}}(\tau) \psi_{\vec{k}}(\tau)\,.
        \end{split}
        \end{equation}
        Here $\xi_{\vec{k}}$ describes the fermion dispersion, and the FS is defined by $\xi_{\vec{k}}=0$. $r$ is the tuning parameter for accessing the quantum critical point (QCP).  The bare $\partial_\tau^2$ dynamics of the boson term is dropped because it is irrelevant compared to the generated Landau damping. $S$ is the form factor of the Yukawa coupling. For example, in the nematic critical point $S_{\vec{k},\vec{q}}=\cos(2\theta_k)$ where $\theta_k$ is the angular coordinate on the FS.  Most of our conclusions will be unsensitive to the form factor, so we will set $S_{\vec{k},\vec{q}}=1$ for simplicity. Finally, the dispersion of the boson is $\omega_{\vec{q}}^2=\vn{q}^{z_b-1}$, where $z_b$ is the dynamical exponent of the boson. For most of the discussion, we will consider $z_b=3$ as required by locality of the boson kinetic term (which is $\vec{q}^2$), but we will comment on generic $2<z_b<3$ when applicable. The coupling $g^2$ has the dimension of energy, and we assume that it is much smaller than the Fermi energy $g^2\ll k_Fv_F$.

     The theory \eqref{eq:Lagrangian} will be handled using the Migdal-Eliashberg framework, which can be formally justified by the Yukawa-SYK model \cite{Iesterlis2021,HGuo2022a,EEAldape2022,ZDShi2022,ZDShi2023} or the small-$(z_b-2)$-large-$N$ expansion \cite{DFMross2010}. In this work, we will follow the Yukawa-SYK approach, which utilizes the random couplings inspired by the Sachdev-Ye-Kitaev model \cite{SSachdev1993,AYKitaev2015,JMaldacena2016c} to obtain a large $N$ expansion of the system. We introduce $N$ flavors of fermions $\psi_i$ and bosons $\phi_i$, and replace the Yukawa coupling $g$ by Gaussian random couplings $g_{ijl}$. The Lagrangian now becomes
     \begin{equation}\label{eq:SYK_Lagrangian}
     \begin{split}
       \mathcal{L}&=\sum_{i}\psi_i^\dagger(\partial_\tau+\hat{\varepsilon}_k)\psi_i+\frac{1}{2}\sum_{i}\phi_i\left(\hat{q}^2+r\right)\phi_i
       \\
       &+\sum_{ijl}\frac{g_{ijl}}{N}\psi_i^\dagger \psi_j \phi_l\,.
     \end{split}
     \end{equation} Here the couplings $g_{ijl}$ satisfy $g_{ijl}=g_{jil}^*$, $\overline{g_{ijl}}=0$ and $\overline{|g_{ijl}|^2}=g^2$.  We focus on the normal state of the theory, so we suppress pairing by using complex-valued coupling $g_{ijl}$. Assuming the system self-averages, we can perform the Gaussian average over $g_{ijl}$ with a single replica, and the theory can be rewritten using $G$-$\Sigma$ bi-local field theory \cite{Iesterlis2021,HGuo2022a}. We introduce the fermionic Green's function $G(x_1,x_2)$ and the bosonic Green's function $D(x_1,x_2)$, given by the definition:
     \begin{equation}\label{}
       G(x_1,x_2)=-\frac{1}{N}\sum_{i}\psi_i(x_1)\psi_i^\dagger(x_2)\,,
     \end{equation}
     \begin{equation}\label{}
       D(x_1,x_2)=\frac{1}{N}\sum_{i}\phi_i(x_1)\phi_i(x_2)\,.
     \end{equation} Here $x_1,x_2$ denote spacetime coordinates. To enforce these constraints, we introduce Lagrangian multipliers $\Sigma(x_1,x_2)$ and $\Pi(x_1,x_2)$ which can be interpreted as self-energies. The resulting $G$-$\Sigma$ action then becomes

     \begin{equation}\label{eq:S_Gsigma_clean}
\begin{split}
   & \frac{1}{N}S[G,\Sigma,D,\Pi] = -\ln\det\left(\left(\partial_\tau+\varepsilon_k-\mu\right)\delta(x-x')+\Sigma\right) \\&+\frac{1}{2} \ln\det\left(\left(-\partial_\tau^2+|q|^{2}+m_b^2\right)\delta(x-x')-\Pi\right)  \\
   &-\Tr\left(\Sigma\cdot G\right)+\frac{1}{2}\Tr\left(\Pi\cdot D\right)+\frac{g^2}{2}\Tr\left((GD)\cdot G\right)\,.
\end{split}
\end{equation} Here $\delta(x-x')$ denotes a spacetime delta function. Here $\det$ means functional determinant, and the $\Tr$ notation is given by $\Tr(\Sigma\cdot G)=\int\rd^3 x_1 \rd^3 x_2 \Sigma(x_1,x_2)G(x_2,x_1)$. In the last term, $(GD)$ means the point-wise product $G(x_1,x_2)D(x_1,x_2)$.

    The large $N$ saddle point of \eqref{eq:S_Gsigma_clean} is the Eliashberg equation
    \begin{equation}\label{eq:MET}
      \begin{split}
         G(i\omega,\vec{k}) & =\frac{1}{i\omega-\xi_{\vec{k}}-\Sigma(i\omega,\vec{k})}\,, \\
         D(i\Omega,\vec{q})  &= \frac{1}{\vn{q}^{2}+r-\Pi(i\Omega,\vec{q})}\,, \\
         \Sigma(\tau,\vec{r})  & =g^2 G(\tau,\vec{r})D(\tau,\vec{r})\,, \\
         \Pi(\tau,\vec{r})  &=-g^2 G(\tau,\vec{r})G(-\tau,-\vec{r})\,.
\end{split}
\end{equation} Here we have assumed the saddle point to be translational invariant $G(x_1,x_2)=G(x_1-x_2)$, and we Fourier transformed the Green's function with $G(k)=\int\rd^3x G(x)\exp(-ik\cdot x)$, where $k=(i\omega,\vec{k})$, $x=(-i\tau,\vec{r})$ and $k\cdot x=\vec{k}\cdot\vec{r}-\omega\tau$. Here $\tau$ is imaginary time and $\omega$ is Matsubara frequency. For most of the paper, we will work with frequencies on the imaginary axis, which is signaled writing $i\omega$ in the function arguments (for example $G(i\omega,\vec{k})$). Since we often work with zero temperature where the Matsubara frequencies becomes effective continuous, we will write $i\omega$ instead of $i\omega_n$ to simplify our notation.
For most of the paper, we will assume the fermion dispersion $\xi_{\vec{k}}$ to be isotropic but not necessarily parabolic.

\begin{figure}[htb]
  \centering
  \includegraphics[width=0.95\columnwidth]{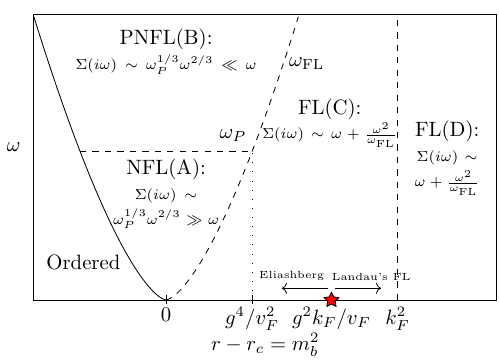}
  \caption{Saddle-point phase diagram near the Ising-nematic quantum critical point in terms of external frequency $\omega$ and the tuning parameter $r$ at zero temperature. The quantum critical point corresponds to $m_b^2=r-r_c=0$. When $r<r_c$ the system orders at $T=0$. The transition to the ordered phase is marked by a solid line and other crossovers are marked by dashed lines. The dotted line connecting $\omega_P$ and $g^4/v_F^2$ is a guide to eye.
  The regimes of interest are labeled A, B, C, D. A is the quantum critical NFL regime with $z_b=3$ boson. B is a perturbative NFL where the boson still has $z_b=3$, but the fermion self energy $\Sigma(i\omega)$ is small compared to $\omega$. The crossover scale between A and B is $\omega=\omega_P$ where $\omega_P$ is defined in Eq.\eqref{eq:omegaP_main}. C is a Fermi liquid regime where the boson $\phi$ acquires a mass term $0<m_b\ll k_F$. D is a Fermi liquid regime where the boson mass $m_b\gtrsim k_F$. The crossover between C and A, B is defined by the curve $\omega=\omega_\text{FL}=m_b^3/\gamma$ \eqref{eq:omegaFL_main}. The intersection between $\omega=\omega_\text{FL}$ and $\omega=\omega_P$ occurs at $m_b\sim g^2/v_F$. In the FL regime C, there is an additional scale $(m_b)_\text{ME}\sim \sqrt{g^2 k_F/v_F}$, which marks the boundary of the applicability of Eliashberg theory.  }\label{fig:pd}
\end{figure}

The NFL regime of the saddle point equations \eqref{eq:MET} has been analyzed in \cite{Iesterlis2021} and \cite{HGuo2022a}. In Appendix.\ref{app:saddle} we reproduce the analysis and extend it to the FL regimes. We obtain four different regimes in terms of the tuning parameter $r$ (bare boson mass) and energy scale $\omega$, as shown in the zero-temperature phase diagram Fig.~\ref{fig:pd}. The solutions are summarized below.

   \textbf{The non-Fermi liquid (A) and the perturbative non-Fermi liquid (B):}
  In these two regimes, the boson is gapless, with the propagator
  \begin{equation}\label{eq:Dprop}
    D(i\Omega\neq 0,\vn{q})=\frac{1}{\vn{q}^2+\gamma\frac{|\Omega|}{\vn{q}}}\,, \gamma=\frac{\calN g^2}{v_F}\,,
  \end{equation} where the Matsubara frequency is nonzero $i\Omega\neq 0$. Here $\calN=k_F/(2\pi v_F)$ is the density of states on the FS. Eq.\eqref{eq:Dprop} describes a gapless boson with Landau damping. When the Matsubara frequency is zero, the boson acquires a thermal mass due to irrelevant operators of the QCP
  \begin{equation}\label{}
    D(i\Omega=0,\vn{q})=\frac{1}{\vn{q}^2+\Delta(T)^2}\,.
  \end{equation} We note that the thermal mass cuts off the IR divergence of the finite-$T$ fermion self-energy \cite{Iesterlis2021} due to $\Omega=0$ bosons. When $z_b=3$, the thermal mass scales as $\Delta(T)^2\sim T\ln(1/T)$ \cite{SAHartnoll2014,AJMillis1993,Iesterlis2021}.

  The fermion self-energy is only a function of frequency, and can be decomposed as $\Sigma(i\omega)=\Sigma_Q(i\omega)+\Sigma_T(i\omega)$. The quantum part $\Sigma_Q$ arises from the bosons carrying non-zero Matsubara frequency, which reads
    \begin{equation}\label{eq:SigmaQ_main}
    \begin{split}
      &\Sigma_Q(i\omega)=-ic_f|\omega|^{2/3} \sgn \omega\,,\quad (T=0)\\
      &=-\frac{2i\sgn \omega}{3}c_f (2\pi T)^{2/3}H_{1/3}\left(\frac{|\omega|}{2\pi T}-\frac{1}{2}\right)\,,\quad (T>0)\\
      c_f&=\frac{g^2}{2\pi\sqrt{3}v_F \gamma^{1/3}}\,.
    \end{split}
    \end{equation}Here $H_{1/3}(x)=\sum_{m=1}^{x} m^{-1/3}$ is the Harmonic number. 

  The thermal part $\Sigma_T$ is due to the boson carrying zero Matsubara frequency:
   \begin{equation}\label{eq:SigmaT}
       \Sigma_T(i\omega)=-i\sgn(\omega)\frac{\Gamma(T)}{2}\,,\quad \Gamma(T)=\frac{g^2 T}{2v_F \Delta(T)}\,.
     \end{equation}

    From dimensional analysis, the coefficient $c_f$ can be translated to an energy scale
    \begin{equation}\label{eq:omegaP_main}
      \omega_P=c_f^3=\frac{g^4}{12\sqrt{3}\pi^2k_F v_F}\,,
    \end{equation} and then $\Sigma_Q(i\omega,T=0)=-i\sgn \omega |\omega|^{2/3}\omega_P^{1/3}$.

    At $T=0$ and $\omega>\omega_P$, the self-energy is smaller than the bare frequency term $\Sigma_Q(i\omega)<|\omega|$, and the system is dubbed a perturbative NFL (PNFL), as the typical dispersion is the same as that of FL $\xi\sim \omega$. When $\omega<\omega_P$, the self-energy becomes larger than the bare frequency term $|\Sigma_Q(i\omega)|>|\omega|$, and the system becomes a NFL. The typical dispersion is then $\xi\sim |\omega|^{2/3}$. The NFL and the PNFL regimes are labeled as A and B in Fig.~\ref{fig:pd}, respectively.

    At finite but low temperatures, the thermal self-energy $\Sigma_T$ dominates over the quantum part $\Sigma_Q$ at the lowest Matsubara frequency. When $\Gamma(T)<\omega_P$, there exists a frequency window $\omega_T<\omega<\omega_P$ where $|\Sigma_Q|>|\Sigma_T|$ and $|\Sigma_Q|>|\omega|$ so that $|\omega|^{2/3}$ self-energy is significant. On the other hand, when $\Gamma(T)>\omega_P$, the $|\omega|^{2/3}$ self-energy is overshadowed  by thermal fluctuations, because at the frequency scale when $|\Sigma_Q|>|\Sigma_T|$, we already enter the PNFL where $|\Sigma_Q|<|\omega|$.

    Finally, if we consider the boson dispersion to be $\vn{q}^{z_b-1}$ with $2\leq z_b<3$, we would obtain the fermion self-energy $\Sigma(i\omega)\propto |\omega|^{2/z_b}$ at $T=0$ (see Appendix~\ref{app:saddle}). However, our computation does not apply to finite $T$ because the thermal mass $\Delta(T)$ of the $2<z_b<3$ theory is qualitatively different.

  \textbf{The small-angle scattering Fermi liquid (C):}
When the boson acquires a finite mass $m_b^2=r-\Pi(0,0)$, the system enters the FL regime:
  \begin{equation}\label{}
      D(i\Omega,\vn{q})=\frac{1}{\vn{q}^2+m_b^2+\gamma\frac{|\Omega|}{\vn{q}}}\,.
    \end{equation}
    The boson mass will receive a thermal correction proportional to $\delta m_b^2\sim T^2$ \cite{SAHartnoll2014,Iesterlis2021}, which we ignore. Here we assume $m_b\ll k_F$, so that the boson still mediates small-angle scattering.

    The FL regime is associated with an energy scale $\omega_{\text{FL}}$ below which the effect of Landau damping is weak.
    \begin{equation}\label{eq:omegaFL_main}
      \omega_{\text{FL}}=\frac{m_b^3}{\gamma}=\frac{2\pi m_b^3 v_F^2}{g^2 k_F}\,.
    \end{equation} The fermion self-energy is $(T=0)$
    \begin{equation}\label{eq:SigmaFLC_main}
  \Sigma(i\omega)=(-i\omega)c_f'\left(\pi+\frac{|\omega|}{\omega_{\text{FL}}}\ln\left(\frac{|\omega|}{\omega_{\text{FL}}}\right)\right),
\end{equation} with
\begin{equation}\label{eq:cfp_main}
  c_f'=\frac{g^2\calN}{2\pi k_F m_b}=\frac{g^2}{(2\pi)^2v_F m_b}\,.
\end{equation} The logarithmic in the second term of Eq.\eqref{eq:SigmaFLC_main} is a feature of 2D  Landau damping \cite{AVChubukov2003}.

  \textbf{The large-angle scattering Fermi liquid (D):}
 As we continue to increase the boson mass $m_b$ to be comparable to $k_F$, we enter the regime where the boson mediates large-angle scattering. The boson dispersion can then be ignored in the boson propagator:
  \begin{equation}\label{}
      D(i\Omega,\vn{q})=\frac{1}{m_b^2+\gamma\frac{|\Omega|k_F}{\vn{q}\sqrt{k_F^2-\vn{q}^2/4}}}\,.
    \end{equation} Here the form of the Landau damping is different because the boson is also sensitive to $2k_F$ singularities.

    The fermion self-energy reads
    \begin{equation}\label{eq:SigmaFLD_main}
     \Sigma(i\omega)=c_f'(-i\omega)\left(\pi+\frac{|\omega|}{\omega_\text{FL}}\ln\frac{|\omega|}{\omega_{\text{FL}}\sqrt{e}}\right)\,.
    \end{equation} This form is qualitatively similar to the small-angle scattering FL (see Eq.\eqref{eq:SigmaFLC_main}), but the exact parameters are different (some factors of $m_b$ are replaced by $k_F$):
    \begin{equation}\label{eq:omegaFLD_main}
      \omega_\text{FL}=\frac{k_F m_b^2}{\gamma}\,,\quad c_f'=\frac{g^2 \calN }{2\pi m_b^2}\,.
    \end{equation} Here we use the same notation $\omega_\text{FL}$ and $c_f'$ in regimes C and D, and it should not cause confusion according to context.

    \textbf{Crossover between different regimes:}
        We discuss the crossover between the four regimes A, B, C, D at zero temperature.

        Suppose we start from the large-angle scattering FL (D) and decrease the boson mass $m_b$. The crossover to the small-angle scattering FL (C) happens at $m_b\sim k_F$.

        In the small-angle scattering FL (C), the system can potentially cross over to two NFL regimes A and B. This is seen in Eq.\eqref{eq:SigmaFLC_main} where the damping term is suppressed by the linear term in $\omega$ by a factor of $|\omega|/\omega_{\text{FL}}$, with $\omega_{\text{FL}}$ defined in \eqref{eq:omegaFL_main}. The damping term becomes important when $\omega\sim\omega_\text{FL}$, i.e. $\omega\sim m_b^3$. When $\omega>\omega_\text{FL}$, the fermion self-energy crosses over to the $|\omega|^{2/3}$ form of a NFL. However, at the scale of $\omega\sim \omega_\text{FL}$, the parameter $c_f'$  in Eq.\eqref{eq:SigmaFLC_main}, which characterizes the relative magnitude of self-energy compared to the bare $i\omega$ term, is undetermined. Using Eq.\eqref{eq:omegaFL_main}, \eqref{eq:cfp_main} and \eqref{eq:omegaP_main}, we find $c_f'\sim (\omega_P/\omega_\text{FL})^{1/3}$. When $c_f'<1$, the system crosses from FL (C) to PNFL (B), and when $c_f'>1$, the system crosses from FL (C) to NFL (A). Our conclusion here differs from Refs. \cite{YGindikin2024,SLi2023} which only assumed a single crossover.

        Finally, when we enter the NFL regimes, we can replace $m_b$ by $|\gamma\omega|^{1/3}$, but the crossover between NFL (A) and PNFL (B) remains and is given by $\omega\sim \omega_P$.

        The crossovers above are summarized in the phase diagram Fig.~\ref{fig:pd}.

        \textbf{Applicability of the Eliashberg Theory}  Before concluding the section, we comment on the applicability of the Eliashberg theory. One feature of the Eliashberg theory is that the self-energy $\Sigma=\Sigma(i\omega)$ is momentum independent, implying that the effective mass $m^*$ and the quasiparticle residue $Z$ satisfy the relation $Zm^*/m=m$, which holds in the vicinity of the QCP \cite{DLMaslov2010a}. This relation ceases to be true at a distance from the QCP when we move into the FL phase, and beyond this point the Eliashberg theory is no longer accurate. In Appendix.~\ref{app:saddle}, we express this point in terms of the boson mass, and it is located at $m_b\sim (m_b)_\text{ME}=\sqrt{g^2k_F/v_F}$, which sits inside the FL (C) regime and marked by star in Fig.~\ref{fig:pd}. In Sec.~\ref{sec:eft}, we show that when $m_b>(m_b)_\text{ME}$ we recover Landau's FL theory. Therefore, our result for the FL (D) regime based on the Eliashberg theory is only qualitatively correct.

\section{The Fluctuation Spectrum of the 2+1D critical FS}\label{sec:fluctuation}

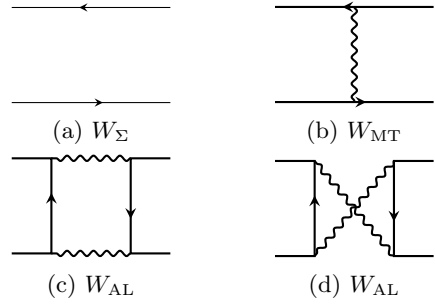
\begin{figure}
  \centering
  \begin{subfigure}[t]{0.4\columnwidth}
  \begin{tikzpicture}[scale=1.5][baseline={([yshift=-4pt]current bounding box.center)}]
                     \draw[ mid arrow] (40pt,12pt)--(0pt,12pt);
                     \draw[ mid arrow] (0pt,-12pt)--(40pt,-12pt);
                     \end{tikzpicture}
  \caption{$W_\Sigma$}
  \end{subfigure}
  \begin{subfigure}[t]{0.4\columnwidth}
  \begin{tikzpicture}[scale=1.5][baseline={([yshift=-4pt]current bounding box.center)}]
                     \draw[thick, mid arrow] (40pt,12pt)--(0pt,12pt);
                     \draw[thick, mid arrow] (0pt,-12pt)--(40pt,-12pt);
                     \draw[thick, boson] (20pt,12pt)--(20pt,-12pt);
                     \end{tikzpicture}
  \caption{$W_\text{MT}$}
  \end{subfigure}

  \begin{subfigure}[b]{0.4\columnwidth}
  \begin{tikzpicture}[scale=1.5][baseline={([yshift=-4pt]current bounding box.center)}]
                     \draw[thick, mid arrow] (40pt,12pt)--(30pt,12pt)--(30pt,-12pt)--(40pt,-12pt);
                     \draw[thick, mid arrow] (0pt,-12pt)--(10pt,-12pt)--(10pt,12pt)--(0pt,12pt);
                     \draw[thick, boson] (30pt,12pt)--(10pt,12pt);
                     \draw[thick, boson] (30pt,-12pt)--(10pt,-12pt);
                     \end{tikzpicture}
                     \caption{$W_\text{AL}$}
  \end{subfigure}
  \begin{subfigure}[b]{0.4\columnwidth}
                     \begin{tikzpicture}[scale=1.5][baseline={([yshift=-4pt]current bounding box.center)}]
                     \draw[thick, mid arrow] (40pt,12pt)--(30pt,12pt)--(30pt,-12pt)--(40pt,-12pt);
                     \draw[thick, mid arrow] (0pt,-12pt)--(10pt,-12pt)--(10pt,12pt)--(0pt,12pt);
                     \draw[thick, boson] (30pt,12pt)--(10pt,-12pt);
                     \draw[thick, boson] (30pt,-12pt)--(10pt,12pt);
                     \end{tikzpicture}
                     \caption{$W_\text{AL}$}
  \end{subfigure}
  \caption{Feynman diagrams generating $K_\text{BS}$. }\label{fig:KBS_diag}
\end{figure}

We now study the fluctuation spectrum near the ME saddle point. We expand the bilocal variables $G,\Sigma,D,\Pi$ around the saddle point \eqref{eq:MET}, and study the leading-order Gaussian fluctuation. After integrating out $\Sigma,D,\Pi$, the quadratic fluctuation action is
\begin{equation}\label{eq:S=GKBSG}
\begin{split}
      S[\delta G]=-\frac{N}{2}&\int_{x_1,x_2,x_3,x_4} \delta G(x_2,x_1)K_\text{BS}(x_1,x_2;x_3,x_4)\\
      &\times\delta G(x_3,x_4)\,.
\end{split}
\end{equation}  Here $\delta G(x_1,x_2)$ is the fluctuation of the fermion bilinear around the solution of the ME equations \eqref{eq:MET}.
The fluctuation kernel $K_\text{BS}$ generates the Feynman diagrams that appear in the Bethe-Salpeter equation. It can be decomposed into three parts:
\begin{equation}\label{eq:KBS}
       K_\text{BS}=W_\Sigma^{-1}-W_\text{MT}-W_\text{AL}.
     \end{equation}  Here $W_\Sigma$, $W_\text{MT}$ and $W_\text{AL}$ are four-point functions that generate the density-of-states (DOS), Maki-Thompson (MT) and Aslamazov-Larkin (AL) diagrams, respectively (see Fig.~\ref{fig:KBS_diag}). They can be conveniently defined in real space as ($\delta$ is spacetime $\delta$-function)
     \begin{align}
       W_\Sigma(x_1,x_2;x_3,x_4)&=G(x_1,x_3)G(x_4,x_2) \label{eq:WSigma}\,, \\
       W_\text{MT}(x_1,x_2;x_3,x_4)&=g^2 D(x_3,x_4)\delta(x_1,x_3)\delta(x_2,x_4)\,, \label{eq:WMT}\\
       W_\text{AL}(x_1,x_2;x_3,x_4)&=-g^4G(x_1,x_2)G(x_4,x_3)\label{eq:WAL}\\
       \times[D(x_1,x_3)D(x_2,x_4)&+D(x_1,x_4)D(x_2,x_3)]\nonumber\,.
     \end{align}
      The expressions above are derived in \cite{HGuo2022a} and reviewed in Appendix.~\ref{app:KBS}.

 It is also convenient to Fourier transform two-point fluctuations in real space to momentum space using the center-of-mass (CoM) 3-momentum $p$ and relative 3-momentum $k$,
\begin{equation}\label{}
\begin{split}
  \delta G(k;p)&=\int\rd^3 x_1 \rd^3 x_2 \delta G(x_1,x_2)\\
  &\times\exp(-ip\cdot (x_1+x_2)/2-ik\cdot (x_1-x_2))\,.
\end{split}
\end{equation} Because of translational symmetry, $p$ is conserved by $K_\text{BS}$. For notational clarity we will suppress $p$ in $\delta G(k;p)$ unless otherwise mentioned. The detailed form of $K_\text{BS}$ in momentum space will be specified later.  

Since $K_\text{BS}$ is the kernel of the Gaussian fluctuation of the bilocal fields $\delta G$, its inverse can be related to fermion four-point functions or linear response function of fermion bilinears.  For example, given two operators $\hat{A}=\int_{x_1,x_2}A(x_1,x_2)\psi^\dagger(x_1)\psi(x_2)$ and $\hat{B}=\int_{x_3,x_4} B(x_3,x_4)\psi^\dagger(x_3)\psi(x_4)$, we have (see Appendix.~\ref{app:KBS} for a derivation)
\begin{equation}\label{eq:<AB>}
  -\braket{\hat{A}\hat{B}}=\int_{x_1,x_2,x_3,x_4} A(x_2,x_1)K_\text{BS}^{-1}(x_1,x_2;x_3,x_4)B(x_3,x_4)\,.
\end{equation} The necessary condition for the stability of the ME saddle point is that Eq.\eqref{eq:<AB>} satisfy constraints for two-point response functions such as unitarity and causality for all hermitian operators $\hat{A},\hat{B}$. 

To identify the low-energy fluctuation of the system, we therefore need to compute the eigenvalues and eigenfunctions of $K_\text{BS}$. This is the goal of the rest of the section.

\subsection{Inner product for the Diagonalization Problem: Hint from Ward identities}

   The first step of diagonalizing $K_\text{BS}$ is to define a suitable inner product for the two-point functions. This is because, since $\delta G$ appears as a dummy variable in the path integral, the form of $K_\text{BS}$ can change if we choose another parameterization by $\delta G=\tilde{M} \delta \tilde{G}$ where $\tilde{M}$ is an arbitrary invertible linear operator. Therefore, we need to invoke some more physical arguments to motivate a suitable parameterization, which is equivalent to defining an inner product.

  The standard inner product
  \begin{equation}\label{eq:inner_std}
    \left(A|B\right)=\int_{x_1,x_2} A(x_2,x_1)B(x_1,x_2)
  \end{equation} is not a good choice because $K_\text{BS}$ does not have an obvious eigenvalue under $\left(\cdot|\cdot\right)$. It is useful to apply an auxiliary operator $M$ to $K_\text{BS}$, and consider the operator $L=K_\text{BS}M$ (this is equivalent to reparameterizing $\delta G$). As we shall see in the following, with a good choice of $M$, $L$ can have exact eigenvalues and eigenvectors.

    The natural way to look for exact eigenvectors is through conservation laws, and we consider the Ward identities associated with charge conservation and momentum conservation. We go to the momentum space where $p=(i\Omega,\vec{p})$ denotes the CoM 3-momentum and $k=(i\omega,\vec{k})$ denotes the relative 3-momentum. Due to space-time translational invariance, $K_\text{BS}$ is block diagonal with blocks labeled by $p$, and different $p$ can be considered separately.
    We consider the simplified case $p=(i\Omega,0)$ (the homogeneous limit, which will be assumed for the rest of the section), the Ward identities take the form \cite{HGuo2022a,TToyoda1987,TToyoda1989,TToyoda2001}:
    \begin{equation}\label{eq:Ward}
       K_\text{BS}[(iG(i\omega+i\Omega/2,\vec{k})-iG(i\omega-i\Omega/2,\vec{k}))\Gamma_\alpha]=\Omega\Gamma_\alpha\,.
     \end{equation} Here $\Gamma_\alpha=1,\vec{k}$ is the charge and momentum vertex, respectively. In the rest of the paper, we will consider $\Omega>0$, which can be utilized to compute retarded response functions.

      We propose that the correct operator $L$ to diagonalize should be
     \begin{equation}\label{}
       L=K_\text{BS}\circ M-\Omega I\,,
     \end{equation} where $\circ$ denotes functional composition, $I$ is the identity operator and $M$ attaches the Green's function factor as in \eqref{eq:Ward} ($F$ is a test two-point function):
     \begin{equation}\label{}
       M[F](i\omega,\vec{k})=(iG(i\omega+i\Omega/2,\vec{k})-iG(i\omega-i\Omega/2,\vec{k}))F(i\omega,\vec{k})\,.
     \end{equation} With the above definitions, the Ward identities imply that $L$ has exact zero modes $1$ and $\vec{k}$. We will call $L$ the kinetic operator because as we will show later, its eigenvalues can be interpreted as the eigenvalue of the collision integral of a kinetic equation.

    We therefore transform the diagonalization of $K_\text{BS}$ into the diagonalization of $L$.
    Since $K_\text{BS}$ is symmetric under the standard inner product \eqref{eq:inner_std}, $\left(A|K_\text{BS}|B\right)=\left(B|K_\text{BS}|A\right)$, we need to define the new inner product $\braket{\cdot|\cdot}$ so that this property also holds for $L$.
     This implies that $M$ should also be the kernel of the inner product:
     \begin{equation}\label{eq:innerprod}
     \begin{split}
       &\braket{A|B}=\int\frac{\rd\omega\rd^2\vec{k}}{(2\pi)^3}A(i\omega,\vec{k})\\
       &\times(iG(i\omega+i\Omega/2,\vec{k})-iG(i\omega-i\Omega/2,\vec{k}))
       B(i\omega,\vec{k})\,.
     \end{split}
     \end{equation} By construction, it satisfies $\braket{A|L|B}=\braket{B|L|A}$.
     An additional nice property of the definition is that the response functions \eqref{eq:<AB>} of fermion bilinears take an invariant form: \begin{equation}\label{eq:<AB>2}
                                          -\braket{\hat{A}\hat{B}}=\left(A|K_\text{BS}^{-1}|B\right)=\left(A|ML^{-1}|B\right)=\braket{A|L^{-1}|B}\,.
                                        \end{equation}

      Finally, let us discuss in what sense is the inner product \eqref{eq:innerprod} positive. It is positive when we compute the 3-momentum integrals in the ``Eliashberg" way, by computing the $\vec{k}$-integral first and the $\omega$-integral next, and analytically continue the result to the real axis of $\Omega$.

      Near the QCP, the exchange of integration order does not affect the IR physics but may introduce a contact term error due to UV divergence. In Appendix~\ref{app:exchange}, we show that the error appears only in computation of the fermion bubbles but not in the vertex corrections. This implies that the inner product \eqref{eq:innerprod} can be used to calculate the eigenvalues of $L$, but the contact term must be corrected when computing the linear response functions (see Appendix~\ref{app:exchange}). In Sec.~\ref{sec:LDtoPK}, we show that this procedure is justified when the boson mass is below the threshold $m_b<(m_b)_{\text{ME}}= \sqrt{g^2 k_F/v_F}$, which covers the NFL (A) and the PNFL (B) regimes and also part of the FL (C) regime.

        Assuming the $\xi_k$-integral is carried out first, the inner product \eqref{eq:innerprod} simplifies in the good metal limit $k_F v_F\gg \Sigma$ which we assume in our problem.  The integral over $\xi_{\vec{k}}$ can be done with contour method, and at low energy the dominant contribution is from the poles of the Green's functions in \eqref{eq:innerprod}. This implies that the integral is only non-zero when $-\Omega/2<\omega<\Omega/2$, so the frequency domain becomes finite, and we separate out the IR physics from the UV parts. In Appendix~\ref{app:inner}, we show that \eqref{eq:innerprod} can be made positive up to a factor of $\Omega$, provided that $A$ and $B$ are form factors of Hermitian operators.

\subsection{Hierarchy of the kinetic operator}\label{sec:Lexpand}

   In this part, we derive the explicit form of the kinetic operator $L$ and propose an expansion scheme for it. We start by writing the three parts of Eq.\eqref{eq:KBS} in momentum space.
$W_\Sigma$ generates the density-of-states Feynman diagrams. It is diagonal in the momentum space ($p$ is CoM 3-momentum and $k$ is relative 3-momentum)
\begin{equation}\label{}
     W_\Sigma[F](k)=G(k+p/2)G(k-p/2)F(k)\,.
   \end{equation}

   $W_\text{MT}$ generates  the Maki-Thompson diagram
    \begin{equation}\label{eq:WMTk}
  W_\text{MT}[F](k)=g^2 \int\frac{\rd^3k'}{(2\pi)^3}D(k-k')F(k')\,.
    \end{equation}

    $W_\text{AL}$ generates the Aslamazov-Larkin diagrams
    \begin{equation}\label{eq:WGAL}
    \begin{split}
      &W_{\text{AL}}[F](k_1;p)      =-g^4 \int \frac{\rd^3 q\rd^3 k_2}{(2\pi)^6}  F(k_2)G(k_1-q)\\
      &\times\left(G(k_2-q)+G(k_2+q)\right)D(q+p/2)D(q-p/2)\,.
    \end{split}
    \end{equation}

    Next, we express the kinetic operator $L$ in the momentum space.
    We combine $W_\Sigma$ and $M$ to write
    \begin{equation}\label{}
    \begin{split}
      &W_\Sigma^{-1}M=iG^{-1}(k-i\Omega/2)-iG^{-1}(k+i\Omega/2)\\
      &=\Omega+i\Sigma(k+i\Omega/2)-i\Sigma(k-i\Omega/2)\,.
    \end{split}
    \end{equation} Here we added a 3-vector $k$ to a scalar $i\Omega/2$, and it is understood that only the frequency component is added. The $\Omega$ term will be cancelled by the definition of $L$.  The self-energy can be rewritten using the saddle point equation
    \begin{equation}\label{}
      \Sigma(k\pm i\Omega/2)=g^2\int\frac{\rd^3 k'}{(2\pi)^3} D(k-k')G(k'\pm i\Omega/2)\,.
    \end{equation} We therefore combine $W_\Sigma$ and $W_\text{MT}$ to write
    \begin{equation}\label{eq:LMT}
    \begin{split}
      &L_{\text{MT+DOS}}[F]=(W_\Sigma^{-1}-W_\text{MT})\circ M [F]-\Omega F\\
      &=g^2\int\frac{\rd^3k'}{(2\pi)^3}D(k-k')\left[iG(k'+i\Omega/2)-iG(k'-i\Omega/2)\right]\\
      &\times\left[F(k)-F(k')\right]\,.
    \end{split}
    \end{equation} The remaining Aslamazov-Larkin part is
    \begin{equation}\label{eq:LAL}
    \begin{split}
      &L_\text{AL}[F](k_1)=g^4 \int \frac{\rd^3 q\rd^3 k_2}{(2\pi)^6}\left(G(k_2-q)+G(k_2+q)\right)\\
      &\times G(k_1-q)D(q+i\Omega/2)D(q-i\Omega/2)\\
      &\times \left[iG(k_2+i\Omega/2)-iG(k_2-i\Omega/2)\right]F(k_2)\,.
    \end{split}
    \end{equation}

    We note in passing that Eq.\eqref{eq:LMT} bears some formal similarity with the collision term of the quantum kinetic equation \cite{AKamenev2023}, where $F(k)$ plays the role of the fermionic distribution function. The $F(k)$ term can be interpreted as the departure term of the collision integral and the $F(k')$  term can be interpreted as the arrival term. Following this analogy, Eq.\eqref{eq:LAL} should come from a separate kinetic equation for the boson, by solving the bosonic distribution in terms of the fermionic one and inserting it into the fermionic collision integral.

    To proceed, we now assume that the FS is circular and switch to the angular harmonic basis by setting $$F(k)=F(i\omega,\xi_k)e^{im\theta_k},$$ where $\xi_k$ is the dispersion of the 2-momentum $\vec{k}$ measured from the FS, and $\theta_k$ denotes its direction.

    The $m$-th harmonic component of $L$ is then defined to be
    \begin{equation}\label{}
      L_m[F](i\omega,\xi)=\int_0^{2\pi}\frac{\rd \theta_k}{2\pi} e^{-im\theta} L[F e^{im\theta_k}](i\omega,\xi,\theta)\,.
    \end{equation}

    The angular integrals are carried out in Appendix.~\ref{app:Lint}, and we obtain the kinetic operator $L_m=L_{\text{MT+DOS},m}+L_\text{AL,m}$ in the angular harmonics basis:
    \begin{widetext}
     \begin{equation}\label{eq:LMTT}
        \begin{split}
          L_{\text{MT+DOS},m}[F]&(i\omega,\xi)=g^2\int_{-\infty}^{\infty}\frac{\rd \omega'}{2\pi}\frac{\calN \rd \xi'}{2\pi}\int_0^{\infty}\vn{q}\rd\vn{q}J(\vn{k},\vn{k'},\vn{q})D(\vn{q},i\omega-i\omega')\\
          &\times 2\left[iG(i\omega'+i\Omega/2,\xi')-iG(i\omega'-i\Omega/2,\xi')\right]\left[F(i\omega,\xi)-F(i\omega',\xi')T_m\left(\frac{\vn{k}^2+\vn{k'}^2-\vn{q}^2}{2\vn{k}\vn{k'}}\right)\right]\,.
        \end{split}
        \end{equation}
        \vspace{-2.0em}
        \begin{equation}\label{eq:LALT}
        \begin{split}
        L_{\text{AL},m}[F]&(i\omega_1,\xi_1)=\frac{g^4}{2}(2\pi)^2\int_{-\infty}^{\infty}\calN^3\frac{\rd \nu}{2\pi}\frac{\rd \omega_2}{2\pi}\frac{\rd \xi_2}{2\pi}\frac{\rd \xi'}{2\pi}\frac{\rd \xi''}{2\pi}\int_0^\infty \vn{q}\rd \vn{q} J(\vn{k_1},\vn{k'},\vn{q}) J(\vn{k_2},\vn{k''},\vn{q})\\
        &\times D(\vn{q},i\nu+i\Omega/2)D(\vn{q},i\nu-i\Omega/2)\times 4T_m\left(\frac{\vn{q}^2+\vn{k_1}^2-\vn{k'}^2}{2 \vn{q} \vn{k_1}}\right)T_m\left(\frac{\vn{q}^2+\vn{k_2}^2-\vn{k''}^2}{2 \vn{q} \vn{k_2}}\right)\\
        &\times \left[G(i\omega_1-i\nu,\xi')+(-1)^mG(i\omega_1+i\nu,\xi')\right]\left[G(i\omega_2-i\nu,\xi'')+(-1)^mG(i\omega_2+i\nu,\xi'')\right]\\
        &\times i(G(i\omega_2+i\Omega/2,\xi_2)-G(i\omega_2-i\Omega/2,\xi_2))F(i\omega_2,\xi_2)\,.
        \end{split}
        \end{equation}
     \end{widetext}
      Here, $\calN=k_F/(2\pi v_F)$ is the density of states near the FS. The fermionic momenta $\vn{k}$ are related to the dispersion $\xi$ with the same label (primed or subscripted) as $\vn{k}=k_F+\xi/v_F+\calO(\xi^2)$. $J(\vn{k},\vn{k'},\vn{q})=2/({\sqrt{(\vn{k}+\vn{k'})^2-\vn{q}^2}\sqrt{\vn{q}^2-(\vn{k}-\vn{k'})^2}})$ is the Jacobian from angular integration. $T_m$ is the Chebyshev polynomial $T_m(\cos\theta)=\cos m\theta$.

       To proceed, we apply approximations to Eqs.\eqref{eq:LMTT} and \eqref{eq:LALT}. We focus on the regimes A, B, C in the phase diagram (Fig.~\ref{fig:pd}) where small-angle scattering is dominant and  Eliashberg theory applies.  In regimes A,B,C, there are two small parameters $\xi/(k_Fv_F)$ and $\vn{q}/k_F$, and for Eliashberg theory to be valid the first is much smaller than the second. This motivates us to consider a double expansion in the two parameters. We choose to expand the Chebyshev polynomials in Eqs.\eqref{eq:LMTT} and \eqref{eq:LALT} in $\vn{q}$ and $\xi$.  Since $\xi/(k_F v_F)$ is not larger than $\vn{q}^2/k_F^2$, we choose to expand first in $\xi$ and then in $\vn{q}$, which yields the double expansion below:
     \begin{alignat}{4}\label{eq:Lexpand}
       L_m=&~L^{(0)}_m\quad+\quad&&~L^{(1)}_m\quad+\quad&&~L^{(2)}_m+\dots \nonumber\\[-2pt]
           &~~\verteq  &&~~\verteq  &&~~\verteq \nonumber\\[-4pt]
           &\delta_q^{0}L_m^{(0)}&&\delta_q^{0}L_m^{(1)}&&\delta_q^{0}L_m^{(2)}\\[-4pt]
           &~~+&&~~+&&~~+\nonumber\\[-4pt]
           &\delta_q^{1}L_m^{(0)}&&\delta_q^{1}L_m^{(1)}&&\delta_q^{1}L_m^{(2)}\nonumber\\[-4pt]
           &~~+&&~~+&&~~+\nonumber\\[-5pt]
           &~~~\vertdots&&~~~\vertdots&&~~~\vertdots\nonumber
     \end{alignat}Here, the horizontal direction corresponds to the $\xi/(k_F v_F)$ expansion and the vertical direction corresponds to the $\vn{q}^2/k_F^2$ expansion. The first column $L_m^{(0)}$ assumes the excitations involved in scattering are exactly on the FS (since $\xi$'s are set to zero), and higher-order terms ($L_m^{(1)}$ and so on) allow the excitations to be off the Fermi surface. The first row (terms prefixed by $\delta_q^0$) describes the forward scattering processes (since $\vn{q}$ is set to zero), and the second row (terms prefixed by $\delta_q^1$) describes the effects of small-angle scattering. The explicit form of each term in Eq.\eqref{eq:Lexpand} is discussed in Appendix.~\ref{app:Lint}.

    \subsection{The forward-scattering limit}

    In the expansion \eqref{eq:Lexpand}, the top-left term $\delta_q^0 L_m^{(0)}$ is the leading contribution. It can be obtained from Eqs.\eqref{eq:LMTT} and \eqref{eq:LALT} by taking the $\vn{q}\to 0$ and $\xi\to 0$ limit by setting $T_m=1$ in Eq.\eqref{eq:LMTT} and $T_mT_m=\frac{(-1)^m+1}{2}$ in Eq.\eqref{eq:LALT}. It describes the forward-scattering limit of excitations exactly on the FS.

    In this limit, we expect the local density on the FS to be exactly conserved. Indeed, $\delta_q^0 L_m^{(0)}$ contains a zero mode $F(i\omega,\xi)=1$ for every angular harmonics $m$ as shown by explicit computation in Appendix.~\ref{app:rough}. After Fourier transforming $m$ to $\theta$, we conclude that the function $F(\theta)$ that depends only on the angular coordinate $\theta$ on the FS is a zero mode of $\delta_q^0 L_m^{(0)}$, which can be interpreted as the local density on the FS or parameterizing the shape fluctuation of the FS. Therefore, the ${\rm LU(1)}$ symmetry \cite{DVElse2021a} emerges at the leading order of the fluctuation spectrum.

    However, the conservation of $F(\theta)$ will be broken by higher order perturbations such as $\delta_q^1 L_m$, as they describe the effects of finite-angle scattering. Since in the critical FS small-angle scattering is dominant, the breaking of ${\rm LU(1)}$ will be weak and $F(\theta)$ can still be soft modes. This investigation will be carried out in detail in the next section.

    Before closing the section, we comment on the spectrum of $\delta_q^0 L_m^{(0)}$. Apart from the soft mode $F(i\omega,\xi)=1$, $\delta_q^0 L_m^{(0)}$ contains non-zero eigenvalues which are studied in the appendix.~\ref{app:rough}. This is done by expanding $F(i\omega,\xi)$ as a polynomial in $\xi$ and transforming $\delta_q^0 L_m^{(0)}$ into a block matrix functional that acts only in the frequency domain $\omega\in[-\Omega/2,\Omega/2]$, which can be diagonalized numerically.  We find that the non-zero eigenvalues scale with $\Omega$ in the same way as the self-energies. For example, in the NFL regime (A) and the NFL regime (B), the eigenvalues scale as $\lambda_\Omega\sim \Omega^{2/3}$, where the numerical prefactor varies across different eigenvectors. In the FL regimes, the eigenvalues scale as $\lambda_\Omega\sim \Omega+\Omega^2$.

\section{The Soft modes of the Critical Fermi Surface}\label{sec:soft}

    In this section, we analyze the soft modes of the fluctuation spectrum. At the leading order of the $\vn{q}$-$\xi$ double expansion, the soft modes appear as zero modes, having the form of the constant function $F(i\omega,\xi)=1$ for every angular harmonics $m$. The goal of this section is to compute the effects of small-angle scattering and resolve the soft-mode eigenvalues.

    On the technical side, the analysis that we perform here is nothing more than eigenvalue perturbation theory, which is similar to the spirit of \cite{PJLedwith2019}. The $F=1$ zero mode can be represented by a ket $\ket{1}$ and satisfies the leading order eigenvalue equation
    \begin{equation}\label{}
      \delta_q^0 L_m^{(0)}\ket{1}=0\,.
    \end{equation} The scattering effects can be included via perturbation theory. For example, the first perturbation to include is $\delta_q^1 L_m^{(0)}$, and the shift of the eigenvalue is
    \begin{equation}\label{eq:first_perturb}
      \lambda_m =\frac{\braket{1|\delta_q^1 L_m^{(0)}|1}}{\braket{1|1}}\,.
    \end{equation}

    Although the emergent ${\rm LU(1)}$ symmetry is broken by finite-angle scattering, we still expect the UV symmetries, that is, particle and momentum conservation, to be preserved. Accordingly, the $m=0$ and the $m=1$ harmonics should remain exact zero modes.

   \subsection{Random-walk on the FS}\label{sec:random_walk}

    Before presenting results of the perturbation analysis, it is useful to have a qualitative picture of scattering events that occur on or close to the FS. Here we present a picture of random-walk on the FS which can capture some aspects of the perturbation analysis.

\begin{figure}
      \centering
      \begin{subfigure}[t]{0.4\columnwidth}
      \begin{tikzpicture}

\draw[thick] (0,0) circle (30pt);

\fill (120:30pt) circle (2pt) node[above left] {A(B')}; 
\fill (60:30pt) circle (2pt) node[above right] {B(A')}; 

\draw[thick, <->, bend right=45] (120:27pt) to (60:27pt);

\end{tikzpicture}
    \caption{Non head-on configuration: No phase space}
      \end{subfigure}
      \begin{subfigure}[t]{0.4\columnwidth}
      \begin{tikzpicture}

\draw[thick] (0,0) circle (30pt);

\fill (120:30pt) circle (2pt) node[above left] {A}; 
\fill (60:30pt) circle (2pt) node[above right] {A'}; 

\fill (120:-30pt) circle (2pt) node[above left] {B}; 
\fill (60:-30pt) circle (2pt) node[above right] {B'}; 

\draw[thick,mid arrow] (120:30pt)--(60:30pt);
\draw[thick,mid arrow] (120:-30pt)--(60:-30pt);

\end{tikzpicture}
    \caption{Head-on configuration: Large phase space, but only relaxes even parity deformations.}
      \end{subfigure}
      \caption{Kinematic constraint for 2-particle scattering $A,B\rightarrow A',B'$ on the FS. }\label{fig:kinematics}
    \end{figure}
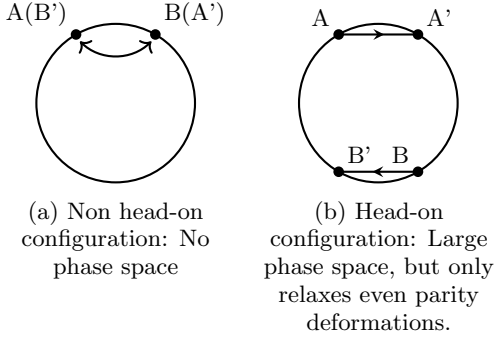

    As a preparation, we first discuss the kinematic constraint \cite{DLMaslov2011,HKPal2012,PJLedwith2019,PLedwith2019,XWang2019,HGuo2022a,SLi2023,YGindikin2024,JHofmann2023,JHofmann2022,ENilsson2024} of two-particle scattering on a circular FS. As shown in Fig.~\ref{fig:kinematics}, the combination of Pauli exclusion and momentum conservation implies that if the two initial momenta on the FS $\vec{k}_A,\vec{k}_B$ satisfy $\vec{k}_A+\vec{k}_B\neq 0$, there is no phase space available for scattering, as the two particles can only forward scatter or exchange between themselves. The exception is head-on scattering with $\vec{k}_A+\vec{k}_B=0$, in which case any head-on pair can scatter to any head-on pair.

      We model the evolution of the local density $n_\theta$ on the FS through a random walk process. It is important to note that the random walk is correlated due to head-on scattering.
      The simpler case is to consider even-parity deformations of the FS where $n_{\theta}=n_{\theta+\pi}$, which corresponds to even angular harmonics $m$. In this case, the two particles of the head-on pair can be identified as a single particle by inversion symmetry, and the random walk becomes effectively uncorrelated. The statistical average of uncorrelated random walks is therefore regular diffusion, which is effectively described by the equation:
      \begin{equation}\label{}
        \partial_t n_\theta = -D_\text{even} \partial_\theta^2 n_\theta\,.
      \end{equation} Here the diffusion constant $D_\text{even}$ can therefore be estimated by multiplying the typical scattering rate (fermion self-energy $\Sigma(i\Omega)$) with the typical angular step ($q^2/k_F^2$ with $q$ being typical boson momentum). Therefore, we expect the eigenvalue to scale as
      \begin{equation}\label{eq:lambda_even_estimate}
        \lambda_m^\text{even}\sim m^2\times \Sigma(i\Omega)\times \left\langle\frac{q^2}{k_F^2}\right\rangle\,.
      \end{equation} Here the bracket denotes a suitable average over the boson Green's function $D$. This agrees with explicit computation in the next subsection.

      The situation for the odd-parity deformations $n_\theta=-n_{\theta+\pi}$ corresponding to odd $m$ is more complicated, and the simple random-walk picture above does not apply. Due to the fact that head-on pairs are all even under inversion, the head-on scattering does not relax any odd-parity deformations. In the previous work of the author \cite{HGuo2022a}, it was conjectured that the angular diffusion dynamics might be described by a fourth diffusion term $\partial_t n_\theta=D \partial_\theta^4 n_\theta$, where $D\propto \braket{q^4/k_F^4}$. This turns out to be incorrect because the head-on scattering argument holds to all orders in $q$, and therefore $q^4$ cannot survive.

      The resolution of the problem turns out to be intertwining the angular and the radial relaxations. In terms of perturbation theory, we have to go out of the subspace of the soft modes (functions $F(\theta)$ of angle only) and include the radial deformations which can depend on $\xi_k$. This is achieved through a second-order perturbation theory, and the result can be power counted as
      \begin{equation}\label{eq:lambda_odd_estimate}
        \lambda_m^{\text{odd}}\sim m^{\eta'}\times \Sigma(i\Omega)\times  \left\langle\frac{q^2}{k_F^2}\right\rangle\times  \left\langle\frac{\xi_k^2}{k_F^2v_F^2}\right\rangle\,.
      \end{equation} Here the exponent $\eta'$ has to be determined from the calculation, and the result turns out to be $\eta'=6$.

      Furthermore, the angular diffusion can be substantiated by computing the Pomeranchuk susceptibility of the FS, which explicitly takes the form of a diffusion propagator. This will be shown at the end of the section.

    After the qualitative discussion, we proceed to explicit computations below.

    \subsection{The even-$m$ soft modes}

    We first consider the eigenvalues of the even-$m$ soft modes. We shall apply perturbation theory in the vertical directions of Eq.\eqref{eq:Lexpand}, which introduce effects of small-angle scattering. We first consider the first-order perturbation theory, given by the term $\delta_q^1 L_m^{(0)}$. The explicit expression is
    \begin{equation}\label{}
    \begin{split}
          \delta_q^1 L_{m}^{(0)}[F](i\omega)&=2g^2\calN\int_{-\Omega/2}^{\Omega/2}\frac{\rd \omega'}{2\pi}\int_0^{\infty}\frac{\rd\vn{q}}{k_F}\\
          &\times D(\vn{q},i\omega-i\omega')\frac{m^2\vn{q}^2}{2k_F^2}F(i\omega')\,.
    \end{split}
    \end{equation} This comes from expanding Eq.\eqref{eq:LMTT} to zeroth order in $\xi$ and then first order in $\vn{q}^2$. Eq.\eqref{eq:LALT} does not contribute when $F(i\omega)$ is an even function in $\omega$.

    According to first-order perturbation theory, the soft mode eigenvalue is given by
    \begin{equation}\label{eq:lambda_even}
    \begin{split}
      &\lambda_m^{\text{even}}(i\Omega)=\frac{\braket{1|\delta_q^1 L_m^{(0)}|1}}{\braket{1|1}}=\frac{2g^2\calN}{\Omega/(2\pi)}\int_{-\Omega/2}^{\Omega/2}\frac{\rd\omega\rd \omega'}{(2\pi)^2}\\
      &\times\int_0^\infty\frac{\rd\vn{q}}{k_F}D(\vn{q},i\omega-i\omega')\frac{m^2\vn{q}^2}{2k_F^2}\,.
    \end{split}
    \end{equation} Here $\ket{1}$ means the constant function $F(i\omega)=1$, which has norm $\braket{1|1}=\Omega/(2\pi)$. The $m^2$ dependence comes from expanding the Chebyshev polynomials and translates to a regular diffusion. We see that Eq.\eqref{eq:lambda_even} agrees with Eq.\eqref{eq:lambda_even_estimate} based on the random-walk picture. As a sanity check $\lambda_0^{\text{even}}=0$ in accordance with charge conservation. 

    We now explicitly evaluate Eq.\eqref{eq:lambda_even} in regimes A,B,C:

      \textbf{The NFL (A) and the PNFL (B)}:
        First, we consider the zero temperature effect first. We substitute the boson propagator
        $$
        D(\vn{q},i\Omega)=\frac{1}{\vn{q}^2+\gamma|\Omega|/\vn{q}}\,.
        $$ The integral in Eq.\eqref{eq:lambda_even} is UV divergent and will be regularized by dimensional regularization: We multiply the integrand by $\vn{q}^{-\eta}$, which makes the integral convergent when $1<\eta<2$, and then we continue to result with $\eta\to 0$:
        \begin{equation}\label{eq:lambda_even_AB}
          \lambda_m^\text{even,$T$=0}(i\Omega)=-\frac{3\sqrt{3}m^2}{14}\frac{\left(c_f\Omega^{2/3}\right)^2}{k_F v_F}\,.
        \end{equation}
        If we instead use a hard cutoff on the $\vn{q}$ integral, there will be additional terms proportional to $\Omega$ that depends on the cutoff. However, the $\Omega^{4/3}$ dependence in \eqref{eq:lambda_even_AB} is universal.

        Eq.\eqref{eq:lambda_even_AB} can be generalized to finite temperature, because the boson propagator in \eqref{eq:lambda_even} has the same form at finite temperature, and we just need to replace the frequency integral by Matsubara summation (we need to exclude the term $\omega=\omega'$ which is treated separately below):
        \begin{equation}\label{eq:lambda_even_AB_T}
        \begin{split}
          &\lambda_m^{\text{even},T>0}(i\Omega)=-\frac{2 m^2}{\sqrt{3}k_F v_F}c_f^2(2\pi T)^{4/3}\\
          &\times\left[H_{-1/3}\left(\frac{\Omega}{2\pi T}-1\right)-\frac{2\pi T}{\Omega}H_{-4/3}\left(\frac{\Omega}{2\pi T}-1\right)\right]\,,
        \end{split}
        \end{equation}
         where $H_{r}(M)=\sum_{n=1}^{M}n^{-r}$ is the harmonic number. This result is derived in Appendix.~\ref{app:finiteT}. In the $\Omega\ll T$ limit, the above result becomes
\begin{equation}\label{eq:lambda_even_AB_T>}
  \lambda_m^{\text{even},T>0}(i\Omega=0)=\frac{m^2 c_f^2 (2\pi T)^{4/3}}{k_F v_F}\times \frac{\zeta \left(\frac{4}{3}\right) \Gamma
   \left(\frac{7}{3}\right)}{2^{1/3} \sqrt{3} \pi ^{4/3}}\,.
\end{equation}

         However, because of the presence of thermal fluctuation in bosons carrying zero Matsubara frequency, there is an additional contribution that violates the $\Omega/T$ scaling. This term can be captured by substituting the boson propagator
        $$
        D_\text{th}(\vn{q},i\Omega)=\frac{1}{\vn{q}^2+\Delta(T)^2}2\pi T \delta(\Omega)\,.
        $$
        \begin{equation}\label{eq:lambda_even_AB_thermal}
          \lambda_m^{\text{even,thermal}}(i\Omega)=-\frac{m^2}{4} \frac{g^2 T \Delta(T)}{k_F^2 v_F}\,.
        \end{equation} As before, the cutoff-dependent term $\propto \Omega$ is implicit. Since $\Delta(T)\sim T^{1/2}\ln^{1/2}(1/T)$ at low temperatures \cite{SAHartnoll2014,AJMillis1993,Iesterlis2021}, thermal fluctuation is subdominant compared to Eq.\eqref{eq:lambda_even_AB_T>}.

    The total eigenvalue is the sum of the two $\lambda_m^{\text{even,total}}=\lambda_m^{\text{even},T}+\lambda_m^{\text{even},thermal}\approx \lambda_m^{\text{even},T}$.

      \textbf{The small-angle scattering FL (C)}:

      In the small-angle scattering FL, we substitute the boson propagator
      \begin{equation}\label{}
        D(\vn{q},i\Omega)=\frac{1}{\vn{q}^2+m_b^2+\gamma|\Omega|/\vn{q}}\,.\nonumber
      \end{equation} We first evaluate the $\vn{q}$-integral in \eqref{eq:lambda_even} using dimensional regularization and the contour method used to obtain Eq.\eqref{eq:DintFL}.
      The cutoff-independent part of \eqref{eq:lambda_even} is
      \begin{equation}\label{eq:lambda_even_C}
     \lambda_m^\text{even}=-c_f' \frac{m^2 m_b^2}{2k_F^2}\left(\pi\Omega+\frac{\Omega^2-4\pi^2 T^2}{3\omega_{\text{FL}}}\right)\,.
    \end{equation} As usual, there is an additional cutoff-dependent term, linear in $\Omega$.

    The above results are derived under the assumption that the angular harmonic number $m$ is not too large, so that the Taylor expansion of the Chebyshev polynomial is valid. Therefore, the validity condition is $m\ll m^*$, where $m^*\approx k_F/\vn{q}$. The typical value of $\vn{q}$ is read off from the boson propagator. For example, we should use $\vn{q}\sim (\gamma\max(|\Omega|,T))^{1/3}, \Delta(T),m_b $ for Eqs.\eqref{eq:lambda_even_AB}, \eqref{eq:lambda_even_AB_T} and \eqref{eq:lambda_even_C}, respectively.

    \textbf{The large-angle scattering FL (D)}: If the boson mass $m_b$ is comparable to $k_F$, the scaling of the eigenvalues with $m$ is qualitatively different. We should substitute the boson propagator
    $$
    D(i\Omega,\vn{q})=\frac{1}{m_b^2+\gamma\frac{|\Omega|k_F}{\vn{q}\sqrt{k_F^2-\vn{q}^2/4}}}\,.
    $$ Although strictly speaking, Eliashberg theory does not apply in this regime, we still perform a qualitative analysis of this regime in Appendix.~\ref{app:softD}, which yields
    \begin{equation}\label{}
      \lambda_m^\text{even}(i\Omega)\sim A_0 \Omega+B_0 \Omega^2\ln m\,, m\gg 1.
    \end{equation} Here the $\ln m$ term arises from $\vn{q}=2k_F$ singularity being cutoff by the angular form factor. Although Eliashberg theory is not accurate in this regime, the result is still consistent with Ref.~\cite{PJLedwith2019} with directly computed the collision integral of a FL. Therefore, deep in the FL regime the even-$m$ fluctuations are actually not soft as they scale similarly with the FL self-energy.

    Finally, the crossover from the FL regime (C) to the NFL and the PNFL regimes (A,B) can be obtained by substituting $c_f'\sim \max(\Omega,T)^{-1/3}$ and $m_b\sim \max(\Omega,T)^{1/3}$ in Eq.\eqref{eq:lambda_even_C}, and the crossover from (C) to (D) can be obtained by substituting $m_b\sim k_F/m$, but the $\ln m$ term cannot be recovered as it involves $2k_F$ singularity.

    \subsection{The odd-$m$ soft modes}

    The soft modes of odd angular harmonics $m$ are more subtle due to cancelations. If we try to apply the first-order perturbation theory as that for even $m$ harmonics, we discover that $\delta_q^1 L_m^{(0)}\ket{1}=0$.

    In fact, we are able to show that $L_m^{(0)}\ket{1}=0$ to all orders in $\vn{q}$. To compute $L_m^{(0)}[1]$, we substitute $F(i\omega,\xi)=1$ in Eqs.\eqref{eq:LMTT} and \eqref{eq:LALT}, and set $\vn{k}=k_F$ in the Chebyshev polynomials. The remaining $\xi$ and $\omega$ integrals can be partially evaluated using the saddle equations (Eq.\eqref{eq:Pi=JGG} in the appendix). We obtain
    \begin{equation}\label{}
    \begin{split}
      &L_{m}^{(0)}[1](i\omega,\xi)=g^2\calN\int\frac{\rd \omega'}{2\pi}\frac{\rd \xi'}{2\pi}\int \vn{q}\rd\vn{q} J(\vn{k},\vn{k'},\vn{q})\\
      &\times D(\vn{q},i\omega-i\omega')2[iG(i\omega'+i\Omega/2,\xi')-iG(i\omega'-i\Omega/2,\xi')]\\
      &\times \left[1-T_m\left(1-\frac{\vn{q}^2}{2k_F}\right)-2T_m^2\left(\frac{\vn{q}}{2k_F}\right)\right]=0\,.
    \end{split}
    \end{equation} The term in the last bracket vanishes due to trignometric identities written in terms of Chebyshev polynomials. As discussed in Sec.~\ref{sec:random_walk}, this cancellation is due to the fact that only head-on collision has large phase space for relaxation, but head-on collision can only relax even $m$ modes.


    To properly resolve the eigenvalues of the odd-$m$ soft modes, we need to consider momenta not exactly on the FS, which are captured by the operators $L_m^{(1)}$ and $L_m^{(2)}$.  Because the boson $\phi$ is real, there is a particle-hole symmetry $(\omega,\xi)\to(-\omega,-\xi)$ near the FS, under which $L_m^{(0)}$ is even and $L_m^{(1)}$ is odd, so first-order perturbation $\braket{1|L_m^{(1)}|1}$ vanishes identically. A non-zero answer requires perturbing second order in $\xi$ and first order in $\vn{q}^2$ :
    \begin{equation}\label{eq:lambda_odd0}
      \lambda_m^{\text{odd}}=\frac{\delta_q^1}{\braket{1|1}}\left[\braket{1|L_m^{(2)}|1}-\braket{1|L_m^{(1)}\frac{1}{L_m^{(0)}}L_m^{(1)}|1}\right]\,.
    \end{equation} The reason to include first order in $\vn{q}^2$ perturbation is because the zeroth order forward scattering does not lead to relaxation.  Eq.\eqref{eq:lambda_odd0} can be computed analytically as described in Appendix.~\ref{app:oddm}. The result reads
    \begin{equation}\label{eq:lambda_odd}
    \begin{split}
      &\lambda_m^\text{odd}=\frac{2g^2\calN}{\Omega/(2\pi)} \int_{-\Omega/2}^{\Omega/2}\frac{\rd \omega\rd \omega'}{(2\pi)^2}\int_{-\infty}^\infty\frac{\rd \xi\rd \xi'}{(2\pi)^2}\int_0^{\infty}\frac{\rd\vn{q}}{k_F}\\
      &\times D(\vn{q},i\omega-i\omega')\left[iG(i\omega+i\Omega/2,\xi)-iG(i\omega-i\Omega/2,\xi)\right]\\
      &\times\left[iG(i\omega'+i\Omega/2,\xi')-iG(i\omega'-i\Omega/2,\xi')\right]\\
      &\times\frac{\vn{q}^2}{k_F^2} \frac{m^2(m^2-1)^2(\xi+\xi')^2}{8k_F^2 v_F^2}\,.
    \end{split}
    \end{equation}

    We observe that Eq.\eqref{eq:lambda_odd} vanishes for $m=1$, which corresponds to momentum conservation.

    We now evaluate Eq.\eqref{eq:lambda_odd} in different regimes. We compute the $\xi$-integrals in Eq.\eqref{eq:lambda_odd} first:
    \begin{equation}\label{eq:lambda_odd_eval}
    \begin{split}
      &\lambda_m^\text{odd}=\frac{2g^2\calN}{\Omega/(2\pi)} \int_{-\Omega/2}^{\Omega/2}\frac{\rd \omega\rd \omega'}{(2\pi)^2}\int_0^{\infty}\frac{\rd\vn{q}}{k_F}D(\vn{q},i\omega-i\omega')\\
      &\times\frac{\vn{q}^2}{k_F^2} \frac{m^2(m^2-1)^2}{4k_F^2 v_F^2}\left[\frac{A_+^2+A_-^2}{2}+\frac{A_++A_-}{2}\frac{A_+'+A_-'}{2}\right]\,.
    \end{split}
    \end{equation} Here $A_\pm=A(i\omega\pm i\Omega/2)$,  $A_\pm'=A(i\omega'\pm i\Omega/2)$ and $A(i\omega)=i\omega-\Sigma(i\omega)$. Recall that we assume $\Omega>0$ and the $\xi$ integrals restrict $-\Omega/2<\omega<\Omega/2$.

       \textbf{The NFL (A):}
        The boson propagator is
        $$
        D(\vn{q},i\Omega)=\left\{
                            \begin{array}{ll}
                              \frac{1}{\vn{q}^2+\gamma\frac{|\Omega|}{\vn{q}}}, & \hbox{$\Omega\neq 0$;} \\
                              \frac{1}{\vn{q}^2+\Delta(T)^2}, & \hbox{$\Omega=0$.}
                            \end{array}
                          \right.
        $$
        We first consider the regime $\Omega\gg T$, where the effects of the thermal mass $\Delta(T)$ and the thermal self-energy $\Sigma_T$ can be ignored. In the NFL regime, the self-energy is much larger than the bare frequency term, therefore in Eq.\eqref{eq:lambda_odd_eval}, we substitute $A(i\omega)\approx -\Sigma_Q(i\omega)$ where $\Sigma_Q(i\omega)$ is given by Eq.\eqref{eq:SigmaQ_main}. We obtain
    \begin{equation}\label{eq:lambda_odd_A_Omega}
      \lambda_m^{\text{odd}}(i\Omega)=\frac{c_f^4 \Omega^{8/3}}{k_F^3v_F^3} m^2(m^2-1)^2\times 0.077056\,.
    \end{equation}  If we try to continue this result to $\Omega\ll T$, we expect it to be proportional to $T^{8/3}$. However, it is subdominant to the effects of thermal fluctuation as we compute below.

    In the opposite regime $\Omega\ll T$, the effect of boson thermal mass becomes important. The thermal part of the fermion self-energy also becomes more important than the quantum part. Substituting $A(i\omega)\approx -\Sigma_T(i\omega)$ where $\Sigma_T$ is given by Eq.\eqref{eq:SigmaT}, we obtain
    \begin{equation}\label{eq:lambda_odd_A_T}
      \lambda_m^{\text{odd, thermal}}(i\Omega)=\frac{g^6 m^2 \left(m^2-1\right)^2 T^3}{128 \Delta(T)  k_F^4
   v_F^5}\,.
    \end{equation} Since $\Delta(T)\sim T^{1/2}\ln^{1/2}(1/T)$, this part is more dominant than Eq.\eqref{eq:lambda_odd_A_Omega} at zero frequency and low temperatures.

      \textbf{The PNFL (B):}
       In the perturbative NFL regime, the boson propagator is the same as in the NFL regime, except that the quantum part of the self-energy $\Sigma_Q$ is much smaller than the bare frequency term. At finite $T$, there is an additional competition between the thermal self-energy $\Sigma_T$ and the bare $i\omega$ term.  Here we assume that $\Sigma_T\ll i\omega$ in the PNFL regime. Therefore, we can substitute $A(i\omega)\approx i\omega$ in Eq.\eqref{eq:lambda_odd_eval}. Then the form of the integrand in Eq.\eqref{eq:lambda_odd_eval} is independent of temperature, and this allows evaluating \eqref{eq:lambda_odd_eval} by replacing the frequency integral with Matsubara sum. The result is (see Appendix.~\ref{app:finiteT})
       \begin{equation}\label{eq:lambda_odd_B_main}
  \lambda_m^{\text{odd}}(i\Omega)=\frac{c_f^2(2\pi T)^{10/3}}{k_F^3 v_F^3} \times m^2(m^2-1)^2{\mathcal L}\left(\frac{\Omega}{2\pi T}\right)\,,
\end{equation} where
\begin{equation}\label{}
\begin{split}
  &{\mathcal L}(x)=\frac{1}{12\sqrt{3}x}\big[x(5x^2-2)H_{-1/3}(x-1)\\
  &+(2-9x^2)H_{-4/3}(x-1)+6x H_{-7/3}(x-1)\\&-2H_{-10/3}(x-1)\big]\,.
\end{split}
\end{equation} The limiting cases are
    \begin{equation}\label{eq:lambda_odd_B_omega}
      \lambda_m^{\text{odd}}(i\Omega)=\frac{c_f^2 \Omega^{10/3}}{k_F^3 v_F^3}m^2(m^2-1)^2\times \frac{249\sqrt{3}}{7280}\,,\quad \Omega\gg T\,,
    \end{equation} and
    \begin{equation}\label{eq:lambda_odd_B_T}
    \begin{split}
      &\lambda_m^{\text{odd}}(i\Omega)=\frac{c_f^2 T^{10/3}}{k_F^3 v_F^3} m^2(m^2-1)^2\\
      &\times \frac{\Gamma(7/3)\left(18\pi^2\zeta(4/3)+35 \zeta(10/3)\right)}{27\sqrt{3}}\,,\quad \Omega\ll T.
    \end{split}
    \end{equation} 

       \textbf{The small-angle scattering FL (C):}
    In this regime, we substitute the boson propagator
    $$
        D(\vn{q},i\Omega)=\frac{1}{\vn{q}^2+m_b^2+\gamma|\Omega|/\vn{q}}
    $$into Eq.\eqref{eq:lambda_odd_eval}. In the FL regime, the damping part $|\omega|^2\ln|\omega|$  of the fermion self-energy\eqref{eq:SigmaFLC_main} is small compared to the $i\omega$ term, but there is a renormalization of the quasiparticle residue and therefore we substitute $A(i\omega)=(1+\pi c_f')i\omega$.
We obtain
\begin{equation}\label{eq:lambda_odd_C}
\begin{split}
  &\lambda_m^{\text{odd}}(i\Omega)=c_f' \frac{m^2(m^2-1)^2 m_b^2}{120 k_F^2} \frac{(1+\pi c_f')^2}{k_F^2 v_F^2}\\
  &\times\left[10\pi\Omega(\Omega^2-\pi^2 T^2)+\frac{\left(3\Omega^2-8\pi^2 T^2\right)\left(\Omega^2-4\pi^2 T^2\right)}{\omega_\text{FL}}\right]\,.
\end{split}
\end{equation} Here the cutoff dependent terms are dropped, and once reinstated they will modify the coefficients of $\Omega$ and $\Omega^3$ terms (see Appendix.\ref{app:continue}).

\textbf{The large-angle scattering FL (D):} This regime is qualitatively different due to large-angle scattering. We substitute the propagator
$$
D(i\Omega,\vn{q})=\frac{1}{m_b^2+\gamma\frac{|\Omega|k_F}{\vn{q}\sqrt{k_F^2-\vn{q}^2/4}}}\,,
$$ and following the qualitative analysis in Appendix.~\ref{app:softD}, we obtain
\begin{equation}\label{}
  \lambda_m^\text{odd}(i\Omega)\sim A_1 \Omega^3+B_1 \Omega^4 (m^2-1)^2 \ln m\,.
\end{equation} This is consistent with the result of \cite{PJLedwith2019}.

The crossover from the FL regime (C) to the NFL regime (A) and the PNFL regime (B) can also be obtained by substituting $m_b\sim \Omega^{1/3}$ and $c_f'\sim \Omega^{-1/3}$. Due to the $(1+\pi c_f')$ factor in Eq.\eqref{eq:lambda_odd_C}, there can be two crossovers depending on whether $c_f'\gg 1$ (from C to A) or $c_f'\ll 1$ (from C to B). Finally, the crossover between the two FL regimes (C) $\to$ (D) can be obtained by setting $m_b\sim k_F/m$, up to logarithmic terms.

\subsection{Pomeranchuk susceptibility}

        To properly interpret the physical meaning of the soft mode eigenvalues we obtained, it is helpful to look at the associated correlation function of fermion bilinears. Due to rotation symmetry, the eigenfunctions are found to be $\cos m\theta$ and $\sin m\theta$ where $\theta$ is the angular coordinate on the FS. These eigenfunctions therefore correspond to the Pomeranchuk order parameters of the FS.
\begin{equation}\label{}
  \hat{P}_m=\sum_{\vec{k}} \cos m\theta_k \psi_{\vec{k}}^\dagger \psi_{\vec{k}}\,.
\end{equation}

The associated correlation function is therefore
\begin{equation}\label{}
  \Pi_m(i\Omega)\equiv -\int_0^\beta\rd \tau e^{i\Omega\tau} \braket{P_m(\tau)P_m(0)}\,.
\end{equation} Using Eq.\eqref{eq:<AB>2}, this becomes
\begin{equation}\label{}
\begin{split}
  &\Pi_m(i\Omega)=\braket{\cos m\theta|\frac{1}{\Omega+L}|\cos m\theta}\\
  &=\frac{\braket{\cos m\theta|\cos m\theta}}{\Omega+\lambda_m(i\Omega)}+\text{const.}
\end{split}
\end{equation} Here $\lambda_m(i\Omega)$ is the soft eigenvalue that we have computed. The final constant term is introduced to correct the error of exchanging integration order (Appendix.~\ref{app:exchange}). Evaluating the inner product, we obtain
\begin{equation}\label{eq:Pim}
  \Pi_m(i\Omega)=\alpha_m \calN \left(\frac{\Omega}{\Omega+\lambda_m(i\Omega)}-1\right)\,,
\end{equation} where $\alpha_0=1$, $\alpha_{m\neq 0}=1/2$ and $\calN$ is the fermion density of states. The fixing constant is the $-1$ in the parenthesis.

From these correlation functions, it is also possible to interpret our eigenvalues as describing angular diffusion on the FS, which we discussed earlier in the section. Consider the even-$m$ eigenvalues in Eq.\eqref{eq:lambda_even}  whose $m$-dependence can be written as
\begin{equation}\label{}
  \lambda_m^{\text{even}}(i\Omega)=m^2 D_{\text{even}}(i\Omega).
\end{equation} The corresponding correlation function is then
\begin{equation}\label{eq:Pim_even}
  \Pi_m^{\text{even}}(i\Omega)=-\alpha_m\calN \frac{m^2 D_\text{even}}{\Omega+m^2 D_{\text{even}}}\,.
\end{equation} Once we interpret $m$ as a momentum, Eq.\eqref{eq:Pim_even} is exactly the density-density correlator of a diffusing fermion \cite{AAltland2010} where $D_\text{even}$ can then be interpreted as the diffusion constant. Since $m$ is the Fourier transform of angular coordinate $\theta$, Eq.\eqref{eq:Pim_even} describes the angular diffusion of fermions on the FS. In the NFL regime (A) and the PNFL regime (B), we substitute $D_\text{even}=A \Omega+B\Omega^{4/3}$. Here the $\Omega^{4/3}$ term is universal with coefficient computed in Eq.\eqref{eq:lambda_even_AB}. However, the linear term in $\Omega$ depends on physics at the cutoff scale, and it is expected to arise due to analyticity (see Appendix.~\ref{app:continue}). Therefore, we obtain $\Pi_{m \neq 0}^\text{even}(i\Omega)\sim A'+B'\Omega^{1/3}$ in the regimes A,B, which agrees with \cite{AKlein2018}.


For the odd-$m$ soft modes, the eigenvalues describe an anomalous diffusion process. The eigenvalue in Eq.\eqref{eq:lambda_odd} can be written as
\begin{equation}\label{}
  \lambda_m^{\text{odd}}(i\Omega)=m^2(m^2-1)^2 D_\text{odd}(i\Omega)\,.
\end{equation} The associated Pomeranchuk susceptibility is
\begin{equation}\label{}
  \Pi_{m}^{\text{odd}}(i\Omega) = -\alpha_m \calN \frac{m^2(m^2-1)^2 D_\text{odd}}{\Omega+m^2(m^2-1)^2 D_\text{odd}}\,.
\end{equation}

We also note that the Pomeranchuk susceptibility vanishes when $m=0,1$ because they are related to the conserved charge and momentum, as required by Ward identity.

\section{Stability of the Migdal-Eliashberg Saddle point}\label{sec:stability}

In this section, we apply the soft modes we obtained in Sec.~\ref{sec:soft} and utilize it to diagnose the stability of the ME saddle point.

\subsection{Criterion for stability}

To test the stability of the theory, we consider the retarded Pomeranchuk susceptibilities $\Pi_m^{(R)}(\omega)=\Pi_m(i\Omega\to \omega+i0)$. Unitarity and causality of a bosonic propagator enforces that
\begin{equation}\label{}
  \omega \Im \Pi_m^{(R)}(\omega)<0\,,
\end{equation} which is equivalent to
\begin{equation}\label{eq:Relambda>0}
  \Re\lambda_m(i\Omega\to \omega+i0)>0\,.
\end{equation} This condition also has a simple meaning in terms of the angular diffusion picture, that the diffusion constants must be positive.

We note that when we analytically continue $\lambda_m(i\Omega)$ to real axis, there can be cutoff-dependent terms when $\lambda_m(i\Omega)\propto \Omega^{\alpha}$ with sufficiently large $\alpha$. We analyze these effects in Appendix.~\ref{app:continue}, and we show that these cutoff dependencies only appear in $\Im \lambda_m(\omega)$, and the real part is universal. 

\subsection{Testing the stability}

With the preparations above, we can test the stability of the Migdal-Eliashberg saddle point \eqref{eq:MET} of the critical FS.

\subsubsection{The FL regimes (C) and (D)}

  For the FL regime (C), both $\Re\lambda_m^{\text{even}}(\omega+i0)$ \eqref{eq:lambda_even_C} and $\Re\lambda_m^{\text{odd}}(\omega+i0)$ \eqref{eq:lambda_odd_C} are positive after analytic continuation, and therefore the FL regime is stable. As for the FL regime (D), we could not test the stability directly because we do not evaluate the numerical pre-factors in this regime. However, based on the computation in \cite{PJLedwith2019}, we do not expect any instability to arise in this regime.

 \subsubsection{The PNFL regime (B)}

  Next we focuse on the PNFL regime (B). The even-$m$ eigenvalues are given by \eqref{eq:lambda_even_AB_T} (the quantum part, with limiting cases \eqref{eq:lambda_even_AB} and \eqref{eq:lambda_even_AB_T>}) and \eqref{eq:lambda_even_AB_thermal} (the thermal part). Here, the quantum part is positive after continuation, and the thermal part is negative after continuation. Since the quantum part decays slower as $T\to 0$, the even-$m$ eigenvalues are stable in the low-$T$ limit. One might be concerned that, at higher $T$ when the thermal mass becomes large, the negative thermal part can overwhelm the quantum part, but this does not happen. The quantum part of the eigenvalue \eqref{eq:lambda_even_AB} is calculated under the assumption that the thermal mass can be ignored when the boson carries non-zero Matsubara frequency. If the at higher temperature the thermal mass cannot be ignored, the system just crosses over to a FL and it is stable. In summary, the even-$m$ eigenvalues show no sign of instability in the PNFL and the NFL.

  We move on to the odd-$m$ eigenvalues.  In the PNFL, the soft eigenvalue \eqref{eq:lambda_odd_B_omega} is proportional to $\Omega^{10/3}$ when $\Omega\gg T$, and after analytic continuation it has a positive real part as $\Re(-i\omega)^{10/3}=(1/2)|\omega|^{10/3}$. One can also check that the general expression for arbitrary $\Omega/T$ \eqref{eq:lambda_odd_B_main} is also positive after analytic continuation. Therefore, the PNFL regime is stable.

  \subsubsection{The NFL regime (A)}

  Finally, we examine the stability of the NFL regime (A). The even-$m$ soft modes are the same as the PNFL regime (B) and show no sign of instability.  However, the odd-$m$ eigenvalues \eqref{eq:lambda_odd_A_Omega} are problematic.
  They are proportional to $\Omega^{8/3}$ for $\Omega\gg T$, and after continuation its real part is negative $\Re(-i\omega)^{8/3}=(-1/2)|\omega|^{8/3}$. Since this part satisfies the $\Omega/T$ scaling, so we expect it to scale as $T^{8/3}$ when $\Omega\ll T$. The contribution from thermal fluctuation \eqref{eq:lambda_odd_A_T}, is positive and $\omega$-independent, and it scales as $T^{5/2}$ up to log terms. Therefore, when $\Omega\ll T$ the odd-$m$ eigenvalue is stable, but at sufficiently low $T$ and large $\Omega$ it will be dominated by \eqref{eq:lambda_odd_A_Omega}. Therefore, the critical FS described by the Migdal-Eliashberg saddle point is unstable in the NFL regime at sufficiently low temperatures.

\subsection{Discussion}

    We concluded that the Migdal-Eliashberg saddle point is unstable in the NFL regime. This is a surprising result of our analysis, and it is useful to make clear the assumptions that go into this result:
\begin{enumerate}
  \item The saddle point on which we focus is the Migdal-Eliashberg saddle point \eqref{eq:MET}. It means that the fermion self-energy is given by the rainbow diagram and the boson self-energy is given by the fermion bubble, both with vertex corrections ignored.
  \item The Bethe-Salpeter ladder has the structure given by Eq.\eqref{eq:KBS}. This means that it consists of blocks of Maki-Thompson and Aslamazov-Larkin like Feynman diagrams, concatenated by rails of (full) fermion Green's functions.
  \item Momentum needs to be conserved. This means that the FS should be small enough to suppress umklapp scattering and that the disorder should only enter perturbatively. If the disorder were too large, the saddle point would be dominated by the disorder and flow to a marginal Fermi liquid \cite{HGuo2022a}.
  \item The FS must be circular or convex \update{(and inversion symmetric)} for the leading-order result of the odd-$m$ eigenvalue to cancel due to the kinematic constraint. When the FS is concave, the contribution from $\delta_q^1 L_m^{(0)}$ does not cancel, and we expect that the odd$m$ eigenvalues will scale similarly to the even-$m$ eigenvalues.
   \item The temperature $T$ should be sufficiently low, or the probing frequency $\Omega$ should be sufficiently high to overcome the thermal fluctuation of the boson.
 \end{enumerate}
 The above assumptions are also satisfied by the double large $N$-small-$(z_b-2)$ expansion ($z_b$ is the boson dynamic exponent) \cite{CNayak1994,DFMross2010,ZDShi2023a,ZDShi2024}, so we expect the instability to also appear when the calculation is extrapolated to $N=\mathcal{O}(1)$ and $z_b=3$.

 \update{Our result can be interpreted either as a physical instability  or an inconsistency of the ME theory. Which interpretation is correct needs to be determined by further numerical simulations. We note that the regime where the potential instability/inconsistency appears is where $\Sigma_Q(i\omega)\gg \omega$, which has not been explored in previous numerics \cite{AKlein2020,XYXu2020,Iesterlis2021}.  We will proceed with our discussion assuming that it is a physical instability.}

Previously, the ferromagnetic quantum critical point has been shown to be unstable due to the static spin susceptibility developing more singular corrections than $\vn{q}^2$ \cite{JRech2006} or the Pomeranchuk instability that preempts the QCP (one of the Landau parameters hits $-1$) \cite{AVChubukov2009}. These instabilities do not involve any dissipation, and they are described by the LHS (streaming term + Landau parameter) of the kinetic equation \cite{LDLandau1956,AAAbrikosov1963}. However, the instability that we discover here is related to the dissipation term in the charge channel, which appears on the RHS (collision term) of the kinetic equation.

Since the scaling of the eigenvalues is captured by counting the phase space available for scattering, the same diagram would yield the same scaling if we apply it to FS coupled to emergent gauge field. However, since the Pomeranchuk order parameter is not gauge invariant if the fermions carry gauge charge, additional diagrams due to the gauge coupling must be included and this will be left for future study.


We discuss the implication of possible metallic states due to the instability.

 First, since the unstable odd-$m$ eigenvalues rely on special cancellations related to convex FS, the first example is to consider a concave FS. If that happens, we expect the odd-$m$ eigenvalues will then be comparable to the even-$m$ eigenvalues $\lambda_m^{\text{odd}}\sim\lambda_m^{\text{even}}$, which are free from instability.

If we instead start with a circular/convex FS, which is assumed in many models in the literature, the system can be stabilized by thermal fluctuation as shown in Eq.\eqref{eq:lambda_odd_A_T}. For the system to be stable, the thermal fermion self-energy $\Sigma_T(i\omega)$ needs to be larger than the quantum part $\Sigma_Q(i\omega)$ in the interval $0<\omega<\omega_P$, so that the $\omega^{2/3}$ self-energy is only identifiable when $\Sigma(i\omega)\ll \omega$. This scenario seems to be consistent with the Monte-Carlo simulations \cite{AKlein2020,XYXu2020} and the numerical solution of the  saddle point equations \cite{Iesterlis2021}. This description, however, cannot persist down to zero temperature, and the low-temperature cutoff is set by the irrelevant operator that defines the thermal mass.

\begin{figure}[htb]
  \centering
  \includegraphics[width=0.5\linewidth]{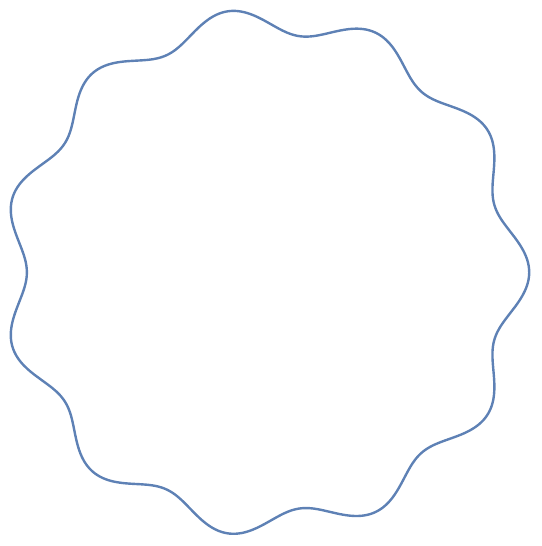 }
  \caption{A  convex FS that undergoes spontaneous deformation and becomes locally concave}\label{fig:fuzzyFS}
\end{figure}

When we decrease the temperature so that the instability becomes significant, the  convex FS can undergo a Pomeranchuk transition and spontaneously deform in some high angular harmonic channels.  The resulting FS may be close to convex at large scales in the momentum space but is concave due to finer structures (see Fig.~\ref{fig:fuzzyFS}).  The exact magnitude of the deformation is, however, not determined by the Gaussian fluctuation analysis, and it requires knowing the higher order $1/N$ effects. One notable feature  is that because the instability happens in the odd angular harmonic channel, the resulting state spontaneously breaks both inversion and time-reversal symmetry.


\begin{figure}[htb]
  \centering
  \includegraphics[width=0.95\columnwidth]{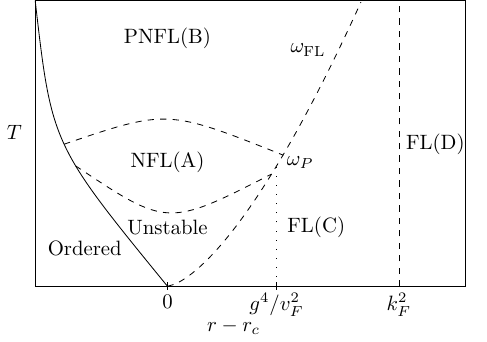}
  \caption{Proposed phase diagram of the Ising-nematic quantum critical point in terms of temperature $T$ and the tuning parameter $r$. The QCP becomes unstable in the low $T$ region. The solid line is a phase transition and dense dashed lines are crossovers.  At finite $T$, the QCP NFL is stabilized by thermal fluctuation. The line $\omega_\text{FL}$ is the crossover line from the FL regime to the NFL regimes, similar to Fig.~\ref{fig:pd}. The scale $\omega_P$ is defined in Eq.\eqref{eq:omegaP_main}, where the vertical dotted line connected to it is a guide to eye. }\label{fig:pdT}
\end{figure}

In Fig.~\ref{fig:pdT}, we show a proposed phase diagram of the scenario above. The low-temperature limit of the QCP is not reachable due to the instability. Above the unstable region, there is a NFL regime which is still labeled by A. However, in this NFL the fermion self-energy is dominated by the thermal fluctuation part $\Sigma_T$ (see Eq.\eqref{eq:SigmaT}). The boundary between the NFL (A) regime and the unstable region is given by $\Sigma_T\sim \omega_P$. When $\Sigma_T\gg \omega_P$, the thermal self-energy $\Sigma_T$ can overcome the problematic quantum part $\Sigma_Q$ for all frequencies in the window $0<\omega<\omega_P$.  At higher temperatures, $\Sigma_T$ can become smaller than $T$, and the system crosses over to the PNFL regime with the boundary to the NFL (A) regime defined by $\Sigma_T\sim T$. The crossover to the FL (C) region is defined by the line $T\sim \omega_\text{FL}=m_b^3/\gamma$. Interestingly, we find these three crossover lines to come close together near the region $m_b\sim g^2/v_F$ and $T\sim\omega_P$. This is because the three crossover lines are all defined by equating the inelastic self-energy with the bare frequency term.

If we demand that the circular/convex FS is stable at zero temperature, we will have to consider theories with different dynamical exponents $z_b$.  We note that in the non-random large $N$ expansion \cite{MAMetlitski2010} $z_b=3$ up to three loop and there can be a non-zero fermion anomalous dimension $\eta_f$. At higher loops $z_b$ also receives correction \cite{THolder2015,THolder2015a}. However, treating $\eta_f$ is beyond the scope of current work because it requires introducing momentum dependence into the self-energy, so we will restrict to $\eta_f=0$.  We can redo the calculation for general boson dynamical exponent $z_b$ by changing the propagator $D^{-1}=\vn{q}^{z_b-1}+\gamma|\Omega|/\vn{q}$. 
 For $2\leq z_b\leq3$, the PNFL has $\lambda_m^\text{even}\propto -\Omega^{4/z_b}$ and $\lambda_m^\text{odd}\propto \Omega^{2+4/z_b}$ and it is always stable. The NFL regime has $\lambda_m^\text{odd}\propto \Omega^{8/z_b}$ and the stability requires $2<z_b<8/3$.  There are known examples that fall into the stability range. The first one is the NFL arising from bandwidth-tuned Mott transition \cite{TSenthil2008}, which has $z_b=2$ and it is marginally stable. Another example is the NFL near a ferrormagnetic QCP studied in \cite{SPRidgway2015}  via functional renormalization group, which yields $\eta_f=0$ and $z_b=13/5$.

We also note some other possible implications. First, the appearance of an instability can always be interpreted as the QCP being actually first order.  Second, the NFL might be unstable towards competing instabilities.
 On the one hand, when the critical boson couples with an even-parity form factor, it is known that the boson facilitates the pairing of fermions \cite{MAMetlitski2015,Iesterlis2021}, and the strongest pairing $T_c$ is found to be comparable with the NFL energy scale $\omega_P$ \cite{YWang2016}.
On the other hand, when the critical boson couples with an odd-parity form factor (such as U(1) gauge field), the system might develop a $2k_F$ charge-density wave order depending on the strength of the interaction \cite{Iesterlis2021}. Finally, there is some calculation suggesting that the FS is destroyed by quantum fluctuation at low energy \cite{DVKhveshchenko1993,DVKhveshchenko1994b}, but this seems only to be possible when the FS is anomaly-free \cite{DCLu2023}.

\section{The Effective Action of the Soft Modes}\label{sec:eft}

In this section, we try to generalize our results in previous sections to finite CoM momentum $p=(i\Omega,\vec{p})$ in the form of a Gaussian effective action. Inspired by the previous discussion at $\vec{p}=0$, we conjecture that the soft modes at finite $\vec{p}$ can be described by the ansatz:
\begin{equation}\label{eq:soft_ansatz}
  \delta G(k;p)=\left[iG(k+p/2)-iG(k-p/2)\right]\phi(\theta_k;p)\,.
\end{equation}  Here $k$ denotes the relative 3-momentum and $p$ denotes the CoM 3-momentum. The soft mode is parameterized by the function $\phi(\theta_k;p)$ which only depends on $k$ through the angle $\theta_k$ on the FS. Substituting this into the Gaussian action \eqref{eq:S=GKBSG}, we therefore obtain the effective action
\begin{equation}\label{eq:action_p}
  S[\phi]=-\frac{N}{2}\int\frac{\rd^3 p}{(2\pi)^3}\braket{\phi(-p)|\Omega+i\vec{v}_k\cdot\vec{p}+L|\phi(p)}
\end{equation} Here similarly to Sec.~\ref{sec:fluctuation}, the inner product is now defined as $\braket{A(-p)|B(p)}=\int_{k}A(k;-p)M(k;p)B(k;p)$ where $M(k;p)=iG(k+p/2)-iG(k-p/2)$.  The kinetic operator is defined in three parts $L=L_\text{DOS}+L_\text{MT}+L_\text{AL}$, where
\begin{equation}\label{}
\begin{split}
  &L_\text{DOS}[F](k;p)=W_\Sigma^{-1}\circ M[F]-\left(\Omega+i\vec{v}_k\cdot\vec{p}\right)F\\
  &=i\left[\Sigma(k+p/2)-\Sigma(k-p/2)\right]F(k;p)\,,
\end{split}
\end{equation}
\begin{equation}\label{}
  L_\text{MT}=-W_\text{MT}\circ M\,,
\end{equation}
\begin{equation}\label{}
  L_\text{AL}=-W_\text{AL}\circ M\,.
\end{equation} Therefore, $L_\text{DOS}$, $L_\text{MT}$ and $L_\text{AL}$ have the same integral expressions as in Eqs. \eqref{eq:LMT} and \eqref{eq:LAL}, except that $\Omega$ is replaced by $p=(i\Omega,\vec{p})$.

\subsection{Emergence of bosonization}

We now evaluate the inner product in Eq.\eqref{eq:action_p} and demonstrate that the action can be related to the action of linear bosonization \cite{FDMHaldane2005,LVDelacretaz2022a,AHoughton2000}. The first hint of the resemblance comes from the free piece $(\Omega+i\vec{v}_k\cdot\vec{p})$:
\begin{equation}\label{}
\begin{split}
  &S_\text{free}=-\frac{N}{2}\int\frac{\rd^3 p}{(2\pi)^3} \int\frac{\rd \omega}{2\pi}\int\frac{\calN \rd \xi_k}{2\pi}\int\rd\theta_{kp} \\
  &\times(\Omega+iv_F p\cos\theta_{kp})\phi(\theta_k;-p)\phi(\theta_k;p)\\
  &\times(iG(k+p/2)-iG(k-p/2))\,.
\end{split}
\end{equation} We only need to integrate over $\omega$ and $\xi_k$ whose dependence all appear in the $iG-iG$ factor. Since the integral is not absolute convergent, we need to compute it by doing the $\omega$-integral first, unlike what we have done in the computation of eigenvalues. The result is simple
\begin{equation}\label{eq:intG-G}
\begin{split}
  &\int\calN\rd\xi_k \int{\frac{\rd \omega}{2\pi}}(iG(k+p/2)-iG(k-p/2))\\
  &=i\int\calN\rd\xi_k \left(n_F(\xi_{\vec{k}+\vec{p}/2})-n_F(\xi_{\vec{k}-\vec{p}/2})\right)\\
  &=-i\calN\vec{v}_k\cdot\vec{p}\,.
\end{split}
\end{equation} Here $n_F(\xi)=\int \rd\omega/(2\pi) G(i\omega,\xi)$ is the occupation number in momentum space. The $\xi$-integral in the second line picks up contribution from the whole BZ, where $n_F$ varies from $1$ to $0$.

The free part of the action becomes
\begin{equation}\label{eq:Sfree}
\begin{split}
  &S_\text{free}=-\frac{N}{2}\int\frac{\rd^3 p}{(2\pi)^3}\int\frac{\calN \rd\theta_k}{2\pi}\left(-i\vec{v}_k\cdot\vec{p}\phi(\theta_k;-p)\right)\\
  &\times\left(\Omega+i\vec{v}_k\cdot \vec{p}\right)\phi(\theta_k;p)\,,
\end{split}
\end{equation} where $|\vec{v}_k|=v_F$ and points to the direction of $\theta_k$. Notice that this is exactly the momentum space version of the bosonized action of a free Fermi gas (see Ref.~\cite{LVDelacretaz2022a}, Eq.(84)). The equation of motion of Eq.\eqref{eq:Sfree} is therefore a linearize collision-less Boltzmann equation $(\Omega+i\vec{v}_k\cdot\vec{p})\phi(\theta_k;p)=0$.  We remind the reader here that the frequency $\Omega$ is on the imaginary axis. Once we rotate back to real-axis $\Omega\to -i\omega+0$, and Fourier transform $p\to x=(t,\vec{x})$, we would obtain $(\partial_t+\vec{v}_k\cdot\nabla)\phi(\theta_k;t,\vec{x})=0$. As is pointed out in Ref.~\cite{LVDelacretaz2022a}, $\phi(\theta_k;t,\vec{x})$ is not directly interpreted as the local density on the FS, but instead we should interpret $n(\theta_k;t,\vec{x})\propto\vec{v}_k\cdot\nabla\phi(\theta_k;t,\vec{x})$ as the local density on the FS.

We continue to include the effect of collisions which are encapsulated in the kinetic operator $L$. We will first analyze a weak coupling limit where we can reproduce the kinetic equation of Landau's FL theory \cite{LDLandau1956,AAAbrikosov1963}. Next, we will move on to the strong coupling regime where we obtain a kinetic equation first derived by Prange and Kadanoff \cite{REPrange1964}, which is later applied to study NFLs \cite{YBKim1995a,SEHan2023,IMandal2022,KRIslam2023}.

\subsection{Landau's kinetic equation}

We first consider a weak coupling limit where FL theory is expected to apply, which corresponds to the regimes (C) or (D) in the phase diagram (Fig.~\ref{fig:pd}). In this regime the Green's function can be written in the form
\begin{equation}\label{eq:GFL}
  G(i\omega,\vec{k})=\frac{Z}{i\omega-\frac{m}{m^*}\xi_{\vec{k}}}+\Phi(i\omega,\vec{k})\,,
\end{equation} where $Z$ is the quasiparticle residue, $m^*$ is the effective mass and $m$ is the bare mass, and $\Phi(i\omega,\vec{k})$ represents the incoherent part.

We compute $\braket{\phi(-p)|L_\text{MT+DOS}|\phi(p)}$ and represent the result in the form of Landau interaction. The expression for  $L_\text{MT+DOS}$ is
\begin{equation}\label{eq:LMT_FL}
\begin{split}
  &L_\text{MT+DOS}[\phi(p)](k)=g^2\int\frac{\rd \omega'}{2\pi}\int\frac{\rd^2 \vec{k'}}{(2\pi)^2} D(k-k')\\
  & \left[iG(i\omega'+i\Omega/2,\vec{k'}+\vec{p}/2)-iG(i\omega'-i\Omega/2,\vec{k'}-\vec{p}/2)\right]\\
  &\times\left[\phi(\theta;p)-\phi(\theta';p)\right]\,.
\end{split}
\end{equation} Here $k=(i\omega,\vec{k})$ and $k'=(i\omega,\vec{k'})$ represent 3-momentum, and $\vec{k}$, $\vec{k'}$ are momenta close to the FS with directions parameterized by $\theta,\theta'$ respectively. We then expand the $iG-iG$ factor to linear order in $\Omega$ and $\vec{p}$, by assuming the important contribution arises from the quasiparticle pole:
\begin{equation}\label{eq:Gexpand_FL}
\begin{split}
   & iG(i\omega'+i\Omega/2,\vec{k'}+\vec{p}/2)-iG(i\omega'-i\Omega/2,\vec{k'}-\vec{p}/2)   \\
     &= \frac{\Omega }{Z} \left[G(k')^2\right]_\Omega+\frac{i\vec{v}_{k'}^*\cdot \vec{p}}{Z}\left[G(k')^2\right]_{p}\,.
\end{split}
\end{equation} Here $\left[G(k')^2\right]_{\Omega,p}$ means sending $p\to 0$ in $G(k')G(k'+p)$ with $p=(i\Omega,0)$ and $p=(0,\vec{p})$ respectively. $\vec{v}^*_{k}=(m/m^*)\vec{v}_k$ is the renormalized velocity. The difference between the two limits is given by the standard identity \cite{AAAbrikosov1963}
\begin{equation}\label{eq:FLG2-G2}
  \left[G(k')^2\right]_p-\left[G(k')^2\right]_\Omega=-\frac{Z^2 m^*}{m} 2\pi\delta(\omega')\delta(\xi')\,,
\end{equation} where $\xi'$ is the (bare) dispersion of $\vec{k'}$. Substituting Eqs.\eqref{eq:Gexpand_FL} and \eqref{eq:FLG2-G2} into Eq.\eqref{eq:LMT_FL}, we obtain
\begin{equation}\label{eq:LMT2_FL}
\begin{split}
  &L_\text{MT+DOS}\left[\phi(p)\right](\omega=0,|\vec{k}|=k_F)\\
  &= g^2 \int\frac{\rd^3 k'}{(2\pi)^3} D(k-k') \left[G(k')^2\right]_\Omega \\
  &\times\frac{\Omega+i\vec{v}^*_{k'}\cdot \vec{p}}{Z}(\phi(\theta;p)-\phi(\theta';p))\\
  &-g^2\calN\int\frac{\rd \theta'}{2\pi} ZD(\theta,\theta') i\vec{v}_{k'}\cdot\vec{p} (\phi(\theta;p)-\phi(\theta';p))\,.
\end{split}
\end{equation} Here the second term comes from Eq.\eqref{eq:FLG2-G2}, where the interaction $D(k,k')$ is projected to $\omega=\omega'=0$ and $\vn{k}=\vn{k'}=k_F$. To proceed, we rewrite the result in the following way
\begin{equation}\label{eq:FLeq}
\begin{split}
  (\Omega+i\vec{v}_k\cdot\vec{p}+L_\text{MT+DOS})\left[\phi(p)\right]\\
  =\left(\frac{\Omega+i\vec{v}_k^*\cdot\vec{p}}{Z}+L_\text{Landau}\right)[\phi(p)]\,.
\end{split}
\end{equation} Here, the free particle term $(\Omega+i\vec{v}_k\cdot\vec{p})$ is renormalized to $(\Omega+i\vec{v}_k^*\cdot\vec{p})/Z$ according to FL theory, and it arises from the terms that depend on $\phi(\theta)$ in Eq.\eqref{eq:LMT2_FL}. By comparing both sides of Eq.\eqref{eq:FLeq}, we can derive the following expression for $Z$ and $m^*$.
\begin{equation}\label{eq:Z_FL}
  \frac{1}{Z}=1+\frac{g^2}{Z}\int\frac{\rd^3 k'}{(2\pi)^3}D(k-k')\left[G(k')^2\right]_\Omega\,,
\end{equation}
\begin{equation}\label{eq:meff_FL}
  \frac{1}{m^*}=\frac{1}{m}-\frac{1}{m}g^2\calN \int\frac{\rd \theta'}{2\pi} ZD(\theta,\theta')\cos(\theta-\theta')\,.
\end{equation} In deriving Eq.\eqref{eq:meff_FL}, we have used the Ward identity for momentum conservation and Galilean invariance to change $\vec{v}_{k'}\cdot\vec{p}\phi(\theta)$ in the third line of Eq.\eqref{eq:LMT2_FL} to $\vec{v}_k \cdot \vec{p}\phi(\theta)$. The results for $Z$ and $m^*$ agree with FL theory if we identify $D(k-k')=Z\Gamma^{\Omega}(k,k')$ where $\Gamma^{\Omega}$ is the fully-antisymmetrized interaction vertex of FL theory \cite{AAAbrikosov1963,DLMaslov2010a}.

Combining everything, we therefore obtain the effective action for a FL to be
\begin{equation}\label{eq:SFL}
\begin{split}
 & S_\text{FL}=-\frac{N}{2}\int\frac{\rd^3 p}{(2\pi)^3}\int\frac{\calN \rd \theta}{2\pi}(-i\vec{v}_k\cdot\vec{p}\phi(\theta;-p))\\
 & \Bigg[\frac{\Omega+i\vec{v}_k^*\cdot\vec{p}}{Z}\phi(\theta;p)+g^2\calN\int\frac{\rd \theta'}{2\pi}ZD(\theta,\theta')i\vec{v}_{k'}\cdot\vec{p}\phi(\theta';p)\\
 &-g^2\int\frac{\rd^3 k'}{(2\pi)^3}D(k-k')[G(k')^2]_\Omega\frac{\Omega+i\vec{v}^*_{k'}\cdot \vec{p}}{Z}\phi(\theta';p)\Bigg]\,.
\end{split}
\end{equation}  At first glance, the terms in the bracket of Eq.\eqref{eq:SFL} is not identical to the kinetic equation of a FL \cite{LDLandau1956,AAAbrikosov1963}, as the $D(\theta,\theta')$ term resembles a Landau interaction but the $D(k-k')$ term seems to be spurious. However, the $D(k-k')$ term is crucial for getting the correct correspondence to FL theory. If we drop the $D(k-k')$ term, we would obtain the Landau interaction to be $f_\text{wrong}(\theta,\theta')=Z^2 D(\theta,\theta')$. As discussed in the last paragraph, to obtain the correct $Z$ and $m_*$ we need to identify $D=Z \Gamma^\Omega$, so we would have $f_\text{wrong}=Z^3 \Gamma^\Omega$, which disagrees with FL theory especially when $Z\ll 1$.

To see how the correct correspondence is established, we take the EoM of \eqref{eq:SFL} by setting the terms in the bracket to zero. Notice that the $D(k-k')$ terms act exactly on $(\Omega+i\vec{v}_{k'}^*\cdot \vec{p})\phi(\theta')$, so the EoM can be simplified by repeated substitution, and we obtain
\begin{equation}\label{}
  \left(\Omega+i\vec{v}_k^*\cdot\vec{p}\right)\phi(\theta;p)+g^2\calN\int\frac{\rd \theta'}{2\pi} \Gamma(\theta,\theta')i\vec{v}_{k'}\cdot \vec{p}\phi(\theta';p)=0\,,
\end{equation} where $\Gamma(\theta,\theta')=\left.\Gamma(k,k')\right|_{|\vec{k}|=\vn{k'}=k_F,\omega=\omega'}$, and $\Gamma(k,k')$ satisfies the integral equation
\begin{equation}\label{eq:Gamma_ladder_FL}
\begin{split}
  &\Gamma(k,k')=Z^2D(k,k')\\
  &+g^2\int\frac{\rd^3 k''}{(2\pi)^3}D(k-k'')\left[G(k'')^2\right]_\Omega \Gamma(k'',k')\,.
\end{split}
\end{equation} This is exactly the ladder sequence considered in \cite{DLMaslov2010a} to correct the RPA level interaction vertex $\Gamma^\Omega$ (in our case $Z^2D$) to the actual interaction vertex $\Gamma^\Omega$. The result of \cite{DLMaslov2010a} shows that $\Gamma(k,k')=Z^2D(k,k')/Z_\Gamma$, where $Z_\Gamma=C Z$ and the coefficient $C$ is an $\mathcal{O}(1)$ number that depends on the form factor of the interaction and  angular averaging schemes. In Appendix.~\ref{app:FLladder}, we review this calculation and find $C\approx 1$ when forward scattering is dominant.

Therefore, the final EoM is
\begin{equation}\label{}
  (\Omega+i\vec{v}_k^*\cdot\vec{p})\phi(\theta;p)+g^2\calN \int\frac{\rd \theta'}{2\pi}ZD(\theta,\theta')i\vec{v}_{k'}\cdot\vec{p}\phi(\theta';p)=0\,.
\end{equation} This is exactly Landau's kinetic equation after identifying $D=Z \Gamma^\Omega$, and the Landau interaction is $f(\theta,\theta')=ZD(\theta,\theta')=Z^2 \Gamma^\Omega(\theta,\theta')$. Furthermore, due to $C\approx 1$ it is also possible to rewrite the action Eq.\eqref{eq:SFL} into a simpler form (see Appendix.~\ref{app:FLladder} for detail)
\begin{equation}\label{eq:SFL_final}
\begin{split}
 & S_\text{FL}=-\frac{N}{2}\int\frac{\rd^3 p}{(2\pi)^3}\int\frac{\calN \rd \theta}{2\pi}(-i\vec{v}_k\cdot\vec{p}\phi(\theta;-p))\\
 & \Bigg[\left(\Omega+i\vec{v}_k^*\cdot\vec{p}\right)\phi(\theta;p)\\
 &+g^2\calN\int\frac{\rd \theta'}{2\pi}ZD(\theta,\theta')i\vec{v}_{k'}\cdot\vec{p}\phi(\theta';p)\Bigg]\,.
\end{split}
\end{equation} Here the $1/Z$ renormalization of the $\Omega+i\vec{v}^*_{k'}\cdot \vec{p}$ is offset by $Z_\Gamma\approx Z$.

Finally, we note that the Aslamazov-Larkin diagrams vanish within FL theory (see Appendix.~\ref{app:FLAL} for detail), so Eq.\eqref{eq:SFL_final} is the full effective action for the FS soft modes in a FL. We note that it has the same form as the Gaussian-level action of coadjoint orbit bosonization \cite{LVDelacretaz2022a}.

\subsection{Moving towards the strongly coupled regime}\label{sec:LDtoPK}

Our next goal is to obtain the effective action in the strongly coupled regime, where FL theory no longer applies. Within FL theory, only the nonanalyticity associated with the particle-hole bubble is considered, and the boson propagator $D$ is assumed to be analytic. As we will demonstrate, the departure from FL theory exactly arises from the Landau-damping term in $D$, which is nonanalytic.

To demonstrate the crossover out of the FL regime, we again study the kinetic operator defined by Eq.\eqref{eq:LMT_FL}. Instead of handling the integral using FL identities, we directly evaluate the integral by computing the $\omega'$-integral first and $\xi'$-integral next. This integration order is only plausible if the inelastic self-energy $\Im\Sigma(i\omega\to \omega+i0)$ of the fermion is much smaller than $|\omega|$, so we will restrict our analysis to regimes (B) and (C) in Fig.~\ref{fig:pd} where this assumption holds. Therefore, we assume fermion Green's function to take the form
\begin{equation}\label{}
  G(i\omega,\xi)=\frac{Z}{i\omega-\xi_k^*}\,,
\end{equation} where $\xi_k^*=(m/m^*)\xi_k$. The boson Green's function is
\begin{equation}\label{}
\begin{split}
   D(i\Omega,\vn{q})&=\frac{1}{\vn{q}^2+m_b^2+\frac{\gamma|\Omega|}{\sqrt{\vn{q}^2+|\Omega|^2/v_F^2}}}\\
  &=\frac{1}{\vn{q}^2+m_b^2}\frac{1}{1+\kappa_q|\Omega|}\,,
\end{split}
\end{equation} where $\kappa_q=\gamma/(\sqrt{\vn{q}^2+|\Omega|^2/v_F^2}(\vn{q}^2+m_b^2))$. Here, we have included the upper cutoff of the $1/\vn{q}$ Landau damping form by changing $1/\vn{q}\to 1/\sqrt{\vn{q}^2+|\Omega|^2/v_F^2}$. Therefore, when $|\Omega|\ll\Lambda_q=v_F\vn{q}$, the Landau damping is important in the boson Green's function. In contrast, when $|\Omega|\gg \Lambda_q$, $D(i\Omega,\vn{q})\approx D(\vn{q})$ becomes approximately static.

We evaluate Eq.\eqref{eq:LMT_FL}, and we obtain
\begin{equation}\label{eq:LMT_F}
\begin{split}
  &L_\text{MT+DOS}[\phi(p)](\omega,\vn{k}=k_F)=\calN g^2\left(\frac{Z m^*}{m}\right)\int\frac{\rd \theta'}{2\pi} \\
  & \frac{1}{\vn{q}^2+m_b^2}\calF(\kappa_{q},\Lambda_{q})\left[\phi(\theta)-\phi(\theta')\right]\,.
\end{split}
\end{equation} Here $\vn{q}$ is related to $\theta,\theta'$ via the relation $\vn{q}=2k_F\sin\frac{\theta-\theta'}{2}$, and $\Lambda_q$ enters as the UV cutoff of the $\xi'$-integral. The function $\calF$ is computed in Appendix.~\ref{app:FLtoME}. We are interested in the following limiting behavior of $\calF$. First, in the weak coupling limit $\kappa_q\to 0$ (as $g\to 0$ or $m_b\to\infty $), we obtain
\begin{equation}\label{}
  \calF(\kappa_q\to0,\Lambda_q)=-i\vec{v}_{k'}^*\cdot\vec{p}\,,
\end{equation}  which recovers the Landau FL behavior.  The second interesting limit happens when $\kappa_q\Lambda_q\gg 1$, where we obtain
\begin{equation}\label{}
\begin{split}
  &\calF(\kappa_q\Lambda_q\to\infty)=\frac{1}{\kappa_q}\left[\ln\left(1+\kappa_q(\omega+\Omega/2)\right)\right.\\
  &\left.+\ln\left(1-\kappa_q(\omega-\Omega/2)\right)\right]\,.
\end{split}
\end{equation} We note this is exactly the result when the integration order is switched to performing the $\xi'$-integral before the $\omega'$ integral:
\begin{equation}\label{}
\begin{split}
   &\calF(\kappa_q\Lambda_q\to\infty)  = \int_{-\Omega/2}^{\Omega/2}\rd \omega' \int_{-\infty}^{\infty}\frac{\rd \xi'}{2\pi} \frac{1}{1+\kappa_q|\omega-\omega'|}\\
   &\times\left[\frac{1}{i\omega'+i\Omega/2-\xi'-\vec{v}_{k'}^*\cdot\vec{p}}-\frac{1}{i\omega'-i\Omega/2-\xi'+\vec{v}_{k'}^*\cdot\vec{p}}\right]\,.
   \end{split}
\end{equation} This is anticipated because the $\kappa_q\Lambda_q\to\infty$ limit corresponds to strong Landau damping in the boson propagator. As shown in Appendix.~\ref{app:exchange}, in this limit it is legitimate to exchange the order of $\xi'$ and $\omega'$-integral.
Notice that in this regime $Zm^*/m\approx 1$ \cite{DLMaslov2010a}, so the analysis of the kinetic operator falls back to the Eliashberg analysis of the $\vec{p}=0$ case.

After obtaining $\calF$, we still need to perform the remaining angular integral over $\theta'$. This is effectively an integral over $\vn{q}$ as $\vn{q}=2k_F\sin\frac{\theta-\theta'}{2}$. As we have seen, the behavior of $\calF$ depends on $\kappa_q\Lambda_q=v_F \gamma/(\vn{q}^2+m_b^2)$. If we are deep in the FL phase where $m_b$ is large, both $\kappa_q$ and $\kappa_q\Lambda_q$ is small and the integral \eqref{eq:LMT_F} falls back to the FL regime. On the other hand, when we approach the QCP and reduce $m_b$, $\kappa_q\Lambda_q$ can become large when $\vn{q}=0$ and $m_b$ becomes small. As we have demonstrated, when $\kappa_q\Lambda_q\gg 1$, the result of the integral is the same as the result of the Eliashberg theory described in Sec.~\ref{sec:fluctuation}.  The crossover point can be determined by setting $\kappa_q\Lambda_q\sim 1$ and $\vn{q}=0$, and we obtain $m_b\sim  (m_b)_\text{ME}=\sqrt{g^2 k_F/v_F}$, which sits inside the FL (C) regime in Fig.~\ref{fig:pd}. Therefore, when $m_b<(m_b)_\text{ME}$, the dominant contribution of the integral arises from $\vn{q}\approx 0$, and we obtain the Eliashberg results. Since the Eliashberg result can be obtained by first performing the $\xi'$-integral, it can be generalized to the case where $\Im \Sigma(i\omega\to \omega+i0)\gg \omega$ as the $\xi'$-integral is insensitive to the inelastic self-energy. Furthermore, since the $\vec{p}$ dependence also appears through shifting $\xi'\to \xi'\pm \vec{v}_{k'}\cdot \vec{p}/2$, it also drops out within the Eliashberg theory.  This argument applies to both $L_\text{MT+DOS}$ and $L_\text{AL}$.



Therefore, we can obtain the kinetic equation in the Eliashberg regime. Since the $\vn{p}$-dependence drops out, we can treat the kinetic operators as if it were at $\vec{p}=0$. Therefore, we can directly import the results of the eigenvalue calculations in Sec.~\ref{sec:soft}, and the effective action is
\begin{equation}\label{eq:SME}
\begin{split}
  &S_\text{Eliashberg}=-\frac{N}{2}\int\frac{\rd^3 p}{(2\pi)^3}\int\frac{\calN \rd\theta_k}{2\pi}\left(\Omega\phi(\theta_k;-p)\right)\\
  &\times\left(\Omega+i\vec{v}_k\cdot \vec{p}+\lambda_{-i\partial_{\theta_k}}\right)\phi(\theta_k;p)\,,
\end{split}
\end{equation} where  $\lambda_{-i\partial_{\theta_k}}$ is the soft-mode eigenvalues $\lambda_m$ that we have computed, but with the $m$-dependence Fourier transformed to $-i\partial_{\theta_k}$. Eq.\eqref{eq:SME} also make the physical meaning the eigenvalue explicit: After analytically continue $\lambda(i\Omega)\to \lambda(\omega+i0)$, the real part corresponds to the eigenvalue of the collision integral, and the imaginary part corresponds to renormalization of the streaming term $\Omega+i\vec{v}_k\cdot \vec{p}$. As noted in Appendix.~\ref{app:continue}, under analytical continuation, the real part is independent of the UV cutoff but the imaginary part depends on it. This is consistent with the picture of a kinetic equation: The dissipation rates involve on-shell excitations which only appears in the low-energy and is cutoff-independent, but the renormalization effects involve virtual excitations which can depend on the UV physics.

An additional caveat is the evaluation of Eq.\eqref{eq:intG-G} in the Eliashberg regime. Strictly speaking Eq.\eqref{eq:intG-G} should still hold, but it is inconsistent with the bosonization interpretation of Eq.\eqref{eq:SME}. Instead, we switch the integration order of $\xi$ and $\omega$ and we obtain the current form of Eq.\eqref{eq:SME}, where the $-i\vec{v}_k\cdot \vec{p}$ factor is replaced by $\Omega$. After that, we should interpret the local density on the FS to be $n_\theta\propto \Omega\phi(\theta)$. This interpretation is consistent with Ref.~\cite{SEHan2023}, and we can reproduce the Pomeranchuk susceptibilities computed in Sec.~\ref{sec:soft}.

The EoM of the above action is closely related to the kinetic equation first derived by Prange and Kadanoff \cite{REPrange1964}, which is originally proposed to treat the electron-phonon problem at high temperature where the electron quasiparticle is not well defined. Similar ideas have also been used to derive quantum Boltzmann equations (QBEs) that apply to NFLs \cite{YBKim1995a,SEHan2023,IMandal2022,KRIslam2023}. There are several key differences between our formalism and the QBE approach:

First, in these QBEs the distribution function $f(\omega,\theta_k;p)$ still depends on the frequency $\omega$, but in our formalism, since we focus only on the soft modes, $\phi(\theta_k;p)$ does not depend on $\omega$.

Second, the counterpart of our soft-mode eigenvalue $\lambda_m$ has been called ``generalized Landau interaction" in the QBE literatures \cite{YBKim1995a,SEHan2023,IMandal2022,KRIslam2023}. However, this part is not properly treated in the QBE literatures because the critical boson has been fixed at thermal equilibrium. Diagrammatically, this is equivalent to retaining only the DOS and MT diagrams but not the AL diagrams, and as a consequence momentum conservation is violated. In our formalism, we have retained all the relevant diagrams, and momentum conservation is satisfied.

Third, the QBE formalism is derived by taking the full Keldysh distribution function $F(\omega,\vec{k};p)$ and integrating over $\xi_k$ to arrive at the generalized distribution function $f(\omega,\theta_k;p)$. This is different from the derivation of Landau's kinetic equation, where $F(\omega,\vec{k};p)$ is first integrated against $\omega$ to obtain the quasiparticle distribution function $n(\vec{k};p)$ \cite{AKamenev2023}. In the literature, it has not been discussed before how these two kinetic equations should be related to each other. Our computation thereby clarifies the crossover between these two kinetic equations, and the important ingredient is the Landau damping term in the boson propagator. When the Landau damping is weak, Landau's kinetic equation is more appropriate, and when the Landau damping becomes strong, the Prange-Kadanoff kinetic equation is more appropriate.

Before concluding the section, we note that it is possible to generalize the effective action beyond the Gaussian order. This can be achieved by expanding the $G$-$\Sigma$ action \eqref{eq:S_Gsigma_clean} beyond the Gaussian order, and then substituting the soft-mode ansatz \eqref{eq:soft_ansatz}.  The structure of the $G$-$\Sigma$ action has been discussed in Ref.~\cite{Iesterlis2021}, where the nonlinear vertices can be divided into two types. The first type comes from expanding the $\ln\det$ terms, and the second type comes from expanding the Yukawa interaction term $\Tr(GD\cdot G)$.

\section{Hydrodynamics}\label{sec:hydro}

In this section, we discuss the implications of our formalism in hydrodynamics. In the hydrodynamic regime, the non-conserved quantities quickly relax through local collisions, and only zero modes or soft modes are left. These conserved and quasi-conserved quantities then can start propagating in space. As a result of this propagation, the current-field relation $\vec{j}=\sigma\vec{E}$ is no longer given by a local conductivity $\sigma$, but instead is described by the non-local conductivity $\sigma(\vec{p})$. Therefore, the non-local conductivity can reveal the physics related to the soft modes.

Computation of the non-local conductivity $\sigma(\vec{p})$ can be done by solving the Boltzmann equation that we derived in Sec.~\ref{sec:eft}. Since we are interested in regimes close to the NFLs, we will use the Eliashberg version Eq.\eqref{eq:SME}. We scale $\phi(\theta)$ by $\Omega$ and interpret it as parameterizing the local density on the FS. We phenomenologically insert a driving electric field at finite $\vec{p}$, and obtain:
\begin{equation}\label{eq:QBE}
  \left(\Omega+i\vec{v}_k\cdot \vec{p}+\lambda_{-i\partial_{\theta_k}}\right)\phi(\theta_k;p)=e\vec{E}_{\vec{p}}\cdot\vec{v}_k\,.
\end{equation} Eq.\eqref{eq:QBE} can be analyzed analogous to FL hydrodynamics \cite{PLedwith2019,QHong2020,SKryhin2023}. We consider the steady state $\Omega=0$ and decompose $\phi$ into angular harmonics
\begin{equation}\label{}
  \phi(\theta_k;p)=\sum_{m}e^{im(\theta_k-\theta_p)} \phi_m(p)\,.
\end{equation} Additionally, in the steady state the current needs to be transverse for charge to be conserved, so we require $\vec{E}_{\vec{p}}\perp\vec{p}$. The Boltzmann equation therefore becomes
\begin{equation}\label{eq:QBE_m}
  \frac{i v_F \vn{p}}{2}\left(\phi_{m-1}+\phi_{m+1}\right)+\lambda_m \phi_m= \frac{e E v_F}{2i}(\delta_{m,1}-\delta_{m,-1})\,.
\end{equation}

To compute the conductivity, we need to derive a relation that relates $\phi_1$ to $E$.
According to the Einstein relation, the electric field $\vec{E}_{\vec{p}}$ plays the same role as the gradient of $\phi_0$, so we set $\phi_0=0$. Due to momentum conservation $\lambda_1=0$, so the $m=1$ equation yields $iz \phi_2=eEv_F/(2i)$, where $z=v_F\vn{p}/2$. Eq.\eqref{eq:QBE_m} can be solved by a continuous fraction formula \cite{QHong2020,SKryhin2023}. This can be seen by defining $\Gamma_m=-iz \phi_{m-1}/\phi_{m}$, then Eq.\eqref{eq:QBE_m} becomes ($m\geq 2$):
\begin{equation}\label{}
  \Gamma_m(\vec{p})=\lambda_m+\frac{z^2}{\Gamma_{m+1}}\,.
\end{equation} By repeat substitution, we therefore obtain
\begin{equation}\label{eq:Gammam}
  \Gamma_m(\vec{p})=\lambda_m+\frac{z^2}{\lambda_{m+1}+\frac{z^2}{\lambda_{m+2}+\dots}}\,.
\end{equation}  Therefore the first harmonics is given by
\begin{equation}\label{}
  \phi_{\pm 1}=\pm \frac{e E v_F}{2i}\frac{\Gamma_2}{z^2}\,.
\end{equation} From this we can compute the transverse current by
\begin{equation}\label{}
  j_\perp = e\calN \int\frac{\rd \theta_k}{2\pi} \phi(\theta_k;p) v_F \sin(\theta_k-\theta_p)\,,
\end{equation} where only the $\phi_{\pm 1}$ harmonics contribute, and we obtain the non-local conductivity $j_\perp=\sigma(\vec{p})E$ to be
\begin{equation}\label{eq:sigmaq}
  \sigma(\vec{p})=\frac{ne^2}{2\pi\calN\vn{p}^2\nu(\vec{p})}
\end{equation}Here $n$ is the total fermion density, and $\nu(\vec{p})$ is the kinematic viscosity, which can be written as
\begin{equation}\label{eq:nuq}
  \nu(\vec{q})=\frac{n}{4\pi\calN^2 \Gamma_2(\vec{p})}\,,
\end{equation} and $\Gamma_2(\vec{p})$ is the effective scattering rate obtained in Eq.\eqref{eq:Gammam}. The wavevector $\vec{p}$ should be understood as a typical wavevector of the external drive, such as inverse sample size.

  In the following, we evaluate $\Gamma_2(\vec{p})$ in different regimes.

    \subsection{Conventional Hydrodynamics Regime }

    The conventional hydrodynamic regime is the ultimate long-wavelength regime, where $\vec{p}\to 0$. In this regime, \begin{equation}\label{eq:Gamma2hydro}
      \Gamma_2=\Gamma_\text{hydro}=\lambda_2\,,
    \end{equation}
   where $\lambda_2$ is the $m=2$ soft-mode eigenvalue computed in Sec.~\ref{sec:soft}.

    \subsection{Tomographic Regime}

    When we go to a shorter length scale $\vn{p}>L_\text{tomo}^{-1}$ which will be specified later, we enter the tomographic regime where all soft modes can propagate and produce an effective scattering rate whose scaling differs from that of each individual eigenvalue.

    We follow the analysis in \cite{QHong2020,SKryhin2023} to solve the continuous fraction \eqref{eq:Gammam}.
    We assume $\lambda_m$ oscillates between two functions $\lambda^{\text{odd}}(m)$ and $\lambda^{\text{even}}(m)$ when $m$ is odd and even, respectively. In \cite{QHong2020,SKryhin2023}, it is shown that we can rewrite the continuous fraction formula for $\Gamma_m$ into a linear recurrence equation in terms of some auxiliary variables $u_m$, and then we can approximate the recurrence equation by a differential equation for $u_m=u(m)$:
    \begin{equation}\label{eq:upp}
      u''-\frac{\lambda^{\text{odd}\prime}}{\lambda^{\text{odd}}}u'-\frac{\lambda^{\text{odd}}\lambda^{\text{even}}}{4z^2}u=0\,,
    \end{equation}   where prime means derivative with respect to $m$. Then $\Gamma_m$ is obtained from $u_m$ by
    \begin{equation}\label{}
      \Gamma_m=-\frac{2z^2}{\lambda^{\text{odd}}}\frac{\rd\ln u}{\rd m}\,.
    \end{equation} The boundary condition for Eq.\eqref{eq:upp} is $\Gamma_m>0$ when $m\to\infty$.

    In regimes A,B,C  we have $\lambda^{\text{even}}_m\approx \gamma m^2$ and $\lambda^{\text{odd}}_m\approx\gamma' m^6$. The prefactor $\gamma,\gamma'$ can be read off from Eqs.\eqref{eq:lambda_even_AB_T>}, \eqref{eq:lambda_even_C},\eqref{eq:lambda_odd_A_T},\eqref{eq:lambda_odd_B_T} and \eqref{eq:lambda_odd_C} in the $\Omega\to0$ limit.
    Eq.\eqref{eq:upp} becomes
    \begin{equation}\label{}
      u''-\frac{6}{m}u'-\frac{\gamma\gamma'}{4z^2}m^8 u=0\,.
    \end{equation} The solution can be expressed in terms of Bessel functions
    \begin{equation}\label{}
      u_m=g^{7/10}\left(I_{7/10}(g)-I_{-7/10}(g)\right),
    \end{equation} where $g=\frac{m^5\sqrt{\gamma\gamma'}}{10z}$ and the linear combination is selected to satisfy the boundary condition.
    The asymptotic behavior of $\Gamma_2$ can then be evaluated by expanding in large $z$, and we obtain
    \begin{equation}\label{eq:Gamma2Circ}
      \Gamma_2=\frac{\Gamma\left(\frac{3}{10}\right)}{2^{4/5}5^{2/5}\Gamma\left(\frac{7}{10}\right)}\frac{\gamma z^{3/5}}{(\gamma\gamma')^{3/10}}-\frac{4\gamma}{3}+\dots.
    \end{equation} Therefore, the length scale below which the tomographic transport becomes important is given by
    \begin{equation}\label{}
      z=\frac{\vn{p}v_F}{2}\gg \sqrt{\gamma\gamma'}\,,
    \end{equation} which implies length scale $\vn{p}^{-1}<L_\text{tomo}=v_F/\sqrt{\gamma\gamma'}$ In this regime the first term of Eq.\eqref{eq:Gamma2Circ} dominates.

    The tomographic transport crosses over to ballistic transport when $\vn{p}$ becomes larger than $L_\text{ball}^{-1}$. Since for small $m$, $\lambda_m^{\text{odd}}\propto m^6$ grows faster than $\lambda_m^{\text{even}}\propto m^2$, there exists a critical $m_c$ above which $\lambda_m^{\text{odd}}\approx \lambda_m^{\text{even}}$.  When the harmonics above $m_c$ play a dominant role, the system enters the ballistic transport regime. To find the crossover length scale between the tomographic regime and the ballistic regime, we repeat the exercise of solving for $\Gamma_2$ but set $\lambda_m^{\text{odd}}=\lambda_m^{\text{even}}$, and find $\Gamma_2=z$. Equating this with Eq.\eqref{eq:Gamma2Circ}, we find that the crossover happens at $\vn{p}^{-1}\sim L_\text{ball}=(v_F/\gamma)(\gamma'/\gamma)^{3/4}$.

    In regime D, due to large-angle scattering, the scaling of eigenvalues become $\lambda^{\text{even}}_m=\gamma$ and $\lambda^{\text{odd}}_m=\gamma'm^4$ where $\gamma\propto T^2$ and $\gamma'\propto T^4$ (see Appendix.~\ref{app:softD} or Ref.\cite{PJLedwith2019}, and we have dropped the logarithmic dependence on $m$). This regime has already been analyzed in \cite{QHong2020,SKryhin2023}, and it was found that  $\Gamma_2\propto \gamma z^{1/3}/(\gamma\gamma')^{1/6}$. The crossover length scale to tomographic transport is still $L_\text{tomo}=v_F/\sqrt{\gamma\gamma'}$, but the crossover length scale to the ballistic regime is instead $L_\text{ball}=(v_F/\gamma)(\gamma'/\gamma)^{1/4}$.

\begin{table}[htb]
  \centering
  \begin{tabular}{|c|c|c|c|c|}
    \hline
    Regimes & $\displaystyle G_\text{hydro}$ & $\displaystyle G_\text{tomo}$ & $L_\text{tomo}$ & $L_\text{ball}$ \\
    \hline
    NFL (A) & $\displaystyle W^2 T^{4/3}$ & $\displaystyle W^{7/5}T^{11/60} $ & $T^{-23/12}$ & $T^{-5/6}$\\ 
    \hline
    PNFL (B) & $\displaystyle W^2 T^{4/3}$ & $W^{7/5}T^{-1/15}$ & $T^{-7/3}$  & $T^{1/6}$\\
    \hline
    FL (C) & $\displaystyle W^2 T^2$ & $W^{7/5}T^{1/5}$ & $T^{-3}$ & $T^{-1/2}$  \\
    \hline
    FL (D) & $W^2 T^2$ & $W^{5/3}T$ & $T^{-3}$ & $T^{-3/2}$ \\
    \hline
  \end{tabular}
  \caption{Scaling of the conductance $G$ as a function of temperature $T$ and system cross section length $W$ in the conventional hydrodynamic regime $G_\text{hydro}$ and the tomographic transport regime $G_\text{tomo}$, and the scaling of the crossover length scales $L_\text{tomo}$ and $L_\text{ball}$ as a function of $T$. The conventional hydrodynamic regime happens when $W\gg L_\text{tomo}$, and the tomographic regime sets in when $L_\text{ball}\ll W\ll L_\text{tomo}$. The different regimes of the critical FS are shown in Fig.~\ref{fig:pdT}. We assume the NFL regime (A) is stabilized by thermal fluctuation (see Sec.~\ref{sec:stability}).  }\label{tab:conductance}
\end{table}

    Finally, the non-local conductance $\sigma(\vec{p})$ can be obtained by substituting Eq. \eqref{eq:Gamma2hydro} or \eqref{eq:Gamma2Circ} into Eqs. \eqref{eq:sigmaq} and \eqref{eq:nuq}. The value of $\gamma,\gamma'$ can be read off as the prefactor of the $m$-dependence from the soft-mode eigenvalues at the DC limit (Eqs.\eqref{eq:lambda_even_AB_T>}, \eqref{eq:lambda_even_C},\eqref{eq:lambda_odd_A_T},\eqref{eq:lambda_odd_B_T} and \eqref{eq:lambda_odd_C}). To obtain the actual conductance of a system, we still need to solve the boundary value problem:
    \begin{equation}\label{eq:jE}
      \vec{j}_{\alpha}(\vec{x})=\int\rd^2\vec{x'} \sigma_{\alpha\beta}(\vec{x}-\vec{x'})\vec{E}_\beta(\vec{x'})\,,
    \end{equation} where
    \begin{equation}\label{}
      \sigma_{\alpha\beta}(\vec{x})=\int\rd^2\vec{p}e^{i\vec{p}\cdot(\vec{x}-\vec{x'})}\sigma(\vec{p})\left(\delta_{\alpha\beta}-\frac{\vec{p}_\alpha\vec{p}_\beta}{\vn{p}^2}\right)\,.
    \end{equation} Eq.\eqref{eq:jE} needs to be solved in the actual geometry where transport experiment is performed, and the electric field $\vec{E}(\vec{x})$ needs to be determined self-consistently according to the boundary condition.  This type of analysis has been done in the literature \cite{HGuo2017a,PLedwith2019,QHong2020,XHuang2022a,SKryhin2023} and we will not carry it out here. However, the scaling of the conductance can be estimated for simple geometry by substituting $G\sim \sigma(1/W)$, where $W$ is the cross section length of the geometry.

     We consider the scenario of a NFL at finite temperature stabilized by thermal fluctuation, which is discussed in Sec.~\ref{sec:stability} with a proposed phase diagram in Fig.~\ref{fig:pdT}.
     Applying the estimate of $G$ to different regimes of the phase diagram , we obtain the scaling of the conductance $G(W,T)$ in terms of temperature $T$ and cross-section length $W$, as shown in Table.~\ref{tab:conductance}. In Table.~\ref{tab:conductance}, we also list the crossover length scales for entering and exiting the tomographic regime.  When $W\gg L_\text{tomo}$, the system is in the conventional hydrodynamics regime with conductance given by $G_\text{hydro}$. When $L_\text{ball}\ll W \ll L_\text{tomo}$, the system enters the tomographic transport regime and the conductance is given by $G_\text{tomo}$. We also note that the crossover of the conductance $G$ between different regimes of the phase diagram follows directly the crossover of the eigenvalues as discussed in Sec.~\ref{sec:soft}.

    Based on the results presented above, we propose the hydrodynamic and the tomographic transport as an ideal transport probe for the translational-invariant critical FS. It has several advantages:

    1.~It introduces a natural breaking of the translational invariance through the boundary of the system and therefore resolves the issue of infinite Drude conductivity.

    2.~It probes exactly the soft modes of the critical FS, which parameterize deformation of the FS, and they are theoretically the low-energy degrees of freedom. The conductances $G_\text{hydro}$ and $G_\text{tomo}$ sensitively depends on the fluctuation eigenvalues of the soft modes, and combined together they can sharply differentiate different regimes near the QCP as the exponents of $T$ are very different.

    More importantly, it does not suffer from the subtleties that plagued other transport probes such as the optical conductivity. In Appendix.~\ref{sec:conductivity}, we compute the optical conductivity $\sigma(\omega)$ of the system, and we find that a non-zero $\delta\sigma(\omega)$ correction to the Drude peak always requires breaking Galilean invariance of the system and depends on the details of the band dispersion. In particular, for a circular FS, we find that the soft mode does not contribute to $\delta\sigma(\omega)$ at all, and the leading contribution arises from other fast-decaying modes which decay at the rate of single-particle lifetime.

   3.~ Even if we cannot access the quantum critical regimes (A,B) directly, new physics can still be observed in the FL regime (C) proximate to the QCP. In this regime, the correlation length of the interaction boson does not diverge as at the QCP, but it is already parametrically longer than the Fermi wavelength $\lambda_F$. Compared to a system deep in the FL regime (regime D in Fig.~\ref{fig:pd}), the prevalence of small-angle scattering already modifies the fluctuation spectrum of the FS and makes $G_\text{tomo}$ different.

\section{Conclusion}

    The problem of a Fermi surface coupled to a gapless critical boson in 2+1D is a complex problem with a long history. In this paper, we studied this problem by utilizing the recent Yukawa-SYK formulation \cite{DChowdhury2022a,Iesterlis2021,HGuo2022a,AAPatel2023,HGuo2024,ZDShi2022,ZDShi2023} of the problem, which provided a large $N$ expansion that admits the Migdal-Eliashberg equations (Eq.\eqref{eq:MET}) as the saddle point.

    In Sec.~\ref{sec:fluctuation}, we investigated the $1/N$ fluctuation of the Yukawa-SYK model. The spirit of our analysis is parallel to the 0+1D SYK model \cite{AYKitaev2015,JMaldacena2016c}. The first step is to define a suitable inner product on the fluctuations. In the SYK model, the inner product was motivated by the conformal symmetry, and in our case it is motivated by the conservation laws, i.e. Ward identities (see Eq.\eqref{eq:innerprod}). Using this inner product, we developed a double expansion of the fluctuation kernel, which we call kinetic operator $L$, in terms of the small scattering angle $q/k_F$ and the small fermion dispersion $\xi_k/(k_Fv_F)$ (Eq.\eqref{eq:Lexpand}). The story at this point is again similar to the SYK model: In the SYK model, it was found that the conformal fluctuation kernel contains a large number of zero modes, which needs to be resolved by including correction terms that break conformal symmetry. In our case, we find the leading order kernel $\delta_q^{0}L^{(0)}$ (see Eqs.\eqref{eq:LMT0} and \eqref{eq:LAL0}) which describes forward-scattering limit of excitations on the FS, also contains a large number of zero modes. These (approximate) zero modes are interpreted as the deformations of the FS, which is the object of study in the ersatz FL \cite{DVElse2021a}.

    In Sec.~\ref{sec:soft}, we resolve the eigenvalue of the approximate zero modes and turn them into soft modes. The eigenvalue of the even-parity deformation modes is computed similarly as in the SYK model, by applying a first-order perturbation theory in the correction terms (Eq.\eqref{eq:first_perturb}). The eigenvalue of the even-parity deformation modes can be understood through a simple picture of fermions random-walking on the FS, where the eigenvalue can be identified as the diffusion constant in the angular direction (Eq.\eqref{eq:lambda_even_estimate}).
    This analysis of the odd-parity deformation modes is more complicated than the SYK counterpart due to additional cancelations related to a convex FS, which turns out to require a second-order perturbation involving both the radial and the angular directions (Eq.\eqref{eq:lambda_odd0}).  As a result, the eigenvalues of the odd modes are parametrically smaller than those of the even modes.

    We also stress the importance of considering the entire FS. In a previous work \cite{Iesterlis2021}, a similar analysis is performed for the Yukawa-SYK model, which contains only a single patch of the FS.  It was concluded that there are no soft modes other than the conserved density. However, a point missed there is that after considering the whole FS, each patch contributes a conserved density mode so that in total there is an extensive amount of soft modes. In this work, we carefully analyzed these soft modes and show that two of them (density and momentum) remain exactly conserved but others acquire a finite lifetime.

    In Sec.~\ref{sec:stability}, we address the stability of the Ising-nematic QCP using the soft-mode eigenvalues. Since the eigenvalues can be physically interpreted as dissipation rates, the stability therefore requires them to have positive real parts after analytically continued to real frequencies (retarded branch). In Appendix.~\ref{app:continue}, we show that the real part of the eigenvalues are independent of the UV cutoff after analytic continuation. We find that while the weakly coupled regimes (B,C,D) near the QCP are stable, the strongly coupled NFL regime (A) near the QCP becomes unstable due to the odd-parity soft modes of a convex FS.  However, we note that thermal fluctuation of the critical boson helps stabilize the QCP, leading to the phase diagram proposed in Fig.~\ref{fig:pdT}. The scenario of a NFL with thermal fluctuation has been observed numerically \cite{AKlein2020,XYXu2020}.

    We note that our conclusion of instability here is based on the Yukawa-SYK model at the leading $1/N$ order (Gaussian fluctuation), where the fermion has no anomalous dimension and the boson dynamical critical exponent is kept at $z_b=3$. As noted in Sec.~\ref{sec:stability}, the theory can become stable if $2<z_b<8/3$. It is known that the fermion receives an anomalous dimension starting at three-loop order \cite{MAMetlitski2010} and $z_b$ gets renormalized from $3$ at four-loop \cite{THolder2015,THolder2015a}. Future study around this issue can proceed in two directions. First, we might consider including higher-order terms in $1/N$, which goes beyond the Gaussian fluctuation, and investigate whether these nonlinear terms can stabilize the theory. Second, if the theory is not stabilized, we need to find out the fate of the instability. For example, we can include particle-particle fluctuation channel into the analysis and investigate whether the instability can enhance or compete with superconductivity.

    In Sec.~\ref{sec:eft}, we derive an effective action of the soft modes by projecting the Gaussian $G$-$\Sigma$ action onto the soft-mode manifold. We obtain an effective action which resembles the one obtained from linear bosonization \cite{FDMHaldane2005,AHoughton2000,LVDelacretaz2022a,UMehta2023,SEHan2023} or ersatz Fermi liquid \cite{DVElse2023}, which yields a linearized Boltzmann equation as the equation of motion. Compared to the recent construction based on co-adjoint orbits \cite{LVDelacretaz2022a,UMehta2023}, our effective action not only reproduces the terms corresponding to free fermion, but also includes the soft-mode eigenvalues as a new ingredient which captures scattering effects.  In that regard, our computation provides a transparent microscopic derivation of the bosonization action at the Gaussian order. In the future, it would be interesting to push the comparison to nonlinear orders (see discussion at the end of Sec.~\ref{sec:eft}).

    Another new finding of our theory is the crossover from Landau's kinetic equation \cite{LDLandau1956,AAAbrikosov1963} deep in the FL to the Prange-Kadanoff kinetic equation near the QCP \cite{YBKim1995a,SEHan2023,IMandal2022,KRIslam2023}, and it is at a finite distance from the QCP.   This implies that the zero sound of a FL will change qualitatively before getting close to the critical point, whose detail is left for future study.

    In Sec.~\ref{sec:hydro}. We utilize the derived kinetic equation to compute the non-local conductivity of the system, which shows signatures of hydrodynamic transport.  We find a conventional hydrodynamic regime, where the viscosity is scale-independent, and a tomographic transport regime at shorter length scales, which features a scale-dependent viscosity. We obtain predictions for the conductance $G$ which show scaling behavior with temperature $T$ and system cross section $W$, as shown in Table.~\ref{tab:conductance}. In particular, we predict that a new tomographic regime can emerge in a FL (C) regime, where the correlation length of the critical boson only needs to be much larger than the Fermi wavelength, but does not need to diverge. In this regime, the conductance scales with cross-section length $W$ with an exponent different from a system deep in the FL phase \cite{PLedwith2019,QHong2020,SKryhin2023}, which may be more accessible to near-term experiments.

    For comparison, in Appendix.\ref{sec:conductivity}, we compute the optical conductivity of the system. Although the optical conductivity is an interesting story on its own, it requires a dispersion that breaks Galilean invariance and depends sensitively on the details of the dispersion. Furthermore, it does not detect the soft-mode dynamics, and therefore it might not be a good probe for the low-energy physics.

    All of our discussion above has assumed the FS to be circular or convex. In Appendix.~\ref{app:noncircular}, we give a qualitative discussion on the effects of FS geometry (\update{We still need to assume the FS to be inversion symmetric.}). When the FS is convex, the story is qualitatively similar. However, our results do not apply to concave FS.

    \update{In our computations, we have set the form factor of the Yukawa coupling to $1$, and we briefly comment on the effect of other form factors. In  the computation of self-energies and the fluctuation eigenvalues, the form factors are always squared, and it has been argued in Ref.~\cite{DLMaslov2010a} that angular averaging these squared form factors on the FS leads to physically equivalent results. Therefore, the form factor are not expected to affect the scalings of the soft-mode eigenvalues. However, the form factor can alter the collective responses through the long-range interaction mediated by the critical boson. Diagrammatically, this corresponds to the boson 1PI diagram that appears in the response function of fermion bilinears. In Yukawa-SYK, this diagram is higher order in $1/N$ and neglected. Its effect can be reinstated by the anomaly argument considered in \cite{ZDShi2022,ZDShi2023,DVElse2023,ZDShi2023a}.
    }

    Before we conclude, we also summarize  possible experimental tests of our theory. (a) The Pomeranchuk susceptibilities that we computed in Eq.\eqref{eq:Pim} can potentially be measured by optical probes such as Raman scattering \cite{VKThorsmolle2016,TPDevereaux2007,UKarahasanovic2015}. (b) The instability we found may be interpreted as a Pomeranchuk instability that breaks both inversion and time-reversal (see Sec.~\ref{sec:stability}), which may be probed by second-harmonic generation \cite{MFiebig2005}. (c) In the stable regimes near the QCP as shown in Fig.~\ref{fig:pdT}, we have provided predictions for hydrodynamic transport in Table.~\ref{tab:conductance}, which can be tested in transport experiments \cite{RKrishnaKumar2017}.

\begin{acknowledgments} We thank Dmitrii L. Maslov, Andrey V. Chubukov and Alex Levchenko for the inspiring discussions that initiated this work at the KITP. KITP is supported in part by the National Science Foundation under Grants No. NSF PHY-1748958 and PHY-2309135. We thank Debanjan Chowdhury, Zhengyan Darius Shi, Hart Goldman, Senthil Todadri, Leonid Levitov, J\"org Schmalian, Aavishkar A. Patel, Ilya Esterlis and Subir Sachdev for helpful discussions. Haoyu Guo is supported by the Bethe-Wilkins-KIC postdoctoral fellowship at Cornell University.
\end{acknowledgments}

\appendix

\section{Solutions of the saddle point equations}\label{app:saddle}

We review the solution of the saddle point equations \cite{Iesterlis2021,HGuo2022a} in different regimes of tuning parameter $r$ and energy scale $\omega$, as shown in the phase diagram Fig.~\ref{fig:pd}. We will implement a different integration procedure from \cite{Iesterlis2021,HGuo2022a} that can be generalized to the analysis of the fluctuation spectrum. We start with the boson self-energy in momentum space:
    \begin{equation}\label{}
      \Pi(i\Omega,\vec{q})=-g^2 \int\frac{\rd^2\vec{k}}{(2\pi)^2}\int\frac{\rd \omega}{2\pi} G(i\omega+i\Omega,\vec{k}+\vec{q})G(i\omega,\vec{k})\,.
    \end{equation} We will write frequency integral $\int\rd \omega/(2\pi)$ and Matsubara sum $T\sum_{\omega}$ interchangeably. To utilize the circular shape of the FS, we introduce an auxiliary momentum $\vec{k'}=\vec{k}+\vec{q}$, and rewrite the integral as
    \begin{equation}\label{}
    \begin{split}
      &\Pi(i\Omega,\vec{q})=-g^2(2\pi)^2\int\frac{\rd \omega}{2\pi}\int\calN(\xi)\calN(\xi')\frac{\rd \xi}{2\pi}\frac{\rd \xi'}{2\pi}\\
      &\int \rd \theta\rd \theta' \delta(\vec{k'}-\vec{k}-\vec{q}) G(i\omega+i\Omega,\xi')G(i\omega,\xi)\,.
    \end{split}
    \end{equation} Here $\vec{k}=(\xi,\theta)$ and $\vec{k'}=(\xi',\theta')$ are rewritten using their dispersion $\xi,\xi'$ and angular direction $\theta,\theta'$ respectively. $\calN(\xi)$ is the density of states of the fermion dispersion, and we will approximate it by its value on the FS $\calN(\xi)=\calN=\frac{k_F}{2\pi v_F}$. The angular integral over $\theta,\theta'$ can be evaluated exactly, with the result:
    \begin{equation}\label{eq:Pi=JGG}
      \begin{split}
     &\Pi(i\Omega,\vn{q})=-g^2\calN^2(2\pi)^2\int\frac{\rd \omega}{2\pi}\frac{\rd \xi}{2\pi}\frac{\rd \xi'}{2\pi}  J(\vn{k},\vn{k'},\vn{q})\\
     &\times2 G(i\omega+i\Omega,\xi')G(i\omega,\xi)\,,
     \end{split}
    \end{equation}  where the Jacobian of the delta function is
    \begin{equation}\label{eq:jacobian}
          J(\vn{k},\vn{k'},\vn{q})=\frac{2}{\sqrt{(\vn{k}+\vn{k'})^2-\vn{q}^2}\sqrt{\vn{q}^2-(\vn{k}-\vn{k'})^2}}\,,
    \end{equation} which is inverse proportional to the area of the triangle formed by $\vec{k},\vec{k'},\vec{q}$.

    The fermion self-energy in momentum space is
    \begin{equation}\label{}
      \Sigma(i\omega,\vec{k})=g^2\int\frac{\rd^2\vec{q}}{(2\pi)^2}\int \frac{\rd \Omega}{2\pi} D(i\Omega,\vec{q})G(i\omega+i\Omega,\vec{k}+\vec{q})\,.
    \end{equation}
    We can also perform a similar treatment to the fermion self-energy, the result is
    \begin{equation}\label{eq:Sigma_expr_app}
    \begin{split}
      &\Sigma(i\omega,\xi)=g^2\int\frac{\rd \Omega}{2\pi} \int\frac{\calN \rd\xi'}{2\pi}\int\vn{q}\rd\vn{q} J(\vn{k},\vn{k'},\vn{q})\\
      &\times 2G(i\omega+i\Omega,\xi')D(i\Omega,\vn{q})\,.
      \end{split}
    \end{equation}

    The boson propagator should be substituted with the expression
    \begin{equation}\label{}
      D(i\Omega,\vn{q})=\frac{1}{\vn{q}^2+m_b^2-\bar{\Pi}(i\Omega,\vn{q})}\,,
    \end{equation} where $\bar{\Pi}(i\Omega,\vn{q})=\Pi(i\Omega,\vn{q})-\Pi(0,0)$, and we have exchanged the tuning parameter $r$ to the renormalized boson mass $m_b^2=r+\Pi(0,0)$. Depending on different values of $m_b$, we can access different regimes of critical FS as we detail below.

    \subsection{The non-Fermi liquid (A) and the perturbative non-Fermi liquid (B)}

    The NFL and the PNFL regimes can be accessed by setting $m_b=0$. For momenta near the FS, the Jacobian \eqref{eq:jacobian} can be rewritten as
    \begin{equation}\label{eq:jacobian2}
    \begin{split}
      &J(\vn{k},\vn{k'},\vn{q})=\frac{2}{\sqrt{(2k_F+\xi/v_F+\xi'/v_F)^2-\vn{q}^2}}\\
      &\times \frac{1}{\sqrt{\vn{q}^2-(\xi-\xi')^2/v_F^2}}\,.
    \end{split}
    \end{equation}

    Now we explain how Eq.\eqref{eq:jacobian2} should be approximated. First, if we consider small angle scattering, $\vn{q}\ll k_F$, the first radical in \eqref{eq:jacobian2} can be approximated to $2k_F$. Next, the second radical should be approximated as $\vn{q}$ in Migdal-Eliashberg theory. On the one hand, ME theory assumes that the boson velocity is much smaller than the Fermi velocity $v_F$. Heuristically, the boson has momentum $\vn{q}$ and carries energy $\xi-\xi'$, therefore its velocity is $v_B=|\xi-\xi'|/\vn{q}\ll v_F$, so the ME approximation is equivalent to $\vn{q}$ being larger than $\xi/v_F$. On the other hand, this approximation can also be checked posteriori by looking at the scaling of the saddle point solution, which will be clear later. Therefore, we will proceed with the approximation
    \begin{equation}\label{eq:jacobian3}
      J(\vn{k},\vn{k'},\vn{q})=\frac{1}{k_F\vn{q}}\,.
    \end{equation}

    We can now evaluate the boson self-energy (with $\Pi(0,0)$ subtracted) by performing the $\xi$,$\xi'$ integrals and then evaluate the $\omega$ integral. The $\xi$ integral is evaluated by assuming $\Sigma$ to be $\xi$-independent:
    \begin{equation}\label{eq:dxiG}
      \int\frac{\rd \xi}{2\pi}G(i\omega,\xi)=-\frac{i}{2}\sgn \omega\,.
    \end{equation} The result is
    \begin{equation}\label{eq:barPi}
      \bar{\Pi}(i\Omega,\vn{q})=-\gamma\frac{|\Omega|}{\vn{q}}\,,\quad \gamma=\frac{\calN g^2}{v_F}\,.
    \end{equation} This result is valid at both zero and finite temperatures.

    The self-energy is given by the integral and frequency sum
    \begin{equation}\label{eq:Sigma=DG}
      \Sigma(i\omega)=-i\frac{\calN g^2}{k_F} T\sum_{\Omega} \int\rd\vn{q}D(i\Omega,\vn{q}) \sgn(\omega+\Omega)\,.
     \end{equation} The result is written as two parts $\Sigma(i\omega)=\Sigma_T(i\omega)+\Sigma_Q(i\omega)$. The thermal part is due to the boson carrying zero Matsubara frequency, which must be regularized with a thermal mass $\Delta(T)^2\sim T\ln(1/T)$ \cite{SAHartnoll2014,AJMillis1993,Iesterlis2021}:
     \begin{equation}\label{}
       D(0,\vn{q})=\frac{1}{\vn{q}^2+\Delta(T)^2}\,.
     \end{equation} This yields the thermal part
     \begin{equation}\label{}
       \Sigma_T(i\omega)=-i\sgn(\omega) \frac{g^2 T}{4v_F \Delta(T)}\,.
     \end{equation}

    The quantum part is
    \begin{equation}\label{eq:SigmaQ}
    \begin{split}
      \Sigma_Q(i\omega)&=-ic_f|\omega|^{2/3} \sgn \omega\,,\quad (T=0)\\
      &=-\frac{2i\sgn \omega}{3}c_f (2\pi T)^{2/3}H_{1/3}\left(\frac{|\omega|}{2\pi T}-\frac{1}{2}\right)\,,(T>0)\\
      c_f&=\frac{g^2}{2\pi\sqrt{3}v_F \gamma^{1/3}}\,.
    \end{split}
    \end{equation}Here $H_{1/3}(x)$ is $\verb|HarmonicNumber[x,1/3]|$ in Mathematica.

    At this point, we pause and check the validity of the Jacobian approximation. For thermal boson $\Omega=0$, the typical momentum is $\vn{q}\sim T^{1/2}\ln^{1/2}(1/T)$. For non-thermal boson $\Omega\neq 0$, the typical momentum is $\vn{q}\sim |\Omega|^{1/3}\sim T^{1/3}$. In comparison, the typical dispersion is $\xi\sim |i\omega+\Sigma_Q(i\omega)+\Sigma_T(i\omega)|$. At sufficiently low temperatures $\xi\sim \Sigma_T(i\omega)\sim T^{1/2}\ln^{-1/2}(1/T)$, which is smaller than the typical $\vn{q}$ of both thermal and non-thermal boson.

    We also consider the case where the boson her dynamical exponent $2<z_b< 3$, where we replace the boson propagator by
    \begin{equation}\label{}
      D(i\Omega\neq 0,\vn{q})=\frac{1 }{\vn{q}^{z_b-1}+\gamma |\Omega|/\vn{q}}\,.
    \end{equation} We obtain
    \begin{equation}\label{}
      \Sigma_Q(i\omega,T=0)=-ic_f|\omega|^{2/z_b} \sgn \omega\,,
    \end{equation} where
    \begin{equation}\label{eq:cfz}
      c_f=\frac{g^2}{4\pi v_F \gamma^{1-2/z_b}\sin(2\pi/z_b)}\,.
    \end{equation}

    The thermal fluctuation of the $z_b<3$ theory is different. We recall from Ref.~\cite{SAHartnoll2014} that the boson thermal mass is computed by including the $\phi^4$ term of the boson, and the expression for $\Delta(T)$ is schematically:
    \begin{equation}\label{}
      \Delta(T)^2\sim m_b^2+U T\sum_{i\Omega}\int \rd^2\vec{q} D(i\Omega,\vec{q})\,.
    \end{equation} Here $U$ is the $\phi^4$ coupling.  When $z_b=3$ and $m_b\to 0$, the integral contains a log divergence at $i\Omega=0$ and $\vn{q}\to 0$, leading to the $\Delta(T)^2\sim T\ln(1/T)$ behavior. However, when $z_b<3$ the log divergence disappears, and we therefore expect $\Delta(T)^2 \sim T$.

    We proceed
    \begin{equation}\label{}
      D(i\Omega=0,\vn{q})=\frac{1}{\vn{q}^{z_b-1}+\Delta(T)^2}\,.
    \end{equation}
    and the thermal part is
    \begin{equation}\label{eq:Gammaz}
    \begin{split}
      &\Sigma_T(i\omega)=-i\sgn\omega \frac{\Gamma(T)}{2}\,,\quad\\
       &\Gamma(T)= \frac{g^2 T}{ v_F} \frac{1}{\Delta(T)^{2(z_b-2)/(z_b-1)}}\frac{1}{(z_b-1)\sin\frac{\pi}{z_b-1}}\,.
     \end{split}
    \end{equation} We examine the Jacobian approximation, where typical $\vn{q}\sim |\Delta(T)|^{2/(z_b-1)}$ and typical $\xi\sim T/\Delta(T)^{2(z_b-2)/(z_b-1)}$, therefore the validity of the Jacobian approximation requires $\vn{q}/\xi\sim \Delta(T)^2/T\gg 1$. However, since $\Delta(T)^2\sim T$, we are at the boundary of the applicability. In fact, because the prefactor of $\Delta(T)^2$ is suppressed by the UV cutoff \cite{SAHartnoll2014}, the condition is likely to be violated. Therefore, our calculation only applies to the $2<z_b<3$ theory at zero temperature.

    \subsection{Small-angle scattering Fermi liquid (C)}

    We can enter the FL phase by considering a finite boson mass $m_b$. We first consider the case of $m_b\ll k_F$ meaning the boson mediates small-angle scattering. Now the typical boson mass is of order $m_b$, which is larger than the typical dispersion $\xi$ in the scaling sense; therefore, the Jacobian approximation \eqref{eq:jacobian3} continues to apply. The boson self-energy is still given by Eq.\eqref{eq:barPi} and the fermion self-energy is still given by Eq.\eqref{eq:Sigma=DG}, with
    \begin{equation}\label{}
      D(i\Omega,\vn{q})=\frac{1}{\vn{q}^2+m_b^2+\gamma\frac{|\Omega|}{\vn{q}}}\,.
    \end{equation}

     The $\vn{q}$ integral in \eqref{eq:Sigma=DG} can be evaluated using the contour method. We multiply the integrand by a factor $\vn{q}^{\alpha}$ and calculate the integral with general $\alpha$ and set $\alpha\to 0$ eventually. The integrand
     $$
     \frac{\vn{q}^{\alpha}}{\vn{q}^2+m_b^2+\gamma|\Omega|/\vn{q}}
     $$ has a branch cut along the positive real axis. We can consider the contour connecting $+\infty+i0\to 0\to +\infty-i0$ and closed by the large circle at infinity. The integral along the branch cut is proportional to the desired integral by a factor of $1-e^{2\pi i\alpha}$. The contour integral is calculated by picking up the residues and the result can be expressed using $\verb|RootSum|$ in Mathematica.
        The result is then expanded in $1/m_b$, and the second-order result is
    \begin{equation}\label{eq:DintFL}
    \begin{split}
      &\int_0^{\infty} \rd \vn{q}\frac{1}{\vn{q}^2+m_b^2+\gamma \frac{|\Omega|}{\vn{q}}}\\
      &=\frac{1}{m_b}\left[\frac{\pi}{2}+\frac{|\Omega|}{\omega_{\text{FL}}}\left(\frac{1}{2}+\ln\left(\frac{|\Omega|}{\omega_{\text{FL}}}\right)\right)\right]\,,
    \end{split}
    \end{equation} where
    \begin{equation}\label{eq:omegaFL}
      \omega_{\text{FL}}=\frac{m_b^3}{\gamma}=\frac{2\pi m_b^3 v_F^2}{g^2 k_F}\,.
    \end{equation} The fermion self-energy is then
    \begin{equation}\label{eq:SigmaFLC}
  \Sigma(i\omega)=(-i\omega)c_f'\left(\pi+\frac{|\omega|}{\omega_{\text{FL}}}\ln\left(\frac{|\omega|}{\omega_{\text{FL}}}\right)\right),
\end{equation} with
\begin{equation}\label{eq:cfp}
  c_f'=\frac{g^2\calN}{2\pi k_F m_b}=\frac{g^2}{(2\pi)^2v_F m_b}\,.
\end{equation}

    Eq.\eqref{eq:SigmaFLC} describes a FL self-energy in 2+1D. The linear in $\omega$ term describes a quasiparticle $Z$ factor, and notice that Eq.\eqref{eq:SigmaFLC} is independent of $\xi_k$, meaning that the effective mass $m^*/m=1/Z$. This property is shown to hold in a FL close enough to the QCP \cite{DLMaslov2010a}.  The second term of Eq.\eqref{eq:SigmaFLC}, the $\omega|\omega|\ln|\omega|$ term, describes quasiparticle damping, and it is correctly captured by our integration scheme because by optical theorem damping must involve on-shell particles that reside on the FS. The logarithmic dependence in $|\omega|$ in a special feature due to 2D and Landau damping \cite{AVChubukov2003}: As we expand the integrand in \eqref{eq:DintFL} to second order in $1/m_b^2$, the $1/\vn{q}$ Landau damping factor causes a log divergence that needs to be cut off by $\omega_\text{FL}$.

    Finally, let us discuss when the momentum dependence of $\Sigma(i\omega,\xi)$ on $\xi$ becomes significant. As we see in Eq.\eqref{eq:Sigma_expr_app}, the momentum dependence of $\Sigma(i\omega,\xi)$ is all encoded inside the Jacobian $J(\vn{k},\vn{k'},\vn{q})$. In the current approximation scheme, we assume $|\xi-\xi'|\ll v_F \vn{q}$, and therefore $J$ becomes approximately $\xi$-independent. Therefore, the $\xi$-dependence will emerge when $v_F\vn{q}$ becomes comparable to $|\xi-\xi'|$. To estimate its position in the phase diagram, we substitute $\vn{q}\sim m_b$ and $|\xi-\xi'|\sim \omega\sim \omega_\text{FL}$, and we obtain $m_b\sim (m_b)_\text{ME}=g^2 k_F/v_F$. Therefore, the Eliashberg result is accurate when $m_b\ll (m_b)_\text{ME}$.

    \subsection{Large-angle scattering FL (D)}

    As we further increase the boson mass $m_b$ and it becomes comparable with $k_F$, the boson starts to mediate large-angle scattering. In this regime, the Jacobian approximation \eqref{eq:jacobian3} breaks down because $\vn{q}$ can be comparable to $2k_F$, and instead we should use
    \begin{equation}\label{eq:jacobian4}
      J(\vn{k},\vn{k'},\vn{q})=\frac{2}{\vn{q}\sqrt{4k_F^2-\vn{q}^2}}\,.
    \end{equation} With this new Jacobian, we can redo the integral for $\bar{\Pi}$ \eqref{eq:Pi=JGG}, and the result becomes
    \begin{equation}\label{}
      \bar{\Pi}(i\Omega,\vn{q})=-\gamma\frac{|\Omega|k_F}{\vn{q}\sqrt{k_F^2-\vn{q}^2/4}}\,.
    \end{equation} Notice that the Landau damping contains an additional singularity at $\vn{q}=2k_F$. As we will see shortly, this generates an additional log divergence.

    To proceed with the calculation of fermion self-energy, we neglect the $\vn{q}^2$ in the boson propagator because of large $m_b$:
    \begin{equation}\label{}
      D(i\Omega,\vn{q})=\frac{1}{m_b^2+\gamma\frac{|\Omega|k_F}{\vn{q}\sqrt{k_F^2-\vn{q}^2/4}}}\,.
    \end{equation} Continuing to the fermion self-energy, we have
    \begin{equation}\label{}
    \begin{split}
      \Sigma(i\omega,\xi)&=g^2\int\frac{\rd \Omega}{2\pi} \int\frac{\calN \rd\xi'}{2\pi}\int\vn{q}\rd\vn{q} J(\vn{k},\vn{k'},\vn{q})\\
      &\times 2G(i\omega+i\Omega,\xi')D(i\Omega,\vn{q})\,, \\
      &=g^2\calN \int\frac{\rd \Omega}{2\pi} \int_{0}^{2k_F}\frac{\rd \vn{q}}{\sqrt{k_F^2-\vn{q}^2/4}}\\
      &\times\frac{-i\sgn(\omega+\Omega)}{m_b^2+\gamma\frac{|\Omega|k_F}{\vn{q}\sqrt{k_F^2-\vn{q}^2/4}}}\,.
    \end{split}
    \end{equation}

     We see that if we expand the integrand to second order in $1/m_b^2$, we hit log divergences at both $\vn{q}=0$ and $\vn{q}=2k_F$. Expanding the result to second order in $1/m_b^2$, we obtain
    \begin{equation}\label{eq:SigmaFLD}
     \Sigma(i\omega)=c_f'(-i\omega)\left(\pi+\frac{|\omega|}{\omega_\text{FL}}\ln\frac{|\omega|}{\omega_{\text{FL}}\sqrt{e}}\right)\,.
    \end{equation} This form is qualitatively similar to the small-angle scattering FL (see Eq.\eqref{eq:SigmaFLC}), but the exact parameters are different (some powers of $m_b$ are replaced by $k_F$):
    \begin{equation}\label{eq:omegaFLD}
      \omega_\text{FL}=\frac{k_F m_b^2}{\gamma}\,,\quad c_f'=\frac{g^2 \calN }{2\pi m_b^2}\,.
    \end{equation}

    We note that since $m_b\gg (m_b)_\text{ME}$, the Eliashberg theory is inaccurate in this regime, in the sense that it produces the wrong effective mass, but it still captures the qualitative aspects of inelastic scattering.

\section{Derivation of $K_\text{BS}$}\label{app:KBS}

    In this appendix, we review the derivation of the quadratic fluctuation \eqref{eq:S=GKBSG} around the saddle point \cite{HGuo2022a}. Expanding the action \eqref{eq:S_Gsigma_clean} to first order in the bilocal fields, we obtain
    \begin{widetext}
    \begin{equation}\label{eq:deltaSstar}
  \frac{\delta S}{N}=\Tr\left(\delta\Sigma\cdot(G_*[\Sigma]-G)+\delta G\cdot(\Sigma_*[G]-\Sigma)+\frac{1}{2}\delta\Pi\cdot(D-D_*[\Pi])+\frac{1}{2}\delta D\cdot (\Pi-\Pi_*[D])\right)\,,
\end{equation} where $G_*,\Sigma_*,D_*,\Pi_*$ are the right-hand-side of the saddle point equations \eqref{eq:MET}.
\end{widetext}

  We can further expand \eqref{eq:deltaSstar} to second order around the saddle point to obtain the fluctuations around the saddle point. Define the collective notation $\mathcal{G}_a=(D,G)$ and $\Xi_a=(\Pi,\Sigma)$, where $a=b,f$ denotes boson/fermion. The gaussian fluctuations around the saddle point is described by
  \begin{equation}\label{eq:deltaS1}
  \frac{1}{N}\delta^2 S=\frac{1}{2}\begin{pmatrix}
                        \delta \Xi^T & \delta \calG^T
                      \end{pmatrix}\Lambda
                      \begin{pmatrix}
                        W_\Sigma & -1 \\
                        -1 & W_G
                      \end{pmatrix}
                      \begin{pmatrix}
                        \delta \Xi \\
                        \delta \calG
                      \end{pmatrix}\,,
\end{equation} where $\Lambda=\text{diag}(-1/2,1)$ acts on the $b,f$ indices, and $W_\Sigma$ and $W_G$ are defined by
\begin{equation}\label{}
\begin{split}
  W_\Sigma(x_1,x_2;x_3,x_4)_{a\tilde{a}}&=\frac{\delta \calG_*[\Xi]_a(x_1,x_2)}{\delta \Xi_{\tilde{a}}(x_3,x_4)}\,,\\ W_G(x_1,x_2;x_3,x_4)_{a\tilde{a}}&=\frac{\delta \Xi_*[\calG]_a(x_1,x_2)}{\delta\calG_{\tilde{a}}(x_3,x_4)}\,.
\end{split}
\end{equation}

    Using the saddle point equations \eqref{eq:MET}, $W_\Sigma$ and $W_G$ can be represented by Feynman diagrams
    \begin{widetext}
    \begin{equation}\label{eq:Wsigmagraph}
  W_\Sigma(x_1,x_2;x_3,x_4)=\begin{pmatrix}
                      \begin{tikzpicture}[baseline={([yshift=-4pt]current bounding box.center)}]
                     \draw[thick, boson] (40pt,12pt)--(0pt,12pt);
                     \path[mid triangle] (40pt,12pt)--(0pt,12pt);
                     \draw[thick, boson] (0pt,-12pt)--(40pt,-12pt);
                     \path[ mid triangle] (0pt,-12pt)--(40pt,-12pt);
                     \node at (-5pt,12pt) {\scriptsize $1$};
                     \node at (-5pt,-12pt) {\scriptsize $2$};
                     \node at (48pt,12pt) {\scriptsize $3$};
                     \node at (48pt,-12pt) {\scriptsize $4$};
                     \end{tikzpicture}
                        & 0 \\
                      0 &
                      \begin{tikzpicture}[baseline={([yshift=-4pt]current bounding box.center)}]
                     \draw[thick, mid arrow] (40pt,12pt)--(0pt,12pt);
                     \draw[thick, mid arrow] (0pt,-12pt)--(40pt,-12pt);
                     \node at (-5pt,12pt) {\scriptsize $1$};
                     \node at (-5pt,-12pt) {\scriptsize $2$};
                     \node at (48pt,12pt) {\scriptsize $3$};
                     \node at (48pt,-12pt) {\scriptsize $4$};
                     \end{tikzpicture}
                    \end{pmatrix},
\end{equation}
\begin{equation}\label{eq:WGgraph}
  W_G(x_1,x_2;x_3,x_4)=\begin{pmatrix}
                 0 & -g^2\left(\begin{tikzpicture}[baseline={([yshift=-4pt]current bounding box.center)}]
                                  \draw[thick, dashed] (20pt,12pt)--(0pt,12pt);
                                  \draw[thick, dashed] (0pt,-12pt)--(20pt,-12pt);
                                  \draw[thick, mid arrow] (20pt,12pt)--(20pt,-12pt);
                                  \node at (-5pt,12pt) {\scriptsize $1$};
                                  \node at (-5pt,-12pt) {\scriptsize $2$};
                                  \node at (24pt,12pt) {\scriptsize $3$};
                                  \node at (24pt,-12pt) {\scriptsize $4$};
                               \end{tikzpicture}
                               +
                               \begin{tikzpicture}[baseline={([yshift=-4pt]current bounding box.center)}]
                                  \draw[thick, dashed] (20pt,-12pt)--(0pt,12pt);
                                  \draw[thick, dashed] (0pt,-12pt)--(20pt,12pt);
                                  \draw[thick, mid arrow] (20pt,12pt)--(20pt,-12pt);
                                  \node at (-5pt,12pt) {\scriptsize $1$};
                                  \node at (-5pt,-12pt) {\scriptsize $2$};
                                  \node at (24pt,12pt) {\scriptsize $3$};
                                  \node at (24pt,-12pt) {\scriptsize $4$};
                               \end{tikzpicture}
                               \right) \\
                 \frac{g^2}{2}\left(\begin{tikzpicture}[baseline={([yshift=-4pt]current bounding box.center)}]
                                  \draw[thick, dashed] (20pt,12pt)--(0pt,12pt);
                                  \draw[thick, dashed] (0pt,-12pt)--(20pt,-12pt);
                                  \draw[thick, mid arrow] (0pt,-12pt)--(0pt,12pt);
                                  \node at (-5pt,12pt) {\scriptsize $1$};
                                  \node at (-5pt,-12pt) {\scriptsize $2$};
                                  \node at (24pt,12pt) {\scriptsize $3$};
                                  \node at (24pt,-12pt) {\scriptsize $4$};
                               \end{tikzpicture}+
                                \begin{tikzpicture}[baseline={([yshift=-4pt]current bounding box.center)}]
                                  \draw[thick, dashed] (20pt,-12pt)--(0pt,12pt);
                                  \draw[thick, dashed] (0pt,-12pt)--(20pt,12pt);
                                  \draw[thick, mid arrow] (0pt,-12pt)--(0pt,12pt);
                                  \node at (-5pt,12pt) {\scriptsize $1$};
                                  \node at (-5pt,-12pt) {\scriptsize $2$};
                                  \node at (24pt,12pt) {\scriptsize $3$};
                                  \node at (24pt,-12pt) {\scriptsize $4$};
                               \end{tikzpicture}

                 \right) & g^2\begin{tikzpicture}[baseline={([yshift=-4pt]current bounding box.center)}]
                                  \draw[thick, dashed] (20pt,12pt)--(0pt,12pt);
                                  \draw[thick, dashed] (0pt,-12pt)--(20pt,-12pt);
                                  \draw[thick, boson] (20pt,-12pt)--(20pt,12pt);
                                  \path[mid triangle] (20pt,-12pt)--(20pt,12pt);
                                  \node at (-5pt,12pt) {\scriptsize $1$};
                                  \node at (-5pt,-12pt) {\scriptsize $2$};
                                  \node at (24pt,12pt) {\scriptsize $3$};
                                  \node at (24pt,-12pt) {\scriptsize $4$};
                               \end{tikzpicture}
               \end{pmatrix}\,,
\end{equation}
where a black arrowed line denotes fermion propagator, a wavy arrowed line denotes boson propagator (the arrow denotes momentum), and a dashed line denotes spacetime $\delta$-function. The first entry is boson and the second entry is fermion. Recalling $\Lambda=\mathrm{diag}(-1/2,1)$, we see that $\Lambda W_\Sigma$ and $\Lambda W_G$ are explicitly symmetric as required by quadratic expansion.
\end{widetext}

We now integrate out the self-energies $\delta\Xi$ in Eq.\eqref{eq:deltaS1}, which for Gaussian integrals can be done by minimizing the action with respect to $\delta\Xi$ with $\delta\calG$ fixed, and the result is
\begin{equation}\label{}
  \frac{1}{N}\delta^2 S=-\frac{1 }{2}\delta\calG^T \Lambda(W_\Sigma^{-1}-W_G)\delta \calG
\end{equation} We can further integrate out $\delta D$ in $\delta\calG=(\delta D,\delta G)$, and the result is given by
\begin{equation}\label{eq:deltaS2}
\begin{split}
&\frac{1}{N}\delta^2 S=\\
&-\frac{1}{2}\delta G \left(\underbrace{W_{\Sigma,FF}^{-1}}_{W_\Sigma^{-1}}-\underbrace{W_{G,FF}}_{W_\text{MT}}-\underbrace{W_{G,FB}W_{\Sigma,BB}^{-1}W_{G,BF}}_{W_\text{AL}}\right)\delta G\,.
\end{split}
\end{equation} Here, the subscripts $F,B$ refer to the fermion, boson entries in Eqs.\eqref{eq:Wsigmagraph} and \eqref{eq:WGgraph}, and the notations in the underbraces correspond to the notations in Eq.\eqref{eq:KBS}. Here the operator multiplication are just given by concatenation of Feynman diagrams, and in real space individual terms in Eq.\eqref{eq:deltaS2} reduce to Eqs.\eqref{eq:WSigma}-\eqref{eq:WAL}.

As the subscripts suggest, $W_\Sigma$, $W_\text{MT}$ and $W_\text{AL}$ generate the density-of-states, Maki-Thompson and Aslamazov-Larkin diagrams when expanded perturbatively at low orders. When we evaluate the inverse of $K_\text{BS}$, we obtain a geometric series of the form
\begin{equation}\label{}
  \frac{1}{W_\Sigma-W_\text{MT}-W_\text{AL}}=W_\Sigma+W_\Sigma(W_\text{MT}+W_\text{AL})W_\Sigma+\dots
\end{equation} At first order, $W_\Sigma$ is just a pair of (full) fermion Green's function. If we perturbatively expand in $g^2$ (which is not justified in the NFL regime), we would have
\begin{equation}\label{}
  W_\Sigma=\begin{tikzpicture}[baseline={([yshift=-4pt]current bounding box.center)}]
                     \draw[ mid arrow] (40pt,12pt)--(0pt,12pt);
                     \draw[ mid arrow] (0pt,-12pt)--(40pt,-12pt);
                     \end{tikzpicture}
                     +
                     \begin{tikzpicture}[baseline={([yshift=-9pt]current bounding box.center)}]
                     \draw[ mid arrow] (40pt,12pt)--(0pt,12pt);
                     \draw[ mid arrow] (0pt,-12pt)--(40pt,-12pt);
                     \draw[thick, boson] (35pt,12pt)..controls(20pt,20pt)..(5pt,12pt);
                     \end{tikzpicture}
                     +
                     \begin{tikzpicture}[baseline={([yshift=1pt]current bounding box.center)}]
                     \draw[ mid arrow] (40pt,12pt)--(0pt,12pt);
                     \draw[ mid arrow] (0pt,-12pt)--(40pt,-12pt);
                     \draw[thick, boson] (35pt,-12pt)..controls(20pt,-20pt)..(5pt,-12pt);
                     \end{tikzpicture}+\dots
\end{equation} Here the solid lines are bare fermion propagators. The second order terms are usually called density-of-states diagrams when the left and the right ends are glued by external vertices. Similarly, the Maki-Thompson diagram is generated in
\begin{equation}\label{}
  W_\Sigma W_\text{MT} W_\Sigma=\begin{tikzpicture}[baseline={([yshift=-4pt]current bounding box.center)}]
                     \draw[thick, mid arrow] (40pt,12pt)--(0pt,12pt);
                     \draw[thick, mid arrow] (0pt,-12pt)--(40pt,-12pt);
                     \draw[thick, boson] (20pt,12pt)--(20pt,-12pt);
                     \end{tikzpicture}\,,
\end{equation} and the Aslamazov-Larkin diagram is generated in
\begin{equation}\label{}
  W_\Sigma W_\text{AL} W_\Sigma =
  \begin{tikzpicture}[baseline={([yshift=-4pt]current bounding box.center)}]
                     \draw[thick, mid arrow] (40pt,12pt)--(30pt,12pt)--(30pt,-12pt)--(40pt,-12pt);
                     \draw[thick, mid arrow] (0pt,-12pt)--(10pt,-12pt)--(10pt,12pt)--(0pt,12pt);
                     \draw[thick, boson] (30pt,12pt)--(10pt,12pt);
                     \draw[thick, boson] (30pt,-12pt)--(10pt,-12pt);
                     \end{tikzpicture}+
                     \begin{tikzpicture}[baseline={([yshift=-4pt]current bounding box.center)}]
                     \draw[thick, mid arrow] (40pt,12pt)--(30pt,12pt)--(30pt,-12pt)--(40pt,-12pt);
                     \draw[thick, mid arrow] (0pt,-12pt)--(10pt,-12pt)--(10pt,12pt)--(0pt,12pt);
                     \draw[thick, boson] (30pt,12pt)--(10pt,-12pt);
                     \draw[thick, boson] (30pt,-12pt)--(10pt,12pt);
                     \end{tikzpicture}\,.
\end{equation}

Going to higher orders, the MT and the AL diagrams form the building blocks for the ladder diagram.

Finally, we derive the relation between the correlation functions of fermion bilinears and $K_\text{BS}^{-1}$ (Eq.\eqref{eq:<AB>}). We insert source terms for fermion bilinears into the single particle Hamiltonian:
\begin{equation}\label{}
  \delta h=\int_{x_1,x_2} \psi^\dagger(x_1) \sigma(x_1,x_2) \psi(x_2)\,.
\end{equation} This modifies the $\ln\det$ term in the action to
$$
 \ln\det(\left(\partial_\tau+\varepsilon_k-\mu\right)+\Sigma+\sigma)\,.
$$

We can perform a shift of variable $\Sigma\to \Sigma-\sigma$ to put it as a perturbation term in the action, of the form
\begin{equation}\label{}
  \delta S=\Tr(\sigma\cdot G)
\end{equation} We can therefore evaluate the Gaussian integral over $\delta G$, and the partition function becomes
\begin{equation}\label{}
  Z=e^{-\frac{1}{2}\sigma\cdot K_{\text{BS}}^{-1}\sigma}\,.
\end{equation}

We can now substitute $\sigma(x_1,x_2)=\xi_A A(x_1,x_2)+\xi_B B(x_1,x_2)$, where $A,B$ are form factors of the fermion bilinear of interest. Differentiating with respect to $\xi_A,\xi_B$ yields Eq.\eqref{eq:<AB>}.

\section{Exchange of integration order}\label{app:exchange}

In this appendix we discuss the consequences of exchanging the order of frequency integral and momentum integral. The correct prescription is to perform the frequency integral first (which is truly infinite) and then compute the momentum integral next (which has a physical cutoff such as bandwidth). However, calculations can usually be simplified if the momentum integral is performed first.

As pointed out in \cite{AAAbrikosov1963}, the failure of commuting the two integration order reflects the UV divergence of the integral. According to Fubini's theorem, a double integral can be exchanged if the integral is absolute convergent. Therefore, we examine the integral by counting the superficial degree of divergence in the UV.

We now argue that the error due to exchanging the integration order only happens in the fermion bubble diagram, by counting the degree of divergence. In the fermion bubble diagram, the integral is
\begin{equation}\label{eq:int_bubble}
  \int_{\omega}\int_{\vec{k}}G(i\omega+i\Omega,\vec{k}+\vec{p})G(i\omega,\vec{k})\,.
\end{equation} In the presence of the FS, the momentum integral is factorized into the angular integral $\int_{\theta}$ and the dispersion integral $\int_{\xi_k}$, and only the latter is divergent. Therefore the degree of divergence of Eq.\eqref{eq:int_bubble} is 2 (from $\int_{\omega}\int_{\xi_k}$) - 2 (from 2 factors of $G$)=0. It is marginally divergent. Therefore, we only expect the two integration orders to differ by a constant.

For the Maki-Thompson vertex, we expect no issue when exchanging the limit. The integral looks like
\begin{equation}\label{eq:int_MT}
  \int_{\omega}\int_{\vec{k}}G(i\omega+i\Omega,\vec{k}+\vec{p})G(i\omega,\vec{k})D(\vec{K}-\vec{k},i\nu-i\omega)\,.
\end{equation} At the saddle point, the boson propagator is $D^{-1}(\vec{q},i\Omega)=\vec{q}^2+|\Omega|/\vn{q}$. Due to the presence of FS, the boson momentum $\vec{K}-\vec{k}$ is bounded by $2k_F$, but the Landau damping term still makes $D$ decay at least as $1/|\omega|$ at high frequencies. Therefore the degree of divergence of Eq.\eqref{eq:int_MT} is 2-3=-1, and the integral converges, so the integration order can be exchanged. We note, however, that this exchange of integration order is invalid deep in the FL regime, where $D(\vec{q},i\Omega)=D(\vec{q})$ is a static interaction and does not depend on $|\Omega|$.

We also turn to the Aslamazov-Larkin vertex, with the integral of the form
\begin{equation}\label{eq:int_AL}
\begin{split}
 & \int_{\nu}\int_{\vec{q}}G(i\omega'-i\nu,\vec{k'}-\vec{q})D(\vn{q},i\nu+i\Omega/2)D(\vn{q},i\nu-i\Omega/2)\\
&\times\int_{\omega}\int_{\vec{k}}G(i\omega+i\Omega,\vec{k})G(i\omega,\vec{k})G(i\nu-i\Omega,\vec{q}-\vec{k})
\end{split}
\end{equation} We can analyze it in two steps. First, the inner integral over $GGG$ is convergent because the degree of divergence is $2-3=-1$. Next, the outer integral also converges, with the degree of divergence is $3-5=-2$.

Therefore, at the level of vertex corrections, it is safe to exchange the order of integration. This is also consistent with the fact that our computation can correct reproduce the Ward identities for the charge and the momentum vertices.

The only caution we need is when we encounter fermion-bubble integrals during the evaluation of fermion bilinear correlators. The constant that we need to fix can be easily determined from free-fermion integrals. We have
\begin{equation}\label{}
  \int\rd \xi_k \int \frac{\rd \omega}{2\pi} \frac{1}{i\omega+i\Omega-\xi_k-\vec{v}_k\cdot\vec{p}}\frac{1}{i\omega-\xi_k}=\frac{\vec{v}_k\cdot\vec{p}}{i\Omega-\vec{v}_k\cdot \vec{p}}\,,
\end{equation}
\begin{equation}\label{}
  \int\rd \omega \int\frac{\rd \xi_k}{2\pi}\frac{1}{i\omega+i\Omega-\xi_k-\vec{v}_k\cdot\vec{p}}\frac{1}{i\omega-\xi_k}=\frac{i\Omega}{i\Omega-\vec{v}_k\cdot\vec{p}}\,.
\end{equation} Therefore the constant we need to subtract is $1$.

\section{Positivity of the inner product}\label{app:inner}

    In this appendix we discuss the positivity of the inner product Eq.\eqref{eq:innerprod}. We will show that if we exchange the order momentum and frequency integral, Eq.\eqref{eq:innerprod} is positive up to a factor of $\Omega$. We consider the inner product $\braket{F|F}$ of a two-point function $F$. Since we are interested in CoM momentum $p=(i\Omega,0)$, we assume that its Fourier transform is
\begin{equation}\label{}
\begin{split}
  &F(\tau_1,\vec{r}_1,\tau_2,\vec{r}_2)=\int\frac{\rd^2\vec{k}}{(2\pi)^2}T\sum_{\omega_1} T\sum_{\omega_2}F_{\vec{k}}(i\omega_1,i\omega_2)\\
  &\times e^{-i\omega_1\tau_1+i\omega_2\tau_2+i\vec{k}\cdot(\vec{r}_1-\vec{r}_2)}\,.
\end{split}
\end{equation} Here  we use $\omega_1$, $\omega_2$ to denote the individual frequencies corresponding to the Fourier transform of $\tau_1,\tau_2$.

At fixed CoM frequency $i\Omega$, the norm of $F$ is
\begin{equation}\label{eq:FF1}
\begin{split}
  \braket{F|F}&=T\sum_{\omega} \int_{\vec{k}} F_{\vec{k}}(\omega,\omega+\Omega)\\
  &\times\left[iG(i\omega+i\Omega,\vec{k})-iG(i\omega,\vec{k})\right]F_{\vec{k}}(\omega+\Omega,\omega)\,.
\end{split}
\end{equation} The interesting IR contribution can be captured by first performing the $\vec{k}$ integral, which picks up the poles in the Green's functions. Then the integral is only non-zero when $-\Omega<\omega<0$. Next, we take a step back and perform the Matsubara sum in Eq.\eqref{eq:FF1} but restricts $-\Omega<\omega<0$. We convert the sum to a contour integral in the variable $z=i\omega$. As usual \cite{GDMahan2000}, the branch cut of $F(z,z+i\Omega)$ sits at $\Im z=0$ and $\Im z=-\Omega$. The integration contour can then be deformed to be along $\Im z=-0$ and $\Im z=-\Omega+0$. After continuing $i\Omega\to \Omega_r+i0$, we obtain
\begin{equation}\label{}
\begin{split}
  &\frac{\braket{F|F}}{-i\Omega_r}=\int_{\vec{k}}\int_{-\infty}^{\infty}\frac{\rd z}{2\pi} \frac{n_F(z)-n_F(z+\Omega_r)}{\Omega_r} \\
  &\times\left[iG_R(z+\Omega,\vec{k})-iG_A(z,\vec{k})\right]\\
  &\times F_{AR,\vec{k}}(z,z+\Omega)F_{RA,\vec{k}}(z+\Omega,z)\,.
\end{split}
\end{equation} The subscripts in $F$ denote whether the frequency should be above (R) or below (A) the real axis. Next, we use again the fact that the important contribution can be calculated by doing the momentum integral first. Assuming that the self-energy is momentum-independent (as shown in the saddle point solutions in Appendix.~\ref{app:saddle}), we have
\begin{equation}\label{}
  \int_{\vec{k}} iG_{R/A}(\omega,\vec{k})=\pm\frac{1}{2} \int_{\vec{k}} A(\omega,\vec{k})\,,
\end{equation} where $A$ is the spectral function $A(z,\vec{k})=-2\Im G_R(z,\vec{k})=2\Im G_A(z,\vec{k})$, and therefore we have (the $\vec{k}$-dependence in $F_{\vec{k}}$ is smooth and subdominant to the singular $\vec{k}$-dependence from $G$)
\begin{equation}\label{eq:FF2}
\begin{split}
  &\frac{\braket{F|F}}{-i\Omega_r}=\int_{\vec{k}}\int_{-\infty}^{\infty}\frac{\rd z}{2\pi} F_{RA,\vec{k}}(z+\Omega_r,z) \frac{n_F(z)-n_F(z+\Omega_r)}{\Omega_r} \\ &\frac{A(z,\vec{k})+A(z+\Omega_r,\vec{k})}{2}F_{AR,\vec{k}}(z,z+\Omega_r)\,.
\end{split}
\end{equation}

The last factor in Eq.\eqref{eq:FF2} is positive if the function $F(\tau_1,\tau_2)$ is Hermitian. By Hermitian, we mean that the operator
\begin{equation}\label{}
  \hat{F}=\int_{\tau_1,\tau_2}F(\tau_1,\tau_2)\psi_{\tau_1} \psi_{\tau_2}^\dagger
\end{equation} is Hermitian. The Hermiticity of $\hat{F}$ implies that $F(\tau_1,\tau_2)=F(-\tau_2,-\tau_1)^*$ (remember that Hermitian conjugate in imaginary time is accompanied by reflection). After Fourier transform, we obtain $F(i\omega_1,i\omega_2)=F(-i\omega_2,-i\omega_1)^*$. Analytic continuing to $i\omega_1=\omega_1+i0$, $i\omega_2=\omega_2-i0$, we obtain $F_{RA}(\omega_1,\omega_2)=F_{AR}(\omega_2,\omega_1)^*$.

Therefore, Eq.\eqref{eq:FF2} defines an inner product on the Hermitian fermion bilinears. To simplify the expression for correlation functions, we use the version without the $-i\Omega_r$ factor divided in the main text.


\section{Details of  the derivations of the kinetic operator $L$}\label{app:Lint}

\begin{widetext}
  Here we provide more details on the derivation of the kinetic operator. First, we discuss how Eqs.\eqref{eq:LMTT} and \eqref{eq:LALT} are derived.

    Eq.\eqref{eq:LMT} becomes
    \begin{equation}\label{}
    \begin{split}
      L_{\text{MT+DOS},m}[F](i\omega,\xi_k)&=g^2\int_0^{2\pi}\frac{\rd \theta_k\rd \theta_k'}{2\pi}\int_{-\infty}^{\infty}\frac{\calN \rd \xi_k'}{2\pi}\int_{-\infty}^{\infty}\frac{\rd \omega'}{2\pi}D(\vec{k}-\vec{k'},i\omega-i\omega')\\
      &\times\left[iG(i\omega'+i\Omega/2,\xi_{k}')-iG(i\omega'-i\Omega/2,\xi_k')\right]\\
      &\times\left[F(i\omega,\xi_k)-e^{im(\theta_k'-\theta_k)}F(i\omega',\xi_k')\right]\,.
    \end{split}
    \end{equation} Here $\vec{k'}=(\xi_k',\theta_k')$ and $\vec{k}=(\xi_k,\theta_k)$. We compute the angular integrals by introducing the bosonic momentum $\vec{q}$ and inserting the identity $1=\int\rd^2\vec{q} \delta(\vec{q}=\vec{k}-\vec{k'})$. The integral over $\theta_k$ and $\theta_k'$ can then be calculated by solving the $\delta$-function with $\vn{k}$ and $\vn{k'}$ fixed. For a circular Fermi surface, there are two solutions with
    \begin{equation}\label{}
      \theta_k-\theta_k'=\pm \cos^{-1}\frac{\vn{k}^2+\vn{k'}^2-\vn{q}^2}{2\vn{k}\vn{k'}}\,.
    \end{equation} The Jacobian associated with $\delta$ function is
    \begin{equation}\label{}
          J(\vn{k},\vn{k'},\vn{q})=\frac{2}{\sqrt{(\vn{k}+\vn{k'})^2-\vn{q}^2}\sqrt{\vn{q}^2-(\vn{k}-\vn{k'})^2}}\,,
    \end{equation} which is inverse proportional to the area of the triangle formed by $\vec{k},\vec{k'},\vec{q}$. Finally using the definition of Chebyshev polynomials $T_m(\cos \theta)=\cos m\theta$, we arrive at
    \begin{equation}\label{eq:LMTT_app}
        \begin{split}
          L_{\text{MT+DOS},m}[F]&(i\omega,\xi)=g^2\int\frac{\rd \omega'}{2\pi}\int\frac{\calN \rd \xi'}{2\pi}\int\vn{q}\rd\vn{q}J(\vn{k},\vn{k'},\vn{q})D(\vn{q},i\omega-i\omega')\\
          &\times 2\left[iG(i\omega'+i\Omega/2,\xi')-iG(i\omega'-i\Omega/2,\xi')\right]\left[F(i\omega,\xi)-F(i\omega',\xi')T_m\left(\frac{\vn{k}^2+\vn{k'}^2-\vn{q}^2}{2\vn{k}\vn{k'}}\right)\right]\,.
    \end{split}
    \end{equation}

    Similar manipulations can also be applied to Eq.\eqref{eq:LAL}. We introduce dummy momentum variables $\vec{k'}=\vec{k_1}-\vec{q}$ and $\vec{k''}=\vec{k_2}\mp \vec{q}$. Eq.\eqref{eq:LAL} becomes
    \begin{equation}\label{}
    \begin{split}
      L_{\text{AL},m}[F](i\omega_1,\xi_1)&=g^4(2\pi)^4\int\frac{\rd \theta_1}{2\pi}\int \frac{\rd^2 \vec{q}\rd \nu}{(2\pi)^3}\int\frac{\calN \rd \xi_2 \rd \theta_2}{2\pi}\frac{\calN \rd \xi'\rd \theta'}{2\pi}\frac{\calN \rd \xi''\rd \theta''}{2\pi}D(\vec{q},i\nu+i\Omega/2)D(\vec{q},i\nu-i\Omega/2)\\
      & G(i\omega_1-i\nu,\xi')\delta(\vec{q}=\vec{k_1}-\vec{k'})\left[G(i\omega_2-i\nu,\xi'')\delta(\vec{q}=\vec{k_2}-\vec{k''})+G(i\omega_2+i\nu,\xi'')\delta(\vec{q}=\vec{k''}-\vec{k_2})\right]\\
      &\times \left[iG(\omega_2+i\Omega/2,\xi_2)-iG(\omega_2-i\Omega/2,\xi_2)\right]F(i\omega_2,\xi_2)e^{im(\theta_2-\theta_1)}\,.
    \end{split}
    \end{equation} Here $\vec{k_1}=(\xi_1,\theta_1)$, $\vec{k_2}=(\xi_2,\theta_2)$, $\vec{k'}=(\xi',\theta')$ and $\vec{k''}=(\xi'',\theta'')$. Next, we can compute the integrals over $\theta_1,\theta_2,\theta',\theta''$ by solving the $\delta$-functions. There are two solutions $\pm\bar{\theta}_1$ and $\pm\bar{\theta}_2$ for $\delta(\vec{q}=\vec{k}_1-\vec{k}')$ and $\delta(\vec{q}=\vec{k}_2-\vec{k''})$ respectively, where $\bar{\theta}_1$ and $\bar{\theta}_2$ represent the angles of $\vec{k_1}$ and $\vec{k_2}$ relative to $\vec{q}$ respectively, under the constraint of the $\delta$-functions. The solution of $\delta(\vec{q}=\vec{k''}-\vec{k}_2)$ is $\pi\pm\bar{\theta}_2$. The cosines of $\bar{\theta}_1$ and $\bar{\theta}_2$ are
        \begin{eqnarray}
         \cos\bar{\theta}_1 &=& \frac{\vn{q}^2+\vn{k_1}^2-\vn{k'}^2}{2 \vn{q} \vn{k_1}}\,, \\
          \cos \bar{\theta}_2 &=& \frac{\vn{q}^2+\vn{k_2}^2-\vn{k''}^2}{2 \vn{q} \vn{k_2}} \,.
        \end{eqnarray} After symmetrizing the integral between $\vec{k'}$ and $\vec{k''}$, we obtain:
        \begin{equation}\label{eq:LALT_app}
        \begin{split}
        L_{\text{AL},m}[F]&(i\omega_1,\xi_1)=\frac{g^4}{2}(2\pi)^2\int\calN^3\frac{\rd \nu}{2\pi}\frac{\rd \omega_2}{2\pi}\frac{\rd \xi_2}{2\pi}\frac{\rd \xi'}{2\pi}\frac{\rd \xi''}{2\pi}\int \vn{q}\rd \vn{q} J(\vn{k_1},\vn{k'},\vn{q}) J(\vn{k_2},\vn{k''},\vn{q})\\
        &\times D(\vn{q},i\nu+i\Omega/2)D(\vn{q},i\nu-i\Omega/2)\times 4T_m\left(\frac{\vn{q}^2+\vn{k_1}^2-\vn{k'}^2}{2 \vn{q} \vn{k_1}}\right)T_m\left(\frac{\vn{q}^2+\vn{k_2}^2-\vn{k''}^2}{2 \vn{q} \vn{k_2}}\right)\\
        &\times \left[G(i\omega_1-i\nu,\xi')+(-1)^mG(i\omega_1+i\nu,\xi')\right]\left[G(i\omega_2-i\nu,\xi'')+(-1)^mG(i\omega_2+i\nu,\xi'')\right]\\
        &\times i(G(i\omega_2+i\Omega/2,\xi_2)-G(i\omega_2-i\Omega/2,\xi_2))F(i\omega_2,\xi_2)\,.
        \end{split}
        \end{equation}

        Following the discussion in App.~\ref{app:saddle}, we approximate the Jacobian by Eq.\eqref{eq:jacobian3} or Eq.\eqref{eq:jacobian4} in regimes A,B,C and D respectively. We note the Jacobian should not be expanded in $\xi/(v_F\vn{q})$ because it contains branch point.

        We proceed to expand the Chebyshev polynomials. The first step is to expand in $\xi$'s and the result is
        \begin{equation}\label{eq:TMTexp}
    \begin{split}
      &T_{m}\left(\frac{\vn{k}^2+\vn{k'}^2-\vn{q}^2}{2\vn{k}\vn{k'}}\right)=-\cos \left(2 m \phi _q\right)+\frac{  m \left(\xi '+\xi \right) \cot\phi _q \sin \left(2 m \phi _q\right)}{k_F v_F}\\
      &+\frac{ m \cot ^3\phi _q }{4 k_F^2 v_F^2}
      \Bigg[\sin \left(2 m \phi _q\right) \left[2 (\kappa +1) \left(\xi
   ^2+\xi'^2\right)-2 (\kappa +2) \left(\xi ^2+\xi '^2\right) \sec ^2\phi _q+\left(\xi -\xi '\right)^2 \sec ^4\phi_q\right]\\
   &+2 m \left(\xi '+\xi \right)^2 \tan \phi_q \cos \left(2 m \phi
   _q\right)\Bigg]+\mathcal{O}\left(\frac{\xi^3}{k_F^3 v_F^3}\right)\,,
   \end{split}
    \end{equation}
    \begin{equation}\label{eq:TALexp}
    \begin{split}
      &2T_m\left(\frac{\vn{q}^2+\vn{k_1}^2-\vn{k'}^2}{2 \vn{q} \vn{k_1}}\right)T_m\left(\frac{\vn{q}^2+\vn{k_2}^2-\vn{k''}^2}{2 \vn{q} \vn{k_2}}\right)=
      \cos (2 m \phi_q )+1\\
      &-\frac{ m \csc (2 \phi_q ) \sin (2 m \phi_q ) \left(\xi ''+\xi '+\left(\xi _1+\xi _2\right) \cos (2 \phi_q )\right)}{k_F v_F}-\frac{m \sin ^3\frac{\phi_q }{2} \cos \frac{\phi_q }{2} \csc ^4\phi_q  \sec ^2\phi_q }{8 \left(k_F^2 v_F^2\right)}
      \Bigg[\\
      &-\cos \phi_q  \csc ^2\frac{\phi_q }{2} \sin (2 m \phi_q ) \Big(2 \cos (2 \phi_q ) \left(-(\kappa -1) \left(\xi''^2+\xi'^2\right)+(\kappa +1) \xi _1^2+(\kappa +1) \xi _2^2\right)\\
      &+2 \kappa  \left(\xi''^2+\xi'^2\right)-(\kappa +1) \left(\xi _1^2+\xi _2^2\right) \cos (4 \phi_q )-\left((\kappa -1) \xi _1^2\right)-(\kappa -1) \xi _2^2+4 \xi _2 \xi ''+4 \xi _1 \xi '\Big)\\
      &+8 m \cot \frac{\phi_q }{2} \cos ^2(m \phi_q ) \left(\xi''^2+\xi'^2+2 \cos (2 \phi_q ) \left(\xi _2 \xi ''+\xi _1 \xi '\right)+\left(\xi _1^2+\xi _2^2\right) \cos ^2(2 \phi_q )\right)\\
      &-16 m \cot \frac{\phi_q }{2} \sin ^2(m \phi_q ) \left(\xi '+\xi _1 \cos (2 \phi_q )\right) \left(\xi ''+\xi _2 \cos (2 \phi_q )\right)\Bigg]+\mathcal{O}\left(\frac{\xi^3}{k_F^3 v_F^3}\right)\,.
    \end{split}
    \end{equation} Here $\phi_q$ is
    \begin{equation}\label{}
      \phi_q=\arccos\frac{\vn{q}}{2k_F}\,.
    \end{equation} The fermion momentum $\vn{k}$ is related to dispersion $\xi_k$ via
    \begin{equation}\label{eq:kexpand}
      \vn{k}=k_F+\frac{\xi_k}{v_F}-\frac{\kappa}{2}\frac{\xi_k^2}{k_Fv_F^2}+\frac{\zeta}{2}\frac{\xi_k^3}{k_F^2v_F^3}+\mathcal{O}(\xi_k^4)\,.
    \end{equation} Here we assume the dispersions to be rotationally symmetric but nonparabolic. The nonparabolicity is encoded in the dimensionless parameters $\kappa$ and $\zeta$. In a Galilean invariant system, $\kappa=\zeta=1$.

    Collecting terms at each order of $\xi$'s, we obtain an expansion of $L_m$ as
    \begin{equation}\label{eq:Lexpand1}
      L_m=L_m^{(0)}+L_m^{(1)}+L_m^{(2)}+\dots
    \end{equation}  Further expanding each term in Eqs.\eqref{eq:TMTexp} and \eqref{eq:TALexp} in powers of $\vn{q}^2$, we obtain the double expansion in Eq.\eqref{eq:Lexpand}:
    \begin{equation}\label{eq:TMTexp2}
      \begin{split}
      &T_{m}\left(\frac{\vn{k}^2+\vn{k'}^2-\vn{q}^2}{2\vn{k}\vn{k'}}\right)=
      \underbrace{1}_{\delta_q^0 L_{\text{MT+DOS},m}^{(0)}}\underbrace{-\frac{m^2 \vn{q}^2}{2 k_F^2}}_{\delta_q^1 L_{\text{MT+DOS},m}^{(0)}}+  \underbrace{\frac{m^2 \left(\xi +\xi '\right) \vn{q}^2}{2 k_F^3v_F}}_{\delta_q^1 L_{\text{MT+DOS},m}^{(1)}}\\
   &+\underbrace{\frac{m^2 \left(\xi -\xi '\right)^2}{2 k_F^2 v_F^2}}_{\delta_q^0 L_{\text{MT+DOS},m}^{(2)}}\underbrace{-\frac{\left(m^2 \left(\left(m^2+3
   \kappa +5\right) \xi ^2-2 \left(m^2-4\right) \xi ' \xi +\left(m^2+3 \kappa +5\right) \left(\xi '\right)^2\right)\right) \vn{q}^2}{12
   \left(k_F^4 v_F^2\right)}}_{\delta_q^1 L_{\text{MT+DOS},m}^{(2)}}\,,
   \end{split}
    \end{equation}
    \begin{equation}\label{eq:TALexp2}
    \begin{split}
      &2T_m\left(\frac{\vn{q}^2+\vn{k_1}^2-\vn{k'}^2}{2 \vn{q} \vn{k_1}}\right)T_m\left(\frac{\vn{q}^2+\vn{k_2}^2-\vn{k''}^2}{2 \vn{q} \vn{k_2}}\right)=
      \underbrace{(-1)^m+1}_{\delta_q^0 L_{\text{AL},m}^{(0)}}\underbrace{-\frac{(-1)^m m^2 \vn{q}^2}{2 k_F^2}}_{\delta_q^1 L_{\text{AL},m}^{(0)}}\\
      &+ \underbrace{\frac{(-1)^m m^2 \left(-\xi _1-\xi _2+\xi '+\xi ''\right)}{k_F v_F}}_{\delta_q^0 L_{\text{AL},m}^{(1)}}+\underbrace{\frac{(-1)^m m^2 \left(-\left(\left(\xi '+\xi ''\right) m^2\right)+\left(m^2+2\right) \xi _1+\left(m^2+2\right) \xi _2+\xi '+\xi ''\right) \vn{q}^2}{6 k_F^3 v_F}}_{\delta_q^1 L_{\text{AL},m}^{(1)}}\\
      &+\frac{m^2}{8 k_F^2 v_F^2} \Big[\left(3 \xi _1-3 \xi _2+\xi '-\xi ''\right) \left(\xi _1-\xi _2-\xi '+\xi ''\right)\\
      &+(-1)^m \big(\left(2 m^2+4 \kappa +5\right) \xi _1^2+2 \left(2 m^2+3\right) \left(\xi _2-\xi '\right) \xi _1-2 \left(2 m^2+1\right) \xi '' \xi _1\\
      &+\left(2 m^2+4 \kappa +5\right) \xi _2^2+\left(2 m^2-4 \kappa +1\right) \xi'^2+\left(2 m^2-4 \kappa +1\right) \xi''^2+2 \left(2 m^2-1\right) \xi ' \xi ''\\
      &-2 \xi _2 \left(2 \xi ' m^2+2 \xi '' m^2+\xi '+3 \xi ''\right)\big)\Big]\qquad\Bigg\}\delta_q^0 L_{\text{AL},m}^{(2)}\\
      &+\frac{m^2\vn{q}^2}{96 k_F^4 v_F^2} \Bigg[-3 \left(\xi _1-\xi _2+\xi '-\xi ''\right){}^2\\
      &-(-1)^m \Big(\left(2 \left(m^2+4 \kappa +10\right) m^2+16 \kappa +29\right) \xi _1^2\\
      &+2 \left(-2 \left(\left(m^2+4\right) \xi '+\left(m^2+2\right) \xi ''\right) m^2+\left(2 \left(m^2+8\right) m^2+3\right) \xi _2+13 \xi '+3 \xi ''\right) \xi _1\\
      &+\left(2 \left(m^2+4 \kappa +10\right) m^2+16 \kappa +29\right) \xi _2^2+\left(2 \left(m^2-4 \kappa -2\right) m^2+8 \kappa +5\right) \xi'^2\\
      &+\left(2 \left(m^2-4 \kappa -2\right) m^2+8 \kappa +5\right) \xi''^2+2 \left(2 m^4-8 m^2+3\right) \xi ' \xi ''\\
      &-2 \xi _2 \left(\left(2 m^4+4 m^2-3\right) \xi '+\left(2 m^4+8 m^2-13\right) \xi ''\right)\Big)\Bigg]\qquad\Bigg\}\delta_q^1 L_{\text{AL},m}^{(2)}
    \end{split}
    \end{equation} We can now construct $\delta_q^a L_m^{(b)}$ by replacing the corresponding Chebyshev polynomial factors in Eqs.\eqref{eq:LMTT} and \eqref{eq:LALT} by the corresponding expansion results in Eqs.\eqref{eq:TMTexp2} and \eqref{eq:TALexp2} labeled by braces. For example, for $\delta_q^0 L_m^{(0)}$ in Eqs.\eqref{eq:LMT0} and \eqref{eq:LAL0}, we take the first terms of Eqs.\eqref{eq:TMTexp2} and \eqref{eq:TALexp2} respectively.

\section{Structure and spectrum of the leading order term $\delta_q^0 L_m^{(0)}$}\label{app:rough}\label{sec:dq0L0}

In this appendix, we elaborate on the structure of the leading-order term of the kinetic operator $\delta_q^0 L_m^{(0)}$. This limit describes the effect of strict forward scattering for excitations exactly on the FS.
 By expanding Eqs.\eqref{eq:LMTT} and \eqref{eq:LALT} to zeroth order in $\vn{q}$ and $\xi$, the expression for $\delta_q^0L_m^{(0)}$  is given by the sum of the following (see Appendix.~\ref{app:Lint}):

     \begin{equation}\label{eq:LMTT0}
        \begin{split}
          \delta_q^0 L^{(0)}_{\text{MT+DOS},m}[F]&(i\omega,\xi)=g^2\int_{-\infty}^{\infty}\frac{\rd \omega'}{2\pi}\frac{\calN \rd \xi'}{2\pi}\int\frac{\rd\vn{q}}{k_F}D(\vn{q},i\omega-i\omega')\\
          &\times 2\left[iG(i\omega'+i\Omega/2,\xi')-iG(i\omega'-i\Omega/2,\xi')\right]\left[F(i\omega,\xi)-F(i\omega',\xi')\right]\,.
        \end{split}
        \end{equation}
        \begin{equation}\label{eq:LALT0}
        \begin{split}
        \delta_q^0 L^{(0)}_{\text{AL},m}[F](i\omega_1,\xi_1)&=-g^4(2\pi)^2\left(1+(-1)^m\right)\calN^3\sgn\omega_1\int_{-\infty}^{\infty}\frac{\rd \nu}{2\pi}\frac{\rd \omega_2}{2\pi}\frac{\rd \xi_2}{2\pi}\int_0^\infty \frac{\rd \vn{q}}{k_F^2\vn{q}}\\
        &\times D(\vn{q},i\nu+i\Omega/2)D(\vn{q},i\nu-i\Omega/2) \sgn \omega_2 \theta(|\omega_1|-|\nu|)\theta(|\omega_2|-|\nu|)\\
        &\times i(G(i\omega_2+i\Omega/2,\xi_2)-G(i\omega_2-i\Omega/2,\xi_2))F(i\omega_2,\xi_2)\,.
        \end{split}
        \end{equation}

It can be verified that $F(i\omega,\xi)=1$ is a zero mode of Eqs.\eqref{eq:LMTT0} and \eqref{eq:LALT0}. For Eq.\eqref{eq:LALT0}, the zero mode can be checked by first performing the $\xi_2$-integral, which turns the $iG-iG$ factor to $\theta_{\Omega/2-|\omega_2|}$. The integrand is then an odd function of $\omega_2$ and therefore vanishes. More elaborate discussion of the zero mode is given in the main text.

From now on we focus on understanding the non-zero modes of $\delta_q^0 L_m^{(0)}$. It will be convenient to write the test function $F$ in a polynomial basis in $\xi$, as the following
\begin{equation}\label{eq:Fpoly}
  F(i\omega,\xi)=F_0(i\omega)+\frac{\xi}{A(i\Omega)}F_1(i\omega)+\left(\frac{\xi}{A(i\Omega)}\right)^2 F_2(i\omega)+\dots
\end{equation} Here $A(i\Omega)=i\Omega-\Sigma(i\Omega)$. The normalization $\xi$ by $A(i\Omega)$ makes the matrix element of $L_m$ of the equal scaling when acted on different powers of $\xi$. The hidden assumption of using the ansatz \eqref{eq:Fpoly} is that we are interested in smooth function of $\xi$, which is satisfied by the physical observables such as density or current. We will use the notation $\calH_n$ to denote the Hilbert space of $n$-th monomial in $\xi$.

The structure of $\delta_q^{0}L_m^{(0)}$ can be written as a block matrix in Eq.\eqref{eq:dq0L0block}. Eq.\eqref{eq:dq0L0block} describes the action of $\delta_q^0 L_m^{(0)}$ on the polynomial basis \eqref{eq:Fpoly}. When $\delta_q^0 L_m^{(0)}$ acts on an $n$-th monomial of $\xi$, $F_n(i\omega)(\xi/A(i\Omega))^{n}$, the result contains a piece proportional to $\xi^n$ and a piece which is $\xi$-independent. The part of $\delta_q^0 L_m^{(0)}$ that yields $\xi^n$ is denoted as the restriction $\left.\delta_q^0 L_m^{(0)}\right\vert_{\calH_n}$, appearing in the diagonal entries of \eqref{eq:dq0L0block}. The part of $\delta_q^0 L_m^{(0)}$ that yields the $\xi$-independent result is denoted by $\left.\delta_q^0 L_m^{(0)}\right|_{\calH_n\to\calH_0}$, which appears in the first row. The other blocks are zero.

\begin{equation}\label{eq:dq0L0block}
  \delta_q^{0}L_m^{(0)}\begin{pmatrix}
                         F_0 \\
                         F_1 \\
                         F_2 \\
                         \vdots
                       \end{pmatrix}=\begin{pmatrix}
              \left. \delta_q^{0}L_m^{(0)}\right|_{\calH_0} & \left.\delta_q L_m^{(0)}\right\vert_{\calH_1\to\calH_0} & \left.\delta_q L_m^{(0)}\right\vert_{\calH_2\to\calH_0} & \\
              0 & \left.\delta_q^{0}L_m^{(0)}\right|_{\calH_1} & 0 & \dots\\
              0 & 0 & \left.\delta_q^{0}L_m^{(0)}\right|_{\calH_2} & \\
              &  \vdots & & \ddots
            \end{pmatrix}\begin{pmatrix}
                         F_0 \\
                         F_1 \\
                         F_2 \\
                         \vdots
                       \end{pmatrix}\,.
\end{equation}

The diagonal blocks with $n\geq 1$ have a simple form, because they can only come from the $F(i\omega,\xi)$ term in \eqref{eq:LMT0}:

\begin{equation}\label{eq:Lambdaomega}
  \begin{split}
         & \left.\delta_q^0 L^{(0)}_{m}\right\vert_{\calH_{n\geq 1}}[F](i\omega,\xi)=g^2F(i\omega,\xi)\int_{-\infty}^{\infty}\frac{\rd \omega'}{2\pi}\frac{\calN \rd \xi'}{2\pi}\int\frac{\rd\vn{q}}{k_F}D(\vn{q},i\omega-i\omega')\\
          &\times 2\left[iG(i\omega'+i\Omega/2,\xi')-iG(i\omega'-i\Omega/2,\xi')\right]\\
          &=\underbrace{\left[i\Sigma(i\omega+i\Omega/2)-i\Sigma(i\omega-i\Omega/2)\right]}_{\Lambda_{\omega}}F(i\omega,\xi)\,.
        \end{split}
\end{equation}

 Here in the second line, we have used the saddle-point equations \eqref{eq:MET} to evaluate the integral. This implies that all the eigenvalues of the $n\geq 1$ blocks are labeled by $\Lambda_\omega$ as defined in \eqref{eq:Lambdaomega}. Here the term eigenvalue should be understood in the Jordan form sense because of the off-diagonal entries. However, because upper-triangular matrix is closed under inversion, these eigenvalues can be used to calculate diagonal observables (e.g. $\braket{A|L^{-1}|A}$) as if these were normal eigenvalues.

 For completeness, we also write down the action of the off-diagonal term $\left.\delta_q^0 L_m^{(0)}\right|_{\calH_n\to\calH_0}$, it is given by
 \begin{equation}\label{}
 \begin{split}
   \left.\delta_q^0 L_m^{(0)}\right|_{\calH_n\to\calH_0}[F](i\omega)&=\delta_q^{(0)} L_{m}^{(0)}\left[\left(\frac{\xi}{A(i\omega)}\right)^{n}F(i\omega)\right]-\Lambda(i\omega)\left(\frac{\xi}{A(i\omega)}\right)^{n}F(i\omega)\,.
 \end{split}
 \end{equation} Using Eqs.\eqref{eq:LMTT0}, \eqref{eq:LALT0} and \eqref{eq:Lambdaomega}, it can be verified that the result is independent of $\xi$, therefore it is a mapping from an $n$-th monomial $(\xi/A(i\omega))^n F(i\omega)$ to a $\xi$-independent function.

 Finally, we focus on the zeroth block $\left.\delta_q^0 L_m^{(0)}\right\vert_{\calH_0}$, which maps between $\xi$-independent functions $F(i\omega)$. Apart from the zero mode $F(i\omega)=1$, it also contains a spectrum of non-zero eigenvalues. We will discuss these eigenvalues in different regimes of the phase diagram (Fig.~\ref{fig:pd}). As we shall see, these eigenvalues have the same scaling as the fermion self-energy, and we call them rough modes. We will work exclusively at zero temperature and ignore the thermal fluctuations.



\subsubsection{The NFL regime (A) and the perturbative NFL regime (B)}

    We substitute the boson Green's function
    $$
    D(i\Omega,\vn{q})=\frac{1}{\vn{q}^2+\gamma\frac{|\Omega|}{\vn{q}}}\,,
    $$
    and evaluate the $\vn{q}$ integrals in Eqs.\eqref{eq:LMTT0} and \eqref{eq:LALT0}, we obtain

\begin{equation}\label{eq:LMT0}
          \delta_q^0L_{\text{MT+DOS},m}^{(0)}[F](i\omega)=\frac{2}{3}c_f\int_{-\Omega/2}^{\Omega/2}\rd \omega' \frac{1}{|\omega-\omega'|^{1/3}}\left[F(i\omega)-F(i\omega')\right]\,.
    \end{equation}
    \begin{equation}\label{eq:LAL0}
        \begin{split}
          &\delta_q^0L_{\text{AL},m}^{(0)}[F](i\omega_1)=-\frac{2}{3}c_f\sgn \omega_1\frac{1+(-1)^m}{2}\int_{-\Omega/2}^{\Omega/2}\rd \omega_2\int_{-\Omega/2}^{\Omega/2}\rd\nu\theta(|\omega_1|-|\nu|)\theta(|\omega_2|-|\nu|)\\
          &\times \frac{1}{|\nu^2-\Omega^2/4|^{1/3}\left(|\nu+\Omega/2|^{2/3}+|\nu-\Omega/2|^{2/3}+|\nu^2-\Omega^2/4|^{1/3}\right)}\sgn \omega_2 F(i\omega_2)\,.
        \end{split}
    \end{equation}
 Because of the $\xi$-integral the range of $\omega$ has been bounded to $[-\Omega/2,\Omega/2]$. We observe that Eqs.\eqref{eq:LMT0} and \eqref{eq:LAL0} have a particle hole symmetry $F(i\omega)\to F(-i\omega)$, implying that the eigenvalues can be considered separately for odd and even sectors, i.e. $F(-i\omega)=P F(i\omega)$, $P=\pm 1$. It is easy to diagonalize \eqref{eq:LMT0} and \eqref{eq:LAL0} numerically, and the result can be written as
\begin{equation}\label{}
  \lambda^m_\alpha=\frac{2}{3}c_f \Omega^{2/3}\alpha\,,
\end{equation} where $\alpha$ is a dimensionless factor. Our numerical result is summarized in Table.~\ref{tab:L0Spec}.
\begin{table}[htb]
      \centering
      \begin{tabular}{|c|c|c|c|}
      \hline
      $P$ & $(-1)^m$ & Discrete Spectrum & Continuum Spectrum \\
      \hline
      \multirow{2}{*}{\centering 1} & 1  & \multirow{2}{*}{$\alpha=0$} & \multirow{2}{*}{$\alpha\in\left[1.386,1.890\right]$} \\
      \cline{2-2}
       & -1 & & \\
       \hline
       \multirow{2}{2em}{\centering -1} & 1 & $\alpha=0.856,1.411,1.484$ & $\alpha\in\left[1.498,1.890\right]$ \\
       \cline{2-4}
       & {\centering -1} & $\alpha=1.226,1.449,1.491$ & $\alpha\in\left[1.499,1.890\right]$\\
       \hline
    \end{tabular}
      \caption{Spectrum of $\left.\delta_q^0L_m^{(0)}\right\vert_{\calH_0}$ in regimes A and B in different sectors defined by $P$ and $(-1)^m$. Numerical values are obtained through a 5000 by 5000 discretization of Eqs.\eqref{eq:LMT0} and \eqref{eq:LAL0}.}\label{tab:L0Spec}
    \end{table}

    For the $P=1$ sector, in addition to the zero mode, we found a continuum spectrum with $\alpha\in [1.386,1.890]$. In the $P=-1$ sector, we found 3 discrete modes together with a continuum as shown in Table.~\ref{tab:L0Spec}. The discrete spectrum also depends on whether the angular harmonics $m$ is even or odd.

    However, the particle-hole symmetry only holds for the leading-order term $L_m^{(0)}$. Generically $L_m^{(n)}$ has signature $(-1)^n$ under the particle-hole transformation $(\omega,\xi)\to(-\omega,-\xi)$, so the symmetry is broken for the whole $L_m$. Therefore, after putting back the higher perturbations, the $P=\pm 1 $ modes will hybridize and the spectra merge into one, and some discrete modes will be buried into the continuum. The correct approximate spectrum should be given by Table.~\ref{tab:L0Spec2}.

\begin{table}[htb]
      \centering
      \begin{tabular}{|c|c|c|}
      \hline
      $(-1)^{m}$  & Discrete Spectrum & Continuum Spectrum \\
      \hline
      {\centering 1}   & {$\alpha=0.856$} & {$\alpha\in\left[1.386,1.890\right]$} \\
       \hline
       {\centering -1}  & {$\alpha=1.226$} & {$\alpha\in\left[1.386,1.890\right]$} \\
       \hline
    \end{tabular}
      \caption{The approximate non-soft spectrum of $L_m$ in regimes A and B after breaking particle-hole symmetry.}\label{tab:L0Spec2}
    \end{table}

\subsubsection{The FL regime with small-angle scattering (C)}

    We now substitute the boson propagator
\begin{equation}\label{}
  D(i\Omega,\vec{q})=\frac{1}{\vn{q}^2+m_b^2+\gamma|\Omega|/\vn{q}}\,. \nonumber
\end{equation}

   We perform the $\vn{q}$-integral in \eqref{eq:LMTT0} and \eqref{eq:LALT0} to obtain:
\begin{equation}\label{eq:LMT02}
   \delta_q^0L_{\text{MT+DOS},m}^{(0)}[F](i\omega)=c_f' \int_{-\Omega/2}^{\Omega/2}\rd \omega' \left(\pi+\frac{2|\omega-\omega'|}{\omega_{\text{FL}}}\ln\left(\frac{|\omega-\omega'|}{\omega_{\text{FL}}}\sqrt{e}\right)\right)(F(i\omega)-F(i\omega'))\,,
\end{equation}
\begin{equation}\label{eq:LAL02}
\begin{split}
   &\delta_q^0L_{\text{AL},m}^{(0)}[F](i\omega)=2\frac{c_f'}{\omega_{\text{FL}}}\frac{1+(-1)^{m}}{2}\int_{-\Omega/2}^{\Omega/2}\rd \nu\int_{-\Omega/2}^{\Omega/2} \rd \omega_2 \sgn\omega_1 \theta(|\omega_1|-|\nu|) \theta(|\omega_2|-|\nu|) \\ &\frac{|\nu+\Omega/2|\ln\left(\frac{\gamma|\nu+\Omega/2|}{m_b^3}\sqrt{e}\right)-|\nu-\Omega/2|\ln\left(\frac{\gamma|\nu-\Omega/2|}{m_b^3}\sqrt{e}\right)}{|\nu+\Omega/2|-|\nu-\Omega/2|}\sgn\omega_2 F(\omega_2)\,.
\end{split}
\end{equation}

    Similarly to regimes A and B, we can numerically diagonalize Eqs. \eqref{eq:LMT02} and \eqref{eq:LAL02}, and we found the result can be well fitted by the function
\begin{equation}\label{eq:lambdaFL}
  \lambda_{\alpha}^m=c_f' A_\alpha \frac{\Omega^2}{\omega_{\text{FL}}}\ln\left(\frac{B_\alpha \Omega}{\omega_{\text{FL}}}\right)\,.
\end{equation} The parameters $A_\alpha$ and $B_\alpha$ are summarized in Table.~\ref{fig:L0sepcFL} and Fig.~\ref{fig:AaBa}. We found three discrete modes together with a continuum after considering particle-hole symmetry breaking. The parameters of the discrete modes are listed in Table.~\ref{fig:L0sepcFL} and the continuum parameters are plotted in Fig.~\ref{fig:AaBa} as a function of $\alpha\in[0,1]$.

\begin{table}[htb]
  \centering
  \begin{tabular}{|c|c|c|c|c|c|}
    \hline
      & $(-1)^m$ & \multicolumn{3}{|c|}{Discrete } & Continuum \\
      \hline
    \multirow{2}{*}{\centering $(A_\alpha,B_\alpha)$} & 1 & (0.995,0.981) & (2.000,1.407) & (0.998,0.994) & \multirow{2}{*}{Fig.~\ref{fig:AaBa}} \\
    \cline{2-5}
     & -1 & (0.995,0.981) & (1.219,1.076) & (0.999,0.999) &   \\
    \hline
  \end{tabular}
  \caption{The approximate spectrum of $L_m$ in regime C. The first discrete mode is approximately even and the other two modes are approximately odd under particle-hole symmetry.}\label{fig:L0sepcFL}
\end{table}

\begin{figure}[htb]
  \centering
  \includegraphics[width=0.6\columnwidth]{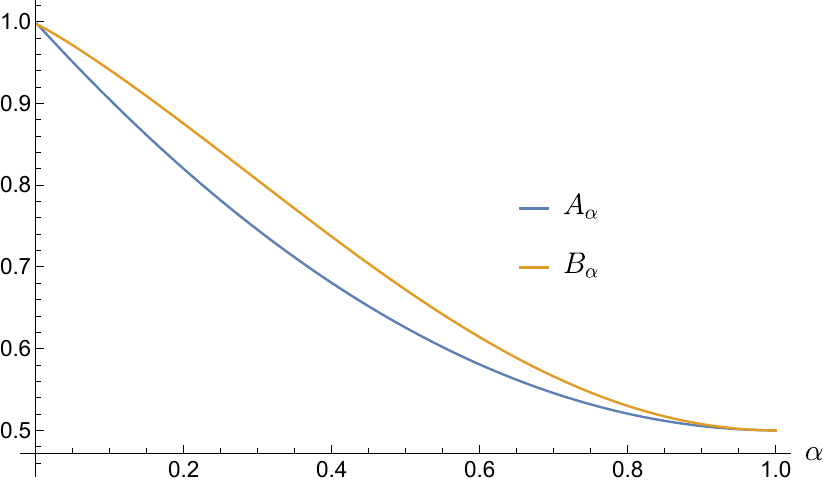}
  \caption{The functions $A_\alpha$ and $B_{\alpha}$ parameterizing the continuum spectrum in regime C.}\label{fig:AaBa}
\end{figure}

\subsubsection{The FL regime with large-angle scattering (D)}

    In this regime, the form of the kinetic operator $L_m$ is complicated because the boson momentum $\vn{q}$ can be as large as $2k_F$. Therefore, we do not have analytical control as we did in regimes A,B,C. However, from the analysis above we see that the nonzero eigenvalues of $L_m$ are not suppressed compared to the self-energy scaling by the small-angle scattering. Therefore, we expect that the nonzero eigenvalues of $L_m$ should scale similarly to the self-energy and be qualitatively similar to regime C.

\section{Computation of odd-$m$ soft eigenvalue}\label{app:oddm}

    In this part, we provide details on the computation of Eq.\eqref{eq:lambda_odd0}.

    Noticing the fact that $G(i\omega,\xi)=-G(-i\omega,-\xi)$, the kinetic operators $L_m$ have a particle hole symmetry under which
    $$
      L_m^{(n)}\to (-1)^n L_m^{(n)}\,.
    $$ Therefore, first order perturbation in $\xi$ $\braket{1|L_m^{(1)}|1}$ vanishes identically and we have to consider the second order expression (same as Eq.\eqref{eq:lambda_odd0})
    \begin{equation}\label{eq:lambda_odd0_app}
      \lambda_m^{\text{odd}}=\frac{\delta_q^1}{\braket{1|1}}\left[\braket{1|L_m^{(2)}|1}-\braket{1|L_m^{(1)}\frac{1}{L_m^{(0)}}L_m^{(1)}|1}\right]\,.
    \end{equation}

    The hard part of evaluating \eqref{eq:lambda_odd0} is to perform the functional inverse of $L_m^{(0)}$. We first compute $L_m^{(1)}\ket{1}$. We obtain
    \begin{equation}\label{eq:L11}
    \begin{split}
      &L_{m}^{(1)}[1](i\omega,\xi)=g^2\calN\int\frac{\rd \omega'}{2\pi}\frac{\rd \xi'}{2\pi}\int \vn{q}\rd\vn{q} J(\vn{k},\vn{k'},\vn{q})D(\vn{q},i\omega-i\omega')\\
      &\times 2[iG(i\omega'+i\Omega/2,\xi')-iG(i\omega'-i\Omega/2,\xi')]\frac{-m}{k_F v_F}\left[\xi \sin(2m\phi_q)\csc(2\phi_q)+\xi'(\cot\phi_q-\csc(2\phi_q))\sin(2m\phi_q)\right]\,.
    \end{split}
    \end{equation} Here the integral related to the AL part \eqref{eq:LALT_app} can be evaluated as follows. First, we read off the expansion of the linear-in-$\xi$ terms from Eq.\eqref{eq:TALexp}. We notice that for odd-$m$, the terms linear in $\xi_2$ or $\xi''$ (only one of them) vanish. This is because the integrand over $\omega_2$ turns out to be odd in $\omega_2$ and cancels itself. Therefore, we only need to retain turns linear in $\xi'$ and $\xi$. Since now there is no additional dependence on $\xi_2,\xi''$ or $\omega_2$ introduced, we can compute the integrals except those that involve $\omega',\xi',\vn{q}$ using the saddle point equation evaluated using \eqref{eq:Pi=JGG}.

    We compare this with $L_m^{(0)}\ket{\xi/(k_F v_F)}$, which is
    \begin{equation}\label{eq:L0xi}
      \begin{split}
      &L_{m}^{(0)}\left[\frac{\xi}{k_F v_F}\right](i\omega,\xi)=g^2\calN\int\frac{\rd \omega'}{2\pi}\frac{\rd \xi'}{2\pi}\int \vn{q}\rd\vn{q} J(\vn{k},\vn{k'},\vn{q})D(\vn{q},i\omega-i\omega')\\
      &\times 2[iG(i\omega'+i\Omega/2,\xi')-iG(i\omega'-i\Omega/2,\xi')]\left[\frac{\xi}{k_F v_F}+\frac{\xi'}{k_F v_F}\cos(2m\phi_q)\right]\,.
    \end{split}
    \end{equation} Here for the same reason as above, the AL parts do not contribute.

    We are interested in the case of small-angle scattering, so we expand Eqs.\eqref{eq:L11} and \eqref{eq:L0xi} to first order in $\vn{q}^2$, yielding
    \begin{equation}\label{eq:L11smallq}
     \begin{split}
      &(\delta_q^0+\delta_q^1)L_{m}^{(1)}[1](i\omega,\xi)=g^2\calN\int\frac{\rd \omega'}{2\pi}\frac{\rd \xi'}{2\pi}\int \vn{q}\rd\vn{q} J(\vn{k},\vn{k'},\vn{q})D(\vn{q},i\omega-i\omega')\\
      &\times 2[iG(i\omega'+i\Omega/2,\xi')-iG(i\omega'-i\Omega/2,\xi')](-m^2)\left[\left(1-\frac{(m^2-1)\vn{q}^2}{6 k_F^2}\right)\frac{\xi}{k_F v_F}-\left(1-\frac{(m^2+2)\vn{q}^2}{6k_F^2}\right)\frac{\xi'}{k_F v_F}\right]\,.
    \end{split}
    \end{equation}
    \begin{equation}\label{eq:L0xismallq}
      \begin{split}
      &(\delta_q^0+\delta_q^1)L_{m}^{(0)}\left[\frac{\xi}{k_F v_F}\right](i\omega,\xi)=g^2\calN\int\frac{\rd \omega'}{2\pi}\frac{\rd \xi'}{2\pi}\int \vn{q}\rd\vn{q} J(\vn{k},\vn{k'},\vn{q})D(\vn{q},i\omega-i\omega')\\
      &\times 2[iG(i\omega'+i\Omega/2,\xi')-iG(i\omega'-i\Omega/2,\xi')]\left[\frac{\xi}{k_F v_F}-\left(1-\frac{m^2\vn{q}^2}{2k_F^2}\right)\frac{\xi'}{k_F v_F}\right]\,.
    \end{split}
    \end{equation} Noticing that the $\vn{q}^0$ order of Eqs.\eqref{eq:L11smallq} and \eqref{eq:L0xismallq} look very similar, we therefore reorganize them into the relation (we drop $\delta_q^0+\delta_q^1$ for clarity)
    \begin{equation}\label{eq:Lh=b}
      \ket{b}\equiv m^2L_m^{(0)}\ket{\frac{\xi}{k_F v_F}}+L_m^{(1)}\ket{1}\,.
    \end{equation} The explicit expression for function $\ket{b}$ is
    \begin{equation}\label{eq:bint}
    \begin{split}
      &b(i\omega,\xi)=g^2\calN\int\frac{\rd \omega'}{2\pi}\frac{\rd \xi'}{2\pi}\int \vn{q}\rd\vn{q} J(\vn{k},\vn{k'},\vn{q})D(\vn{q},i\omega-i\omega')\\
      &\times 2[iG(i\omega'+i\Omega/2,\xi')-iG(i\omega'-i\Omega/2,\xi')]\frac{m^2(m^2-1)\vn{q}^2}{6k_F^2}\left[\frac{\xi}{k_F v_F}+2\frac{\xi'}{k_F v_F}\right]\,.
    \end{split}
    \end{equation}  Notice that $\ket{b}$ is explicitly of order $\vn{q}^2$. Let's rewrite the functional inverse term in Eq.\eqref{eq:lambda_odd0_app} using Eq.\eqref{eq:Lh=b} twice:
    \begin{equation}\label{}
    \begin{split}
      \braket{1|L_m^{(1)}\frac{1}{L_m^{(0)}}L_m^{(1)}|1}&=-m^2 \braket{1|L_m^{(1)}|\frac{\xi}{k_F v_F}}+\braket{1|L_m^{(1)}\frac{1}{L_m^{(0)}}|b}\\
      &=-m^2 \braket{1|L_m^{(1)}|\frac{\xi}{k_F v_F}} -m^2\braket{\frac{\xi}{k_Fv_F}|b}+\braket{b|\frac{1}{L_m^{(0)}}|b}\,.
    \end{split}
    \end{equation} Since we only keep terms up to first order in $\vn{q}^2$, the last term can be dropped.

     So we have
    \begin{equation}\label{}
       \lambda_m^\text{odd}=\frac{1}{\braket{1|1}}\left[\braket{1|\delta_q^{1}L_m^{(2)}|1}+m^2\braket{1|\delta_q^{1}L_m^{(1)}|\frac{\xi}{k_F v_F}}+m^2\braket{\frac{\xi}{k_F v_F}|b}\right]\,.
    \end{equation} Here the second term can be evaluated using Eq.\eqref{eq:L11smallq}, utilizing the fact that the inner product is symmetric. The third term is evaluated using Eq.\eqref{eq:bint}. Finally, the first term is evaluated using the the expansions of $L_m$ from Appendix.~\ref{app:Lint}. To evaluate the AL part of $\delta_q^1 L_m^{(2)}$, we symmetrized the polynomials in $\xi$'s in the integrand in four steps: (1) $(\xi_2,\xi'')\to -(\xi_2,\xi'')$; (2) $(\xi_1,\xi')\to -(\xi_1,\xi')$. These two steps eliminate terms linear in only one of $\xi_2,\xi''$ due to the cancellation explained after Eq.\eqref{eq:L11}. After this step, there is no cross term between $(\xi_1,\xi')$ and $(\xi_2,\xi'')$, and the remaining terms are only quadratic in either $(\xi_1,\xi')$ or $(\xi_2,\xi'')$. (3) Since we are sandwiching $\delta_q^1 L_m^{(2)}$ on the same state $\ket{1}$, we can use the symmetric property of $L$ to set $(\xi_2,\xi'')\to (\xi_1,\xi')$, which then allows us to evaluate the $\omega_2$ and $\nu$ integral using Eq.\eqref{eq:Pi=JGG}. (4) Finally, we symmetrize the resulting expression between $(\xi_1,\xi')$. Combining the results, we obtain Eq.\eqref{eq:lambda_odd} of main text:
    \begin{equation}\label{eq:lambda_odd_app}
    \begin{split}
      &\lambda_m^\text{odd}=\frac{2g^2\calN}{\Omega/(2\pi)} \int_{-\Omega/2}^{\Omega/2}\frac{\rd \omega\rd \omega'}{(2\pi)^2}\int_{-\infty}^\infty\frac{\rd \xi\rd \xi'}{(2\pi)^2}\int_0^{\infty}\frac{\rd\vn{q}}{k_F}\\
      &\times D(\vn{q},i\omega-i\omega')\left[iG(i\omega+i\Omega/2,\xi)-iG(i\omega-i\Omega/2,\xi)\right]\\
      &\times\left[iG(i\omega'+i\Omega/2,\xi')-iG(i\omega'-i\Omega/2,\xi')\right]\times\frac{\vn{q}^2}{k_F^2} \frac{m^2(m^2-1)^2(\xi+\xi')^2}{8k_F^2 v_F^2}\,.
    \end{split}
    \end{equation} Notice here that when we expand $L_m$, the momentum $\vn{k}$ can depend on $\xi_k$ in a complicated way due to the band structure parameters $\kappa,\zeta$ introduced in Eq.\eqref{eq:kexpand}. However, in Eq.\eqref{eq:lambda_odd_app} these dependencies canceled each other.
\end{widetext}
\section{The soft eigenvalues at finite temperature $T$}\label{app:finiteT}

    In this appendix, we demonstrate the calculation of the soft-mode eigenvalues at finite temperature. This computation is possible because the integrand of Eqs.\eqref{eq:lambda_even} and \eqref{eq:lambda_odd_eval} has a temperature-independent form, and we can evaluate the finite-temperature result simply by switching from integral to summation.

    As an example, we compute Eq.\eqref{eq:lambda_even_AB_T}. At finite temperature, it becomes
\begin{equation}\label{}
    \begin{split}
      &\lambda_m^{\text{even}}(i\Omega)=\frac{2g^2\calN}{\Omega/(2\pi)}\int_{-\Omega/2}^{\Omega/2}\frac{\rd\omega\rd \omega'}{(2\pi)^2}\\
      &\times\int_0^\infty\frac{\rd\vn{q}}{k_F}D(\vn{q},i\omega-i\omega')\frac{m^2\vn{q}^2}{2k_F^2}\,.
    \end{split}
    \end{equation} The $\vn{q}$-integral is evaluated using dimensional regularization, with the result
\begin{equation}\label{}
  \int_0^{\infty} \rd\vn{q} \frac{\vn{q}^2}{\vn{q}^2+\gamma\frac{|\Omega|}{\vn{q}}}=-\frac{2\pi\gamma^{1/3}|\Omega|^{1/3}}{3\sqrt{3}}\,.
\end{equation} Also replacing the frequency integral by Matsubara summation, we obtain
\begin{equation}\label{}
  \lambda_m^{\text{even}}=\frac{2g^2\calN }{\Omega/(2\pi)} T^2 \sum_{0<\omega,\omega'<\Omega} \frac{m^2}{2k_F^3}\left(-\frac{2\pi\gamma^{1/3}}{3\sqrt{3}}\right)|\omega-\omega'|^{1/3}\,.
\end{equation} Here we have shifted $\omega,\omega'$ by $\Omega/2$ so that they correspond to fermionic Matsubara frequencies. Let's parameterize $\omega=2\pi T(n-1/2)$ and $\omega'=2\pi T(n'-1/2)$ and $M=\Omega/(2\pi T)$, and then the summation over $n,n'$ runs from 1 to $M$. We can first evaluate the sum over $n'$. With the help of the harmonic number
$$
H_r(M)=\sum_{n=1}^{M}\frac{1}{n^r}\,,
$$ we can write the result as
\begin{equation}\label{}
  \lambda_m^{\text{even}}=-\sum_{n=1}^{M}\frac{4\times 2^{1/3} \pi ^{4/3} \gamma^{1/3} g^2 m^2
   \calN T^{4/3}
   H_{-1/3}(n-1)}{3 \sqrt{3} k_F^3 M}\,.
\end{equation}

The summation over harmonic numbers can be evaluated using the following identities
\begin{widetext}
\begin{eqnarray}
  \sum_{m=1}^{n} H_r(m) &=& (1+n)H_r(n)-H_{r-1}(n)\,, \label{eq:Hsum0} \\
  \sum_{m=1}^{n} m H_r(m) &=& \frac{1}{2}\left[n(n+1)H_r(n)+H_{r-1}(n)-H_{r-2}(n)\right]\,, \\
  \sum_{m=1}^{n} m^2 H_r(m) &=& \frac{1}{6}\left[n(1+n)(1+2n)H_r(n)-H_{r-1}(n)+3H_{r-2}(n)-2H_{r-3}(n)\right]\,.
\end{eqnarray}
\end{widetext}

Using Eq.\eqref{eq:Hsum0}, we obtain Eq.\eqref{eq:lambda_even_AB_T}.

Another example is Eq.\eqref{eq:lambda_odd_B_omega} and Eq.\eqref{eq:lambda_odd_B_T}. The expression for the soft eigenvalue is given by Eq.\eqref{eq:lambda_odd_eval} with $A(i\omega)=i\omega$.
\begin{equation}
    \begin{split}
      &\lambda_m^\text{odd}(i\Omega)=\frac{2g^2\calN}{\Omega/(2\pi)} \int_{-\Omega/2}^{\Omega/2}\frac{\rd \omega\rd \omega'}{(2\pi)^2}\left(-\frac{2\pi \gamma^{1/3}}{3\sqrt{3}}|\omega-\omega'|^{1/3}\right)\\
      &\times\frac{1}{k_F^3} \frac{m^2(m^2-1)^2}{4k_F^2 v_F^2}\times(-1)\left[\omega^2+\Omega^2/4+\omega\omega'\right]\,.
    \end{split}
\end{equation} This integral can be converted to summation and evaluated in a similar way, which yields Eq.\eqref{eq:lambda_odd_B_main}.

\section{Soft modes of the large-angle scattering FL (D)}\label{app:softD}

In this section, we try to give a qualitative discussion of the soft mode eigenvalues in the large-angle scattering FL (D). The expansion we developed in Sec.~\ref{sec:Lexpand} does not apply to this regime and needs to be modified.   As discussed in Appendix.~\ref{app:saddle}, we should substitute the boson propagator
\begin{equation}\label{}
      D(i\Omega,\vn{q})=\frac{1}{m_b^2+\gamma\frac{|\Omega|k_F}{\vn{q}\sqrt{k_F^2-\vn{q}^2/4}}}\,.
\end{equation}

The expression for the eigenvalues also needs to be modified. First, the singularity of the Jacobian near $\vn{q}\sim 2k_F$ should be taken into account. Second, the $m$-dependence is also modified due to the fact $m|q|/k_F\sim \calO(1)$. Therefore, the expression for the soft-mode eigenvalues is replaced by
\begin{equation}\label{eq:lambda_even_D}
    \begin{split}
      &\lambda_m^{\text{even}}(i\Omega)=\frac{2g^2\calN}{\Omega/(2\pi)}\int_{-\Omega/2}^{\Omega/2}\frac{\rd\omega\rd \omega'}{(2\pi)^2}\\
      &\times\int_0^{2k_F}\frac{\rd\vn{q}}{\sqrt{k_F^2-\vn{q}^2/4}}D(\vn{q},i\omega-i\omega')\\
      &\times\left(1-T_m\left(1-\frac{\vn{q}^2}{2k_F^2}\right)\right)\,.
    \end{split}
\end{equation} Here we have re-summed the $m^2\vn{q}^2/k_F^2$ back to the Chebyshev polynomial. The integral is estimated by expanding $D$ in powers of $m_b^2$. The leading order is obtained by setting $D=1/m_b^2$, and the $\vn{q}=2k_F$ singularity is integrable so we obtain a result linear in $\Omega$. However, at the next order we need to set $D\propto |\Omega|/\vn{q}\sqrt{k_F^2-\vn{q}^2/4}$, which make the $\vn{q}=2k_F$ singularity log-divergent. This log needs to be cutoff by the Chebyshev polynomial term and a scale of $\vn{q}/(2k_F)\sim 1-1/m$, and we obtain the estimate
\begin{equation}\label{}
  \lambda_m^\text{even}\sim A_0\Omega+B_0\Omega^2\ln m\,,m\gg 1\,.
\end{equation} At finite $T$, we expect the $\Omega^2$ term will be replaced by $T^2$, but the $\Omega$ term does not change.

  Proceeding to the odd-$m$ soft modes, we have
  \begin{equation}\label{eq:lambda_odd_D_app}
    \begin{split}
      &\lambda_m^\text{odd}=\frac{2g^2\calN}{\Omega/(2\pi)} \int_{-\Omega/2}^{\Omega/2}\frac{\rd \omega\rd \omega'}{(2\pi)^2}\int_{-\infty}^\infty\frac{\rd \xi\rd \xi'}{(2\pi)^2}\\
      &\int_0^{2k_F}\frac{\rd\vn{q}}{\sqrt{k_F^2-\vn{q}^2/4}}{\mathcal K}_m\left(\vn{q}/k_F\right)\frac{(m^2-1)^2(\xi+\xi')^2}{8k_F^2 v_F^2}\\
      &\times D(\vn{q},i\omega-i\omega')\left[iG(i\omega+i\Omega/2,\xi)-iG(i\omega-i\Omega/2,\xi)\right]\\
      &\times\left[iG(i\omega'+i\Omega/2,\xi')-iG(i\omega'-i\Omega/2,\xi')\right]\,.
    \end{split}
    \end{equation} Here $\mathcal{K}_m(\vn{q}/k_F)$ that encodes the $\vn{q}$-dependence, and it reduces to $m^2\vn{q}^2/k_F^2$ at small $\vn{q}$. Its exact form is not important, as long as it is oscillating at large $\vn{q}$. The integral can be estimated similarly as the even case, with the result
    \begin{equation}\label{}
      \lambda_m^\text{odd}\sim A_1\Omega^3+B_1  \Omega^4(m^2-1)^2\ln m \,,
    \end{equation} where the log term is due to the Landau damping term being cut off by the oscillating $\mathcal{K}_m$. At finite $T$, we expect that $\Omega^4$ is replaced by $T^4$.

    Our qualitative estimate here reproduces the results of \cite{PJLedwith2019} which directly computed the collision integral of the Boltzmann equation.

\section{Cutoff dependence in analytic continuation}\label{app:continue}

    Here we discuss the analytic continuation of the soft eigenvalues computed in Sec.~\ref{sec:soft}. At zero temperature, the soft eigenvalues can be written as a sum of powers, and we consider a particular term among them
\begin{equation}\label{}
  \lambda(i\Omega)=A \Omega^{\alpha}\,,\quad (\Omega>0)\,.
\end{equation} Here $A$ is a real constant. As we discussed in Sec.~\ref{sec:soft}, in the computation of $\lambda(i\Omega)$ we used dimensional regularization and the cutoff-dependent terms are dropped. We will see that these cutoffs dependencies are reinstated after we consider analytic continuation.

Given a retarded response function $\chi(z)$ analytic in the upper half plane, it is usually analytically continued to the whole complex plane using the Kramers Kronig relation
\begin{equation}\label{eq:continue1}
  \chi(z)=\int_{-\infty}^{\infty}\frac{\rd \omega}{\pi}\frac{-\Im \chi(\omega)}{z-\omega}\,.
\end{equation} However, there is another similar relation which utilizes $\Re\chi(\omega)$ that is also mathematically correct
\begin{equation}\label{eq:continue2}
  \chi(z)=\int_{-\infty}^{\infty}\frac{\rd \omega}{\pi}\frac{i\Re \chi(\omega)}{z-\omega}\,.
\end{equation}

Although Eq.\eqref{eq:continue1} and Eq.\eqref{eq:continue2} agree in the upper half-plane, they are very different in the lower half-plane. For $z$ in the upper half plane, $\chi(z)$ from Eq.\eqref{eq:continue1} satisfies $\chi(z)=\chi(z^*)^*$, but $\chi(z)$ from Eq.\eqref{eq:continue2} satisfies $\chi(z)=-\chi(z^*)^*$. In the usual physics literature, the response function is defined with the convention that $\chi_R(\omega)=\chi_A^*(\omega)$, so Eq.\eqref{eq:continue1} is most widely used.


However, the soft mode eigenvalue $\lambda(i\Omega)$ does not follow the \eqref{eq:continue1} convention. To deduce its analytic continuation, we can look at how it appears in bosonic correlation functions, which follows the usual analytic continuation convention. Since $\lambda(i\Omega)$ appears in bosonic correlation functions in the form of Eq.\eqref{eq:Pim}, which can be rewritten as
\begin{equation}\label{}
  \Pi(i\Omega)=C\frac{1}{1+\lambda(i\Omega)/\Omega}+D\,.
\end{equation} where $C,D$ are real constants. Therefore it is $h(i\Omega)=\lambda(i\Omega)/\Omega=A\Omega^{\alpha-1}$ that appears as a bosonic self-energy, and we should consider its continuation. This is equivalent to saying that $\lambda(i\Omega)$ should be continued using Eq.\eqref{eq:continue2}.

Under a naive analytic continuation, we would have
\begin{equation}\label{eq:h1}
\begin{split}
  &h_R(\omega)=A (-i\omega+0)^{\alpha-1}\\
  &=\left(\sin\frac{\pi \alpha}{2}+i\cos\frac{\pi\alpha}{2}\sgn \omega\right) A |\omega|^{\alpha-1}\,.
\end{split}
\end{equation}

We try to use Eq.\eqref{eq:h1} and continue back to the Matsubara axis by Kramers-Kronig relation. In particular, we consider
\begin{equation}\label{}
  \tilde{h}(i\Omega)=-\int_{-\Lambda}^{\Lambda} \frac{\rd z}{\pi}\frac{\Im h_R(z)}{i\Omega-z}=\frac{2A}{\pi}\cos\frac{\pi\alpha}{2}\int_0^{\Lambda}\rd z\frac{z^\alpha}{z^2+\Omega^2}\,.
\end{equation} Here $\Lambda$ is the UV cutoff.
The results below assume $\Omega>0$:
When $0<\alpha<1$, we obtain
\begin{equation}\label{}
  \tilde{h}(i\Omega)=A \Omega^{\alpha-1}\,.
\end{equation}
When $1<\alpha<3$, we obtain
\begin{equation}\label{}
  \tilde{h}(i\Omega)=A\Omega^{\alpha-1}+\frac{2A\cos\frac{\pi\alpha}{2}}{\pi(\alpha-1)}\Lambda^{\alpha-1}\,.
\end{equation}
When $3<\alpha<5$, we obtain
\begin{equation}\label{}
  \tilde{h}(i\Omega)=A \Omega ^{\alpha -1}-\frac{2 A \Omega ^2 \Lambda ^{\alpha -3} \cos \frac{\pi
   \alpha }{2}}{\pi  (\alpha -3)}+\frac{2 A \Lambda
   ^{\alpha -1} \cos \frac{\pi  \alpha }{2}}{\pi
   (\alpha -1)}\,.
\end{equation}

As we can see, in addition to the original power-law piece, even powers in $\Omega$  smaller than $\alpha-1$ are generated. However, these terms only affect $\Re h_R(\omega)$ and do not change $\Im h_R(\omega)$ when we continue $\Omega\to -i\omega+0$. Therefore, the real part of $\lambda_R(\omega)$, $\Re \lambda_R(\omega)=\omega \Im h_R(\omega)$ is independent of cutoff, and $\Im \lambda_R(\omega)=-i\omega\Re h_R(\omega)$ can contain odd powers of $\omega$ smaller than $\alpha$, with cutoff-dependent coefficients.

A lesson we learn from this calculation is that the spectral function used to perform analytic continuation ($\Im \chi$ in Eq.\eqref{eq:continue1} or $\Re\chi$ in Eq.\eqref{eq:continue2}) is independent of the UV cutoff of the Kramers-Kronig integral. This is because these spectral functions are defined through the discontinuity of the response function across the real axis. Since the UV cutoff can only generate a regular dependence on the complex frequency $z$, it cannot alternate the discontinuous part of $\chi(z)$.

\section{Vertex computation in the FL regime} \label{app:FLladder}

  In this appendix, we solve the integral equation for the ladder-resummed vertex $\Gamma(k,k')$ introduced in Eq.\eqref{eq:Gamma_ladder_FL}. Our analysis follows Ref.~\cite{DLMaslov2010a}.  Since the model we consider has rotation symmetry, it is convenient to expand the interaction vertices in angular harmonics
  \begin{equation}\label{}
    \Gamma(k,k')=\sum_{l}\Gamma_l(\xi,\omega,\xi',\omega')e^{il(\theta_k-\theta_{k'})}\,,
  \end{equation}
  \begin{equation}\label{}
    D(k,k')=\sum_l D_l(\xi,\omega,\xi',\omega') e^{il(\theta_k-\theta_k')}\,.
  \end{equation} Therefore, Eq.\eqref{eq:Gamma_ladder_FL} becomes
  \begin{equation}\label{eq:Gamma_ladder_app}
  \begin{split}
    &\Gamma_l(\xi,\omega,\xi',\omega')=Z^2 D_l(\xi,\omega,\xi',\omega')+\calN g^2\int\frac{\rd \omega''}{2\pi}\int\rd \xi''\\
    & D_l(\xi,\omega,\xi'',\omega'')\left[G(i\omega'',\xi'')^2\right]_\Omega \Gamma_l(\xi'',\omega'',\xi',\omega')\,.
  \end{split}
  \end{equation} This integral can be evaluated by assuming that both $\Gamma_m$ and $D_m$ do not strongly depend on $\xi''$, so that we only need to integrate over the $[G^2]_\Omega$ factor. The result is
  \begin{equation}\label{}
  \begin{split}
    \int\frac{\rd \xi''}{2\pi} [G^2(i\omega'',\xi'')^2]_\Omega=Z^2\frac{m^*}{m}\delta(\omega'')\,.
  \end{split}
  \end{equation} Here we calculated the integral by dropping the incoherent piece of $G$. Therefore, Eq. \eqref{eq:Gamma_ladder_app} can be closed if we consider $\omega=\omega'=0$\,, and we obtain
  \begin{equation}\label{}
    \Gamma_l=\frac{Z^2 D_l}{1-\frac{Z^2 m^* \calN g^2}{m}D_l}\,.
  \end{equation} Therefore we obtain an $l$-dependent renormalization
  \begin{equation}\label{}
    \frac{1}{Z_\Gamma^{(l)}}=\frac{1}{1-\frac{Z^2 m^* \calN g^2}{m}D_l}\,.
  \end{equation} For comparison, we perform the same computation for the $Z$ factor \eqref{eq:Z_FL} and we obtain
  \begin{equation}\label{}
    \frac{1}{Z}=\frac{1}{1-\frac{Z^2 m^* \calN g^2}{m}D_0}\,.
  \end{equation} To compare $Z_\Gamma^{(l)}$ and $Z$, we look at the explicit form of $D_l$
  \begin{equation}\label{}
  \begin{split}
    &D_l(\xi=0,\omega,\xi
    =0,\omega')=2\int_0^{\pi}\frac{\rd \theta}{\pi}\\
    & \frac{e^{il\theta}}{4k_F^2\sin^2\frac{\theta}{2}+m_b^2+\gamma\frac{|\omega-\omega'|}{2k_F\sin\frac{\theta}{2}}}\,.
  \end{split}
  \end{equation} In the critical FL regime described in \cite{DLMaslov2010a}, we have $m_b\ll k_F$, and forward scattering dominates, so the integral is dominated by contributions near $\theta\approx 0$. Therefore, $D_l$ is only weakly dependent on $l$ for $l\ll k_F/m_b$, therefore $Z_\Gamma^{(l)}\approx Z_\Gamma^{(0)}=Z$.

  The same reasoning can be utilized to simplify the effective action in Eq.\eqref{eq:SFL} to a simpler form in Eq.\eqref{eq:SFL_final}. Introduce the notation $\Psi(\theta)=(\Omega+i\vec{v}_k\cdot\vec{p})\phi(\theta;p)$, the first and the last term in the bracket of Eq.\eqref{eq:SFL} can be written as
  \begin{equation}\label{}
    \frac{\Psi(\theta)}{Z}-g^2\int\frac{\rd^3 k'}{(2\pi)^3}D(k-k')[G(k'^2)]_\Omega \frac{\Psi(\theta')}{Z}
  \end{equation} we can again perform angular harmonics decomposition, and the result is
  \begin{equation}\label{}
    \sum_{l} e^{il\theta}\frac{Z_\Gamma^{(l)}}{Z}\Psi_l\,.
  \end{equation} If $\Psi(\theta)$ only has low angular harmonics, we can approximate $Z_\Gamma^{(l)}\approx Z$, and we obtain Eq.\eqref{eq:SFL_final}.

\section{Vanishing of Aslamazov-Larkin diagram in FL theory} \label{app:FLAL}

    In this appendix, we compute the contribution of Aslamazov-Larkin diagrams and show that they vanish in FL theory. Here we will assume that the boson propagator $D(\vec{q})$ is static. The AL contribution to the interaction vertex is given by
    \begin{equation}\label{}
    \begin{split}
      &\Gamma^{\text{AL}}(k_1,k_2)=-g^4\int\frac{\rd^3 q}{(2\pi)^3} D(q)^2\\
      &\times G(k_1+q)\left[G(k_2+q)+G(k_2-q)\right]\,.
    \end{split}
    \end{equation} Here $q=(i\nu,\vec{q})$. We consider only the quasiparticle part of the fermion Green's function, by setting
    \begin{equation}\label{}
      G(i\omega,\xi)=\frac{Z}{i\omega-\frac{m}{m^*}\xi}\,.
    \end{equation}  We evaluate the $\nu$-integral first, and we obtain
    \begin{equation}\label{}
    \begin{split}
        & \Gamma^{\text{AL}}(k_1,k_2)=-g^4 Z^2 \int\frac{\rd^2\vec{q}}{(2\pi)^2} D(\vec{q})^2  \\
         & \left[\frac{\theta(-\xi_{k_1+q})-\theta(-\xi_{k_2+q})}{i(\omega_2-\omega_1)+\xi^*_{k_1+q}-\xi^*_{k_2+q}}\right.\\
         &+\left.\frac{\theta(-\xi_{k_1+q})-\theta(\xi_{k_2-q})}{i(\omega_1+\omega_2)-\xi^*_{k_1+q}-\xi^*_{k_2-q}}\right]\,.
    \end{split}
    \end{equation}Here $\xi^*_k=(m/m^*)\xi_k$. According to FL theory, we restrict to states exactly on the FS, by setting $\omega_1=\omega_2=\xi_1=\xi_2=0$. We also linearize the dispersion in small $\vn{q}$, and we obtain
    \begin{equation}\label{}
    \begin{split}
        & \Gamma^{\text{AL}}(\theta_1,\theta_2)=-g^4 Z^2 \int\frac{\rd^2\vec{q}}{(2\pi)^2} D(\vec{q})^2  \\
         & \left[\frac{\theta(-\cos\theta_{1q})-\theta(-\cos\theta_{2q})}{ v_F^*\vn{q}\left(\cos\theta_{1q}-\cos\theta_{2q}\right)}\right.\\
         &+\left.\frac{\theta(-\cos\theta_{1q})-\theta(-\cos\theta_{2q})}{-v_F^*\vn{q}\left(\cos\theta_{1q}-\cos\theta_{2q}\right)}\right]=0\,.
    \end{split}
    \end{equation} Here $\theta_{1q}$ is the angle between $\vec{k}_1$ and $\vec{q}$, and $\theta_{2q}$ is the angle between $\vec{k}_2$ and $\vec{q}$.


\section{Crossover from FL to Eliashberg theory} \label{app:FLtoME}

  In this section, we demonstrate the crossover from Landau's FL theory to Eliashberg theory by computing the function $\calF$ defined in Eq.\eqref{eq:LMT_F}. The integral that defines $\calF$ is
\begin{equation}\label{eq:Fint1}
\begin{split}
   &\calF  = \int_{-\Lambda_q}^{\Lambda_q}\frac{\rd \xi'}{2\pi} \int_{-\infty}^{\infty}\rd \omega' \frac{1}{1+\kappa_q|\omega-\omega'|}\\
   &\times\left[\frac{1}{i\omega'+i\Omega/2-\xi'-\vec{v}_{k'}^*\cdot\vec{p}}-\frac{1}{i\omega'-i\Omega/2-\xi'+\vec{v}_{k'}^*\cdot\vec{p}}\right]\,.
   \end{split}
\end{equation} Here the $1/(1+\kappa_q|\omega-\omega'|)$ arises from the boson Green's function, and the rest of the $\vn{q}$ dependence has been factored out (see main text). We also assume that the inelastic fermion self-energy is small and can be neglected. The parameter $\kappa_q$ is implicitly frequency-dependent
\begin{equation}\label{}
  \kappa_q=\frac{\gamma}{\sqrt{\vn{q}^2+|\omega-\omega'|^2/v_F^2}\left(\vn{q}^2+m_b^2\right)}\,.
\end{equation} When $|\omega-\omega'|\ll \Lambda_q=v_F \vn{q}$, we can approximate $\kappa_q$ by a constant the Landau damping non-analyticity is manifest. When $|\omega-\omega'|\gg \Lambda_q$, $\kappa_q\sim 1/|\omega-\omega'|$ and the Landau damping term becomes a constant. In the latter case, the boson Green's function $D(i\Omega,\vn{q})$ is static and therefore analytic in $\omega$ and the integral can be handled using standard FL techniques. Since we are interested in the crossover out of FL theory, we will treat $\kappa_q$ as frequency-independent and implement a cutoff on the dispersion integral using $\Lambda_q$.

We proceed to evaluate Eq.\eqref{eq:Fint1} at $T=0$. We deform the $\omega'$-integral into the complex $z=i\omega'$ plane, and the contribution comes from three branch cuts at $\Im z=-\Omega/2$, $\Omega/2$ and $\omega$. We also analytically continue $\omega$ and $\Omega$ to the real axis by setting $i(\omega+\Omega/2)\to \omega+\Omega/2+i0$ and $i(\omega-\Omega/2)\to \omega-\Omega/2-i0$. The result can therefore be written into three parts $\calF=\calF_1+\calF_2+\calF_3$, where
\begin{equation}\label{}
  \calF_1 =\int_{-\Lambda_q}^{s/2}\rd \xi (-i) \bar{D}_+(\xi-s/2-(\omega-\Omega/2))\,,
\end{equation}
\begin{equation}\label{}
  \calF_2=\int_{-\Lambda_q}^{-s/2}\rd \xi i \bar{D}_{-}(\xi+s/2-(\omega+\Omega/2))\,,
\end{equation}
\begin{equation}\label{}
\begin{split}
  &\calF_3=\int_{-\Lambda_q}^{\Lambda_q} \rd \xi \int_{-\infty}^{0} \frac{\rd z}{2\pi i} \left[\bar{D}_{-}(z)-\bar{D}_{+}(z)\right]\\
  &\times\left[i\bar{G}_+(z+\omega+\Omega/2,\xi+s/2)\right.\\
  &\left.-i\bar{G}_-(z+\omega-\Omega/2,\xi-s/2)\right]\,.
\end{split}
\end{equation} Here $\bar{D}_{\pm}(\omega)=1/(1\mp i\kappa_q \omega)$, $\bar{G}_{\pm}(\omega,\xi)=1/(\omega-\xi\pm i0)$, and $s=\vec{v}_{k'}^{*}\cdot\vec{p}$. $\calF_1$ and $\calF_2$ arises from the pole of the fermion propagator, and $\calF_3$ comes from the branch cut of the Landau damping. We further decompose $\calF_3$ into two parts $\calF_3=\calF_{3a}+\calF_{3b}$, by writing $\bar{G}_{\pm}(\omega,\xi)=\mp i\pi \delta(\omega-\xi)+P\left[1/(\omega-\xi)\right]$, where the delta function piece goes to $\calF_{3a}$ and the principal value goes to $\calF_{3b}$.  We obtain $\calF=\calF_A+\calF_B$, where $\calF_B=\calF_{3b}$ and
\begin{equation}\label{}
\begin{split}
  &\calF_A=\calF_{1}+\calF_{2}+\calF_{3a}\\
  &=\frac{\ln\left(1+\kappa_q\left(\omega+\Omega/2\right)\right)+\ln\left(1-\kappa_q\left(\omega-\Omega/2\right)\right)}{\kappa}\\
  &-\frac{\tanh^{-1}\left[\kappa_q\left(is/2+i\Lambda_q-(\omega-\Omega/2)\right)\right]}{\kappa}\\
  &-\frac{\tanh^{-1}\left[\kappa_q\left(is/2-i\Lambda_q+(\omega+\Omega/2)\right)\right]}{\kappa}\,.
\end{split}
\end{equation} Here we have continued back to the Matsubara frequency. Here $\calF_A$ yields $\kappa_q\to 0$ and $\kappa_q \Lambda_q\to \infty$ limit reported in the main text. The remaining part $\calF_B$ is lengthy and we do not present it, but only to note that it does not contribute in the $\kappa_q\to 0$ or $\kappa_q\Lambda_q\to\infty$ limit.

\section{Optical Conductivity}\label{sec:conductivity}

In this section, we calculate the optical conductivity $\sigma$ at the homogeneous limit $\vec{p}=0$ in a bulk system. Rewriting the Kubo formula derived in \cite{HGuo2022a}, the optical conductivity before analytic continuation can be written as
\begin{equation}\label{}
  \sigma(i\Omega)=\frac{2\pi e^2}{\Omega}\braket{v_k|\frac{1}{L+\Omega}|v_k}\,.
\end{equation} Here $e$ is the fermion charge and $v_k$ is the velocity function.
We apply resolution of identity to the above expression, and we obtain
\begin{equation}\label{}
  \sigma(i\Omega)=\frac{2\pi e^2}{\Omega}\sum_{i} \frac{\braket{v_k|i}\braket{i|v_k}}{\braket{i|i}}\frac{1}{\Omega+\lambda_i}\,.
\end{equation} Therefore, we just need to enumerate all the eigenvectors $\ket{i}$ that has non-zero overlap with the velocity under the inner product \eqref{eq:innerprod}. We will assume the FS to be circular with a non-parabolic dispersion, and in Appendix.~\ref{app:noncircular} we discuss the case of non-circular FS qualitatively.

\subsection{The structure of optical conductivity in a circular FS}

We consider a circular FS with nonparabolic dispersion. We expand $\vn{k}$ as a function of $\xi_k$ around the Fermi level:
\begin{equation}\label{eq:kexpand_main}
      \vn{k}=k_F+\frac{\xi_k}{v_F}-\frac{\kappa}{2}\frac{\xi_k^2}{k_Fv_F^2}+\frac{\zeta}{2}\frac{\xi_k^3}{k_F^2v_F^3}+\mathcal{O}(\xi_k^4)\,.
\end{equation} Here $\kappa$ and $\zeta$ are dimensionless numbers that parameterize the dispersion. In a Galilean invariant system, $\kappa=\zeta=1$. The velocity is then
\begin{equation}\label{eq:vexpand_main}
  v_k=v_F+\frac{\kappa \xi_k}{k_F}+\frac{2\kappa^2-3\zeta}{2k_F^2v_F}\xi_k^2\,.
\end{equation} We will perform calculation to order $\mathcal{O}(\xi^2_k)$.

Now we decompose the velocity $\ket{v_k}$ into eigenvectors of $L$. With rotation symmetry, we only need to consider the eigenvectors of $L_1$. $L_1$ contains a zero mode which is the momentum $\ket{k}$. We decompose the velocity $\ket{v_k}=\ket{v_k^\parallel}+\ket{v_k^\perp}$ which is the projection along the momentum and the orthogonal complement.
The optical conductivity can then be decomposed into two parts:
\begin{equation}\label{}
  \sigma(i\Omega)=\sigma_D(i\Omega)+\sigma_i(i\Omega)\,,
\end{equation} where $\sigma_D$ is the Drude peak related to the conserved momentum, and $\sigma_i$ is the incoherent conductivity:
\begin{equation}\label{eq:sigmaDinner}
  \sigma_D(i\Omega)=\frac{2\pi e^2}{\Omega}\bra{v_k^\parallel}\frac{1}{L+\Omega}\ket{v_k^\parallel}\,,
\end{equation}
\begin{equation}\label{eq:sigmaiinner}
  \sigma_i(i\Omega)=\frac{2\pi e^2}{\Omega}\bra{v_k^\perp}\frac{1}{L+\Omega}\ket{v_k^\perp}\,.
\end{equation}

\subsection{The Drude Peak}

We first consider the Drude peak conductivity $\sigma_D(i\Omega)$. The parallel component $\ket{v_k^\parallel}$ is given by
\begin{equation}\label{}
  \ket{v_k^\parallel}=\ket{k}\frac{\braket{k|v_k}}{\braket{k|k}}\,.
\end{equation} Here $\ket{k}$ is the momentum mode whose functional form is given in Eq.\eqref{eq:kexpand_main}. Since $\ket{k}$ is an exact zero mode of $L$, we have
\begin{equation}\label{}
  \sigma_D(i\Omega)=\frac{2\pi e^2}{\Omega^2}\frac{\braket{v_k^\parallel|k}\braket{k|v_k^\parallel}}{\braket{k|k}}=\frac{2\pi e^2}{\Omega^2}\frac{\braket{v_k|k}\braket{k|v_k}}{\braket{k|k}}\,.
\end{equation} Using Eqs.\eqref{eq:kexpand_main} and \eqref{eq:vexpand_main}, we can calculate the overlap above to be
    \begin{equation}\label{}
      \frac{\braket{v_k|k}\braket{k|v_k}}{\braket{k|k}}=v_F^2\braket{1|1}+\frac{2\kappa^2+2\kappa-3\zeta-1}{k_F^2}\braket{\xi|\xi}+\dots\,,
    \end{equation} where $\dots$ denote terms of higher power in $\xi/(k_Fv_F)$. Here $\braket{1|1}$ and $\braket{\xi|\xi}$ are evaluated according to the inner product \eqref{eq:innerprod}:
    \begin{equation}\label{}
    \begin{split}
      &\braket{1|1}=\int\frac{\rd\omega\rd^2\vec{k}}{(2\pi)^3}i\left[G(i\omega+\Omega/2,\vec{k})-G(i\omega-\Omega/2,\vec{k})\right]\\
      &=\calN \frac{\Omega}{2\pi}\,,
    \end{split}
    \end{equation}
    \begin{equation}\label{}
    \begin{split}
      \braket{\xi|\xi}&=\int \frac{\rd \omega\rd^2 \vec{k}}{(2\pi)^3} i\left[G(i\omega+\Omega/2,\vec{k})-G(i\omega-\Omega/2,\vec{k})\right]\xi_k^2\\
      &=\frac{\calN}{2\pi}\int_{-\Omega/2}^{\Omega/2}\rd \omega \frac{A(i\omega+i\Omega/2)^2+A(i\omega-i\Omega/2)^2}{2}\,.
    \end{split}
    \end{equation} Note here that the $\xi_k$-integrals are evaluated first, and $A(i\omega)=i\omega-\Sigma(i\omega)$.

    Therefore, the Drude peak conductivity is
    \begin{equation}\label{eq:sigmaD_circ}
      \sigma_\text{D}(i\Omega)=\frac{e^2\calN v_F^2}{2}\frac{1}{\Omega}\left[1+\left(2\kappa^2+2\kappa-3\zeta-1\right)\frac{\braket{\xi|\xi}/\braket{1|1}}{v_F^2 k_F^2}\right]\,.
    \end{equation}  Upon analytic continuation $i\Omega=\omega+i0$, the first term generates the usual delta function, and the second term yields the correction to the Drude weight.

    In the PNFL regime and the FL regimes, the correction term is insignificant and does not contribute to dissipation (the real part $\Re\sigma_D(\omega)$). To see this, we utilize the fact that $\Im \Sigma(\omega)\ll \omega$ in these regimes and substitute $A(i\omega)\approx i\alpha\omega$, and we obtain
    \begin{equation}\label{eq:sigmaBCD}
    \begin{split}
      &\sigma_D^{B,C,D}(\omega)=\frac{e^2 \calN v_F^2}{2}\frac{1}{-i\omega}\\
      &\left[1+\left(2\kappa^2+2\kappa-3\zeta-1\right)\frac{\alpha^2(\omega^2+\pi^2 T^2)}{3k_F^2 v_F^2}\right]\,,
    \end{split}
    \end{equation} where $\alpha=(1+\pi c_f')$ in the FL regimes (C,D) ($c_f'$ is defined in Eqs.\eqref{eq:cfp_main} and \eqref{eq:omegaFLD_main}) and $\alpha=1$ in the PNFL regime (B). Here the temperature dependence is obtained by replacing the frequency integral by Matsubara sum, and  we have analytically continued to real-frequency, and we see that the frequency-dependent correction term does not contribute to the real part of $\sigma_D(\omega)$:
    \begin{equation}\label{}
    \begin{split}
      &\Re\sigma_D^{B,C,D}(\omega)=\frac{e^2\calN v_F^2}{2}\pi\delta(\omega)\\
      &\left[1+\left(2\kappa^2+2\kappa-3\zeta-1\right)\frac{\alpha^2\pi^2 T^2}{3k_F^2 v_F^2}\right]\,.
    \end{split}
    \end{equation}

    However, in the NFL regime (A), the correction term becomes non-analytic as $A(i\omega)\approx -\Sigma(i\omega)$. Substituting the self-energies, we obtain (at $T=0$)
    \begin{equation}\label{}
    \begin{split}
      &\sigma_D^{A}(\omega,T=0)=\frac{e^2 \calN v_F^2}{2}\frac{1}{-i\omega}\\
      &\left[1-\left(2\kappa^2+2\kappa-3\zeta-1\right)\frac{c_f^2 z_b}{z_b+2}\left(-i\omega\right)^{4/z_b}\right]\,.
    \end{split}
    \end{equation}  Here we have treated the boson dynamical exponent $z_b$ as a tuning parameter by modifying the Green's function $D^{-1}=\vn{q}^{z_b-1}-\Pi$ to access the NFL regime away from the instability. As discussed in Sec.~\ref{sec:stability}, the NFL regime is stable for $2<z_b<3$. For general $z_b$ the relevant fermion self-energies are calculated in Appendix.~\ref{app:saddle} and $c_f$ is given by Eq.\eqref{eq:cfz}. Due to the thermal mass $\Delta(T)$, the conductivity $\sigma^A_D$ in the NFL regime (A) does not satisfy the $\omega/T$ scaling. The leading temperature dependence is then
    \begin{equation}\label{}
    \begin{split}
      &\sigma_D^{A}(\omega,T>0)=\frac{e^2 \calN v_F^2}{2}\frac{1}{-i\omega}\\
      &\left[1-\left(2\kappa^2+2\kappa-3\zeta-1\right)\frac{\Gamma(T)^2}{4k_F^2 v_F^2}\right]\,,
    \end{split}
    \end{equation} where $\Gamma(T)$ is given in Eq.\eqref{eq:Gammaz} if we consider $2<z_b<3$, or by \eqref{eq:SigmaT} if we consider $z_b=3$.

    The real part is then
    \begin{equation}\label{eq:ResigmaA}
    \begin{split}
      &\Re\sigma_D^{A}(\omega,T=0)=\frac{e^2\calN v_F^2}{2}\\
      &\left[\pi\delta(\omega)-\left(2\kappa^2+2\kappa-3\zeta-1\right)\frac{c_f^2 z_b}{z_b+2}\frac{\sin\frac{2\pi}{z_b}}{k_F^2v_F^2} |\omega|^{4/z_b-1}\right]\,,
    \end{split}
    \end{equation}
    \begin{equation}\label{eq:ResigmaAT}
    \begin{split}
      &\Re \sigma_D^A(\omega\to 0,T>0)=\frac{e^2 \calN v_F^2}{2}\pi\delta(\omega)\\
      &\left[1-\left(2\kappa^2+2\kappa-3\zeta-1\right)\frac{\Gamma(T)^2}{4k_F^2 v_F^2}\right]\,.
    \end{split}
    \end{equation}

    The correction term to the Drude-peak delta function exists only for non-Galilean invariant band structures. This is manifested by the prefactor $(2\kappa^2+2\kappa-3\zeta-1)$, which vanishes at the Galilean symmetric point $\kappa=\zeta=1$.

    The result \eqref{eq:ResigmaA}, only holds in the strongly coupled NFL regime (A), where the self-energy becomes much larger than the bare frequency term. The crossover from the more conventional result \eqref{eq:sigmaBCD} to the new result \eqref{eq:ResigmaA} is similar to the discussion  in Sec.~\ref{sec:model}, where the crossover lines are shown in Fig.~\ref{fig:pd} at $T=0$. As we move from the FL regime (C) to the PNFL regime (B) or the NFL regime (A), we can substitute the parameter $\alpha$ by a frequency-dependent value. When we go from C to B, we have $\alpha\approx 1$, whereas when we go from C to B, $\alpha\approx |\omega|^{-1/3}$, rendering the correction term non-analytic.

    The stability of the theory requires $\Re \sigma(\omega)>0$. For $2<z_b<3$, this implies $2\kappa^2+2\kappa-3\zeta-1<0$, which places a constraint on the dispersion of the fermion. This constraint is satisfied, for example, by the Dirac dispersion where $\kappa=\zeta=0$.

   The structure of our calculation is also formally similar to the memory matrix formalism \cite{XWang2019,SAHartnoll2014}, where the Drude conductivity takes the form
    \begin{equation}\label{eq:sigmaMM}
      \sigma_D(\omega)=\chi_{J,P}\frac{1}{M(\omega)-i\omega \chi_{P,P}}\chi_{P,J}\,.
    \end{equation} Here $P$ is the momentum operator and $J$ is the current operator, $\chi$'s are the susceptibilities and $M$ parameterizes dissipation.  Comparing Eq.\eqref{eq:sigmaDinner} with Eq.\eqref{eq:sigmaMM}, there is a formal similarity if we identify $\chi_{JP}=\braket{v_k|k}$ and $M-i\omega\chi_{P,P}=\Omega+L$. While in the memory matrix formalism the susceptibility $\chi_{JP}$ is assumed to be independent of the external frequency $\omega$, in our formalism the overlap $\braket{v_k|k}$ is explicitly dependent on $\omega$.

    \subsection{The incoherent part}

    Next, we evaluate the incoherent part of the conductivity $\sigma_i$ as defined in Eq.\eqref{eq:sigmaiinner}. It is called incoherent because it does not arise from the coherent center-of-mass motion. The part of $\ket{v_k}$ that is orthogonal to the momentum $\ket{k}$ is
    \begin{equation}\label{eq:vkperp}
      \ket{v_k^\perp}=\ket{v_k}-\ket{k}\frac{\braket{k|v_k}}{\braket{k|k}}
      \approx (\kappa-1)\frac{\ket{\xi}}{k_F}\,.
    \end{equation} Here we have truncated to first order in $\xi/(k_F v_F)$, which is sufficient to calculate $\sigma_i$ to the order of $\xi^2/(k_F v_F)^2$.

    According to the discussion in Sec.~\ref{sec:dq0L0}, $\ket{v_k^\perp}\propto \ket{\xi}$ is not a zero mode of the kinetic operator $L$. However, thanks to the upper-triangular structure of $L$ (see Eq.\eqref{eq:dq0L0block}), the action of $L$ on $\ket{\xi}$ takes a simplified form, in particular,
    \begin{equation}\label{eq:L_xi}
      \bra{\xi}\frac{1}{\Omega+L}\ket{\xi}=\bra{\xi}\frac{1}{\Omega+\Lambda_\omega}\ket{\xi}\,,
    \end{equation} where $\Lambda_\omega=\left[i\Sigma(i\omega+i\Omega/2)-i\Sigma(i\omega-i\Omega/2)\right]$ is an operator diagonal in the frequency domain, defined in Eq.\eqref{eq:Lambdaomega}.

    Combining Eqs.\eqref{eq:sigmaiinner},\eqref{eq:vkperp},\eqref{eq:L_xi} and the definition of the inner product \eqref{eq:innerprod}, we write down the following expression for $\sigma_i(i\Omega)$:
    \begin{equation}\label{eq:sigmai_circ}
    \begin{split}
      \sigma_i(i\Omega)&=\frac{e^2 \calN v_F^2}{2}\left(\frac{\kappa-1}{v_Fk_F}\right)^2\frac{1}{\Omega}\int_{-\Omega/2}^{\Omega/2}\rd \omega\int\frac{\rd \xi}{2\pi}\\
      &\times\frac{(iG(i\omega+i\Omega/2,\xi)-iG(i\omega-i\Omega/2,\xi))\xi^2}{\Omega+\Lambda_\omega}\\
      &=\frac{e^2 \calN v_F^2}{2}\left(\frac{\kappa-1}{v_Fk_F}\right)^2\frac{1}{\Omega}\int_{-\Omega/2}^{\Omega/2}\rd \omega\\&
      \times \frac{1}{2}\frac{A(i\omega+i\Omega/2)^2+A(i\omega-i\Omega/2)^2}{\Omega+\Lambda_\omega}\,,
    \end{split}
    \end{equation}

    Now we evaluate Eq.\eqref{eq:sigmai_circ} in different regimes of Fig.~\ref{fig:pd}. In the FL regimes (C,D) and the PNFL regime (B), we approximate $A(i\omega)=i\alpha\omega$ as before. Since $\Sigma(i\omega)\ll i\omega$, as a corollary we also have $\Lambda_\omega\ll \Omega$, and we can therefore Taylor expand $1/(\Omega+\Lambda_\omega)$ in Eq.\eqref{eq:sigmai_circ}. The leading order term is linear in $\Omega$, which does not contribute to $\Re \sigma_i(\omega)$ after analytic continuation. We therefore focus on the next term,  and in the PNFL regime B we obtain
    \begin{equation}\label{}
      \sigma_{i}^{B}(i\Omega)=\frac{e^2 \calN v_F^2}{2}\left(\frac{\kappa-1}{v_Fk_F}\right)^2 C_B(z_b)c_f\Omega^{2/z_b}\,,
    \end{equation}
    \begin{equation}\label{eq:sigma_Re_B}
      \Re \sigma_i^{B}(\omega)=\frac{e^2 \calN v_F^2}{2}\left(\frac{\kappa-1}{v_Fk_F}\right)^2 C_B(z_b)c_f\cos\frac{\pi}{z_b} |\omega|^{2/z_b}\,.
    \end{equation}
    where
    \begin{equation}\label{}
      C_B(z)=\frac{z (z (2 z+3)+2)}{(z+1) (z+2) (3 z+2)}\,.
    \end{equation} At finite temperatures $T\gg \omega$, we can estimate the result by substituting $\Sigma(i\omega)\propto c_f T^{2/3}$, and we obtain
    \begin{equation}\label{}
      \Re \sigma_i^B(\omega\ll T)\sim \frac{e^2\calN v_F^2}{2}  \left(\frac{\kappa-1}{v_Fk_F}\right)^2\frac{c_f T^{2+2/z_b}}{|\omega|^2}\,.
    \end{equation} Here the $1/|\omega|^2$ factor remains.

    In the FL regimes, we obtain
    \begin{equation}\label{}
      \sigma^{C,D}_i(i\Omega)=-\frac{e^2 \calN v_F^2}{2}\left(\frac{\kappa-1}{v_Fk_F}\right)^2 \frac{c_f' \Omega^2}{\omega_\text{FL}}C_1\left[ \ln\frac{\omega_\text{FL}}{\Omega}+C_2\right]\,,
    \end{equation}
    \begin{equation}\label{}
      \Re \sigma^{C,D}_i(\omega)=\frac{e^2 \calN v_F^2}{2}\left(\frac{\kappa-1}{v_Fk_F}\right)^2 \frac{c_f' |\omega|^2}{\omega_\text{FL}}C_1\left[\ln\frac{\omega_\text{FL}}{|\omega|}+C_2\right]\,.
    \end{equation} Here $C_1=7/30$, while $C_2=-17/60$ in regime (C), and $C_2=13/60$ in regime (D).  At finite temperature, the above result becomes
    \begin{equation}\label{}
      \Re \sigma^{C,D}_i(\omega\ll T)\sim \frac{e^2 \calN v_F^2}{2}\left(\frac{\kappa-1}{v_Fk_F}\right)^2 \frac{c_f' T^4}{\omega_\text{FL} |\omega|^2} \ln\frac{\omega_\text{FL}}{T}\,.
    \end{equation}

    Finally, we consider the NFL regime (A). In this regime, power counting of the integral \eqref{eq:sigmai_circ} is very different, because $\Sigma(i\omega)\gg i\omega$, and we have
    \begin{equation}\label{}
    \begin{split}
      &\sigma_i^A(i\Omega)=\frac{e^2 \calN v_F^2}{2}\left(\frac{\kappa-1}{v_Fk_F}\right)^2\frac{1}{\Omega}\int_{-\Omega/2}^{\Omega/2}\rd \omega \\
      &\times \frac{1}{2}\frac{\Sigma(\omega+\Omega/2)^2+\Sigma(\omega-\Omega/2)^2}{i\Sigma(\omega+\Omega/2)-i\Sigma(\omega-\Omega/2)}\,.
    \end{split}
    \end{equation} The result is
    \begin{equation}\label{eq:sigma_Re_A}
      \Re\sigma_i^A(\omega)= -\frac{e^2 \calN v_F^2}{2}\left(\frac{\kappa-1}{v_Fk_F}\right)^2 c_f \cos\frac{\pi}{z_b}C_A(z_b) |\omega|^{2/z_b}\,.
    \end{equation} Here
    \begin{equation}\label{}
      C_A(z)=\int_{-1/2}^{1/2}\rd x \frac{(1/2+x)^{4/z}+(1/2-x)^{4/z}}{(1/2+x)^{2/z}+(1/2-x)^{2/z}}\,,
    \end{equation} which is a positive function for $2<z<3$. While Eq.\eqref{eq:sigma_Re_A} of the NFL regime A may appear similar to the PNFL result \eqref{eq:sigma_Re_B} at $T=0$, its crossover to finite temperature regime is different. We obtain
    \begin{equation}\label{}
      \Re \sigma_i^A(\omega\ll T)\sim  -\frac{e^2 \calN v_F^2}{2}\left(\frac{\kappa-1}{v_Fk_F}\right)^2 \Gamma(T)\,.
    \end{equation} Here we have replaced the fermion self-energy by the thermal part $\Sigma_T(i\omega)\propto \Gamma(T)$.

    We summarize our results in Table.~\ref{tab:sigma_circ}. The crossover between regimes B,C,D is similar to the discussion in Sec.~\ref{sec:model}, and the crossover lines at $T=0$ is the same as in Fig.~\ref{fig:pd}. The crossover between C,D occurs when the boson mass $m_b\sim k_F$. The crossover from C to A,B happens at the frequency scale $\omega\sim\omega_\text{FL}=m_b^3/\gamma$. Depending on whether $\omega>\omega_P$ or $\omega<\omega_P$ at that point , we crosses over to regime B and A respectively.

   In a previous paper by the author \cite{HGuo2022a}, it was found that the Maki-Thompson and the Aslamazov Larkin diagram cancel in the conductivity computation, leaving only a Drude peak. It was conjectured there that the subleading term would scale as $\delta\sigma(\omega)\propto |\omega|^0$ due to a naive application of the angular diffusion picture. The more careful computation in this work invalidates the previous conjecture, and clarified two crucial points not appreciated before: (a) The angular diffusion picture only apply to the soft modes discussed in Sec.~\ref{sec:soft}, but these soft modes do not contribute to conductivity in a circular FS. (b) A correction to the Drude result requires involving momenta slightly away from the FS, and requires a non-Galilean invariant band dispersion.

    Our results for the incoherent part in the regimes B,C,D $\Re \sigma_i$ agree with recent perturbative computations in Refs.\cite{SLi2023} and \cite{YGindikin2024}. The computations in \cite{SLi2023} and \cite{YGindikin2024} are performed in the FL regime with a finite boson mass $m_b$, and then the result is extrapolated to the QCP by replacing $m_b\propto |\omega|^{1/3}$. However,
     in \cite{SLi2023} and \cite{YGindikin2024} the fermion Green's function is taken to be of the free-particle form with quasiparticle residue $Z=1$, so the extrapolation can only reach the PNFL regime (B). As discussed in Sec.~\ref{sec:model}, if the effect of $Z$ factor (in our notation $c_f'$) is carefully tracked, there can be two crossovers, one is from FL (C) to PNFL (B), and the other one is from FL(C) to NFL (A).

     In the NFL regime (A), we obtain results different from \cite{SLi2023} and \cite{YGindikin2024}. In particular, the Drude peak actually obtains finite correction due to the non-analytic behavior of the overlap between current and momentum $\braket{k|v_k}$, and this correction $\Re\delta\sigma_D\propto |\omega|^{4/z_b-1}$ is actually more singular than the incoherent part $\Re \delta \sigma_i \propto |\omega|^{2/z_b}$. As we see in Table.~\ref{tab:sigma_circ}, the Drude-peak contributions are zero in the regimes B,C,D, and therefore it is missed in Refs.~\cite{SLi2023} and \cite{YGindikin2024}.

    We notice that all the corrections to the optical conductivity require breaking Galilean invariance. In the Galilean invariant limit $\kappa=\zeta=1$, all the correction terms vanish, and we are left with the Drude result $\Re\sigma(\omega)=e^2\calN^2 v_F^2(\pi/2)\delta(\omega)$.

    Our results also indicate that the optical conductivity is not a good probe for the low-energy physics of the critical FS. First, the correction term to the Drude peak is suppressed by higher powers in $1/(k_Fv_F)$, and it depends sensitively on the dispersion away from the FS. Second, in a circular FS,  the only soft mode that overlaps with the current is the momentum. However, since momentum is exactly conserved, the optical conductivity cannot reveal the rich physics related to the soft modes.

    \begin{table}[htb!]
      \centering
      \begin{tabular}{|c|c|c|c|}
        \hline
         Regimes & NFL(A) & PNFL(B) & FL(C,D) \\
        \hline
        \multirow{3}{*}{{$\displaystyle\frac{\Re\delta \sigma_D(\omega\gg T)}{e^2\calN^2 v_F^2}$}} & \multirow{3}{*}{$\displaystyle\frac{c_f^2|\omega|^{4/z_b-1}}{k_F^2 v_F^2}$}
        & \multirow{3}{*}{0} 
        &    \multirow{3}{*}{0}    
        \\
        & & &\\
        &&&\\
        \hline
        \multirow{3}{*}{$\displaystyle\frac{\Re \delta\sigma_D(\omega\ll T)}{e^2\calN^2 v_F^2}$} &
        \multirow{3}{*}{$\displaystyle\frac{\Gamma(T)^2}{k_F^2v_F^2}\delta(\omega)$}
         & \multirow{3}{*}{$\displaystyle\frac{T^2}{k_F^2 v_F^2}\delta(\omega)$}  & \multirow{3}{*}{$\displaystyle(1+\pi c_f')^2\frac{T^2}{k_F^2 v_F^2}\delta(\omega)$}  \\
        &&&\\
        &&&\\
        \hline
        \multirow{3}{*}{{$\displaystyle\frac{\Re \sigma_i(\omega\gg T)}{e^2\calN^2 v_F^2}$}} &
        \multirow{3}{*}{$\displaystyle\frac{c_f|\omega|^{2/z_b}}{k_F^2 v_F^2}$} &
        \multirow{3}{*}{$\displaystyle\frac{c_f|\omega|^{2/z_b}}{k_F^2 v_F^2}$} &
        \multirow{3}{*}{$\displaystyle\frac{c_f'|\omega|^2\ln(|\omega|/\omega_\text{FL})}{\omega_\text{FL}k_F^2v_F^2}$}\\
        &&&\\
        &&&\\
        \hline
        \multirow{3}{*}{{$\displaystyle\frac{\Re \sigma_i(\omega\ll T)}{e^2\calN^2 v_F^2}$}} &
        \multirow{3}{*}{$\displaystyle\frac{\Gamma(T)}{k_F^2 v_F^2}$}&
        \multirow{3}{*}{$\displaystyle\frac{c_fT^{2+2/z_b}}{|\omega|^2k_F^2 v_F^2}$} &
         \multirow{3}{*}{$\displaystyle\frac{c_f'T^4\ln(T/\omega_\text{FL})}{\omega_\text{FL}|\omega|^2k_F^2v_F^2}$}\\
        &&&\\
        &&&\\
        \hline
      \end{tabular}
      \caption{Corrections to the real part of the optical conductivity of a circular FS in different regimes. The zeroth order Drude peak $\Re\sigma_D^{(0)}(\omega)=\pi e^2\calN v_F^2 \delta(\omega)/2$ is subtracted. In the first line, the zeroes mean that $\Re \delta\sigma_D$ is purely imaginary.  Unimportant numerical prefactors are neglected.}\label{tab:sigma_circ}
    \end{table}

\section{Non-circular FS}\label{app:noncircular}

  In this appendix, we comment on the effects of relaxing the circular FS assumption. \update{We still assume the FS to be inversion-symmetric.}

  The self-energies of the system only involve a small patch of the FS, and therefore it is insensitive to the FS geometry. Although our computation for the self-energy in Appendix.~\ref{app:saddle} takes into account the whole FS, the important contribution still arises from a local patch and the result agrees with explicit computations within patch theories \cite{Iesterlis2021}.

   However, in computations which involve the cancellation between self-energy and vertex correction, the FS geometry can play a subtle effect.
    In the following, we discuss the effects on (a) the soft-mode eigenvalues, (b) the hydrodynamics, and (c) the optical conductivity. Since a non-circular FS cannot be treated analytically, our discussion will be qualitative.

  \subsection{Soft-mode eigenvalues}

  According to the discussion in Sec.~\ref{sec:soft}, the soft-mode eigenvalues can be estimated through the random-walk picture on the FS, where the eigenvalue $\lambda_m$ describes the diffusion-like dynamics of the FS deformation.

  The even-$m$ eigenvalues $\lambda_m^{\text{even}}$ describe the relaxation of the FS deformation due to head-on scattering. Since each random-walk event contributes independently to $\lambda_m^{\text{even}}$, its form should be independent to the FS geometry.

\begin{figure}
    \centering
    \includegraphics[width=0.9\columnwidth]{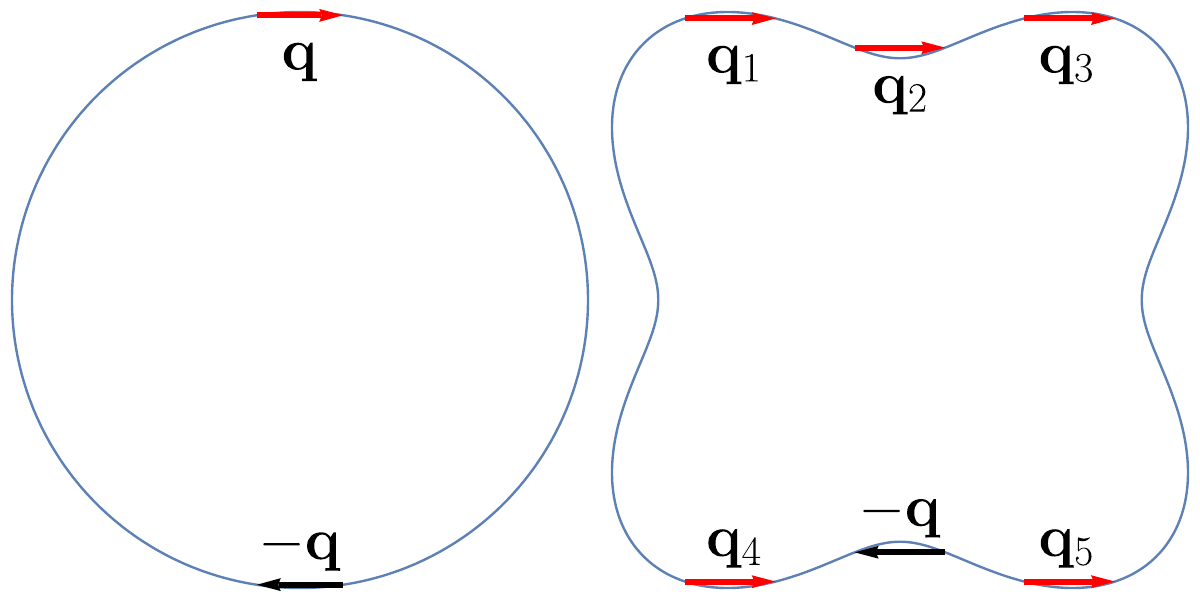}
    \caption{Pairs of scattering events that contribute to $\lambda_m^{\text{odd}}$. For a convex FS, only one pair of scattering events contribute. For a concave FS, multiple scattering events contribute. }\label{fig:convexconcave}
  \end{figure}

  However, the odd-$m$ eigenvalues $\lambda_m^{\text{odd}}$ depend sensitively on the FS geometry. Due to momentum conservation, in computing $\lambda_m^{\text{odd}}$ we need to consider a pair of random walk events with momentum transfer $\vec{q}$ and $\vec{-q}$ which conserve the total momentum. For a convex FS and a fixed $\vec{q}$, there can only be one pair of such random-walk events where a pair of head-on momenta is scattered to another head-on pair (see Fig.~\ref{fig:convexconcave}). The net contribution of this to the odd-harmonic deformation is zero, causing $\lambda_m^{\text{odd}}$ to be parametrically suppressed compared $\lambda_m^{\text{even}}$. However, for a concave FS, there can be multiple scattering events that contribute, as shown in Fig.~\ref{fig:convexconcave}. These different scattering events will not cancel, therefore we expect that $\lambda_m^{\text{odd}}$ will be parametrically similar to $\lambda_m^{\text{even}}$. Therefore, the instability discussed in Sec.~\ref{sec:stability} will not appear.

  \subsection{Hydrodynamics}

  For a convex FS, the hydrodynamics behave similarly to those of a circular FS. Due to the scale separation between $\lambda_m^{\text{odd}}$ and $\lambda_m^{\text{even}}$, there is a tomographic transport regime.

  For a concave FS, the hydrodynamics behave very differently. Since  $\lambda_{m}^{\text{odd}}\approx \lambda_m^{\text{even}}$, the tomographic regime does not exist. If we repeat the computation in Sec.~\ref{sec:hydro} with $\lambda_m^{\text{odd}}\propto \lambda_m^{\text{even}}$, we would find $\Gamma_2\propto \vn{p}v_F$. This is the signal for a ballistic transport regime. Therefore, the concave FS system shows the conventional hydrodynamic to ballistic crossover \cite{HGuo2017a}.

  \subsection{Optical conductivity}

    When the FS is not circular, the velocity $\ket{v_k}$ does not fully overlap with the momentum mode $\ket{k}$, even when we restrict $\vec{k}$ to be exactly on the FS.
   We decompose it as
     \begin{equation}\label{}
       \ket{v_k}=\ket{v_k^\parallel}+\ket{v_{k,\xi=0}^\perp}+\ket{v_{k,\xi}^\perp}\,.
     \end{equation} Here $\ket{v_k^\parallel}$ is still the projection along the momentum, but the orthogonal complement has been further decomposed into a $\xi$-independent part $\ket{v_{k,\xi=0}^\perp}$, and a $\xi$-dependent part $\ket{v_{k,\xi}^\perp}$.

     The total optical conductivity is the sum of the three contributions
     \begin{equation}\label{eq:sigma_noncirc}
      \sigma(\omega)=\sigma_D(\omega)+\sigma_D'(\omega)+\sigma_i(\omega).
    \end{equation}

     In Eq.\eqref{eq:sigma_noncirc}, $\ket{v_k^\parallel}$ and $\ket{v_{k,\xi}^\perp}$ still contribute to $\sigma_D$ and $\sigma_i$ respectively similar to the circular FS case discussed in Appendix.~\ref{sec:conductivity}, but the numerical prefactors are different because of different FS geometry.

     The new contribution is from $\ket{v_{k,\xi=0}^\perp}$. Since rotation symmetry is broken, $\ket{v_{k,\xi=0}^\perp}$ can in principle overlap with other odd-harmonic soft modes, which is not the momentum.
     It leads to a term that looks like a modified Drude peak
     \begin{equation}\label{eq:sigmaDp}
       \sigma_D'(i\Omega)=e^2\calN v_F^2 \frac{1}{\Omega+\lambda^\text{odd}}\,.
     \end{equation} Here $\lambda^\text{odd}$ is the typical odd-harmonic soft-mode eigenvalue computed in Sec.~\ref{sec:soft}.

     In a convex FS, the $\sigma_D'$ term is subleading compared to $\sigma_i$ in the power counting of both $|\omega|$ and $1/(k_Fv_F)$.

     However, in a concave FS, due to $\lambda^\text{odd}\sim \lambda^\text{even}\propto |\omega|^{4/z_b}$, we obtain a more singular contribution that scales like \begin{equation}\label{}
       \sigma_D'(\omega)\sim \frac{\lambda^\text{odd}}{|\omega|^2}\sim \frac{1}{|\omega|^{2-4/z_b}}\,.
     \end{equation} When $z_b=3$, this yields the $|\omega|^{-2/3}$ contribution proposed in \cite{YBKim1994a}. This result agrees with recent computations in \cite{SLi2023} and \cite{YGindikin2024}.


\bibliography{NFL,supp}
\end{document}